\documentclass[12pt, notitlepage]{article}
\usepackage[british]{babel}

\usepackage{amssymb,url,graphicx,csquotes}
\usepackage{hyperref}       % hyperlinks
\usepackage{caption}
\usepackage{amsmath}
\usepackage{newpxtext}
\usepackage{setspace}
\usepackage{multirow}

\usepackage{booktabs}       % professional-quality tables
\usepackage[all]{nowidow}
\setnoclub[2]
\usepackage{chngcntr} %appendix

\usepackage[top=1in,bottom=1in,right=1in,left=1in]{geometry}
\usepackage[utf8]{inputenc}
\providecommand{\keywords}[1]{\small\textbf{\textit{Keywords ---}} #1}
\providecommand{\jelcodes}[1]{\small\textbf{\textit{JEL ---}} #1}

\usepackage{pdfpages}
\usepackage{longtable}
\usepackage{url}            % simple URL typesetting
\usepackage{amsfonts}       % blackboard math symbolss
\usepackage{nicefrac}       % compact symbols for 1/2, etc.
\usepackage{microtype}      % microtypography
\usepackage{graphicx}       % figures
\usepackage{float}
\usepackage{lipsum}		% Can be removed after putting your text content

\usepackage[url=false,isbn=false, doi=false,style=apa,natbib,uniquename=false]{biblatex}
\usepackage{times} % Times New Roman

\begin{document}
\title{Profit Shifting of Multinational Corporations Worldwide}
\author{Javier \textsc{Garcia-Bernardo}\thanks{Institute of Economic Studies, Faculty of Social Sciences, Charles University, Prague, Czechia; Department of Methodology \& Statistics, Utrecht University, the Netherlands; Centre for Complex Systems Studies, Utrecht University, the Netherlands, javier.garcia.bernardo@gmail.com.} \\
Petr  \textsc{Janský}\thanks{Institute of Economic Studies, Faculty of Social Sciences, Charles University, Prague, Czechia, petr.jansky@fsv.cuni.cz.}  \\
%\ }
%\date{\today \\
}
\date{\textit{The working paper first published: May 2021; this revised version: December 2023. \thanks{We are grateful to Diego d'Andria, Michal Bauer, Jozef Baruník, Sebastian Beer, Katarzyna Bilicka, Julie Chytilová, Kimberly Clausing, Dharmika Dharmapala, Tim Dowd, Martin Gregor, Daniel Haberly, Tomáš Havránek, Martin Hearson, Tomáš Křehlík, Dominika Langenmayer, Jan Luksic, Filip Matějka, Ronen Palan, Miroslav Palanský, Nadine Riedel, Caroline Schimanski, Victor Steenbergen, Francis Weyzig and Ludvig Wier for excellent comments. We acknowledge support from from the Czech Science Foundation (CORPTAX, 21-05547M) and the International Centre for Tax and Development. This work was supported by the Cooperation Program at Charles University, research area Economics.}
}}
\maketitle

\begin{abstract} % 100 words for AEJ: EP
\noindent 
We exploit the new country-by-country reporting data of multinational corporations, with unparalleled country coverage, to reveal the distributional consequences of profit shifting.
We estimate that multinational corporations worldwide shifted over \$850 billion in profits in 2017, primarily to countries with effective tax rates below 10\%.
Countries with lower incomes lose a larger share of their total tax revenue due to profit shifting.
We further show that a logarithmic function is better suited for capturing the non-linear relationship between profits and tax rates than linear or quadratic functions.
Our findings highlight effective tax rates’ importance for profit shifting and tax reforms.

\end{abstract}

\bigskip
\bigskip
\noindent \keywords{multinational corporation, corporate taxation, profit shifting, effective tax rate, country-by-country reporting, global development}

\noindent \jelcodes{F23, H25, H26, H32}

%TODOS
%Number of countries: 216 I believe, not 196
%PJ: yes, would be great ot have this right and is it number of countries for which we present estimates of revenue loss/gain (and which method - the number probably differs across methods) or for which we have data that we analyse (but present results for a lower number

\newpage

\onehalfspacing

\section{Introduction}

Corporate tax avoidance contributes to the view of globalisation as inequitable. 
Publicised case studies, such as those based on the Panama and Paradise Papers, have detailed how little some large multinational corporations (MNCs) pay in corporate income tax as a result of their profit shifting to low-tax jurisdictions or tax havens. The case studies are not exceptions, with MNCs estimated to shift up to 40\% (\$600 billion–\$1.1 trillion) of their foreign profits to tax havens such as the Netherlands, Switzerland or Bermuda \citep{clausingEffectProfitShifting2016, torslovMissingProfitsNations2023}. Corporate tax avoidance is problematic because it affects the efficiency and equity of financial markets and societies \citep{slemrodTaxAvoidanceEvasion2002}. For example, large MNCs in the United Kingdom pay lower taxes than domestic firms \citep{bilickaComparingUKTax2019}. Despite recent growth in research interest in tax havens generally \citep{zucmanMissingWealthNations2013,johannesenEndBankSecrecy2014, alstadsaeterTaxEvasionInequality2019} and profit shifting in particular \citep{clausingProfitShiftingTax2020,demooijCostRealEffects2020,guvenenOffshoreProfitShifting2022, laffitteMultinationalsSalesProfit2022,garcia-bernardoDecomposingMultinationalCorporations2022a,vicardProfitShiftingReturns2023,torslovExternalitiesInternationalTax2023}, we still lack reliable estimates on the origin and destination of profit shifting for many countries worldwide, which is necessary to understand the potential effects of international tax reforms.
%A simplified description of such behaviour is that MNCs carry out their activities and generate profits in a variety of countries worldwide, but shift a large share of those profits to tax havens.
%lower tax burdens on top-income earners and higher tax burdens on the middle class

In this paper, we exploit a new dataset and develop a novel methodology to address the question of the distributional consequences of profit shifting of MNCs worldwide. Specifically, we answer the following four research questions: (i) What is the scale of profit shifting? (ii) Which tax havens are the largest? (iii) Which MNCs are the most aggressive in profit shifting? (iv) Which countries lose the most relative to their total tax revenues? These intrinsically linked research questions lack definitive answers due to both data-related and methodological challenges. Profit shifting as a form of tax avoidance cannot be directly observed and it is not clear how economists should estimate it. However, the data utilised as well as the choice of function to model the relationship between tax rates and profits have crucial implications for answers to our research questions. In this paper, we address these challenges by using a new country-by-country reporting (CBCR) dataset with vastly improved country coverage and by modelling the extreme non-linearity of that relationship---two contributions that we now outline before proceeding to our main findings.

%This paper develops a novel methodology and exploits a new dataset to address the question of the scale and distribution of profit shifting of MNCs worldwide. On the one hand, we ask which are the most important tax havens, the amount of profits shifted to them, and what share of the profits reported in tax havens have been shifted there. On the other hand, we address the question of which countries tend to lose more tax revenue from profit shifting, both absolutely and relative to their total tax revenue. In addition to country-level results, we investigate whether profit shifting affects countries differently according to their region or their income level. Moreover, we examine whether MNCs differ in the aggressiveness of their tax planning depending on the country of their headquarters. These intrinsically linked research questions lack definitive answers because of methodological and data challenges. In this paper we address these with two contributions.
%The current lack of definitive answers to these intrinsically linked questions is the result of methodological and data challenges that we address in this paper, with two innovations.

Our first contribution is to pioneer the use of CBCR by MNCs, alongside several other concurrent studies. These unique data were first made available in 2020 thanks to a new regulation, which emerged from the Base Erosion and Profit Shifting project by OECD, that requires all large MNCs to report profits and taxes in every country, including tax havens and low-income countries. For example, the US CBCR data include 25 African countries, while the frequently used US Bureau of Economic Analysis data only include three. 
%The published data aggregate thousands MNC groups with activities across almost 200 foreign countries and with headquarters across 28 countries. 
We provide our own estimates of profit shifting for a total of 214 countries using data from 38 headquarter countries, including the major economies of the United States, China and India.
While CBCR data include the most reliable country-level information on tax payments and profits of MNCs worldwide, CBCR data aggregate small countries into categories (e.g. ``Other Africa''), and might be prone to double-counting of profits due to a lack of clarity in reporting of intercompany dividends and so-called stateless entities. We address the double-counting of profits by developing a method for estimating missing data, disaggregating categories into individual countries, and eliminating double-counting of dividends. 
%In another often used dataset, Orbis, only 17\% of the global profits of MNCs could be located there in 2012 \citep{torslovMissingProfitsNations2023}, whereas the CBCR data includes the countries of 100\% of global profits by definition. 

Our second contribution is a methodological one: we propose to model the extremely non-linear relationship between profit location and MNCs' tax rates using a logarithmic function. We build on literature that confirmed the existence of profit shifting, pioneered by \textcite{grubertTaxesTariffsTransfer1991, hinesFiscalParadiseForeign1994}. The headline specification of that approach assumes a linear semi-elasticity, which \textcite{dowdProfitShiftingMultinationals2017} show to underestimate profit shifting to low-tax jurisdictions. \textcite{dowdProfitShiftingMultinationals2017} instead introduce a quadratic semi-elasticity, which does constitute an improvement. However, we show that not even the quadratic model is capable of capturing the empirically observed extreme non-linearity in the data: 85\% of profit shifted takes place towards countries with tax rates below 10\%. In this paper, we therefore introduce a logarithmic model to fully capture the extreme non-linearity of the semi-elasticity of profits to tax rates. Although estimates of the global scale of profit shifting are similar for linear, quadratic and logarithmic functions as well as a simpler misalignment model measuring the difference between locations of profits and economic activity, they differ substantially at the country level: the logarithmic function and the misalignment model point to profits being shifted relatively more to countries with zero or very low effective tax rates. This naturally has implications for the estimated distributional consequences of profit shifting to tax havens.

We apply the logarithmic model to the CBCR data to establish the scale and distribution of profit shifting in many countries worldwide, revealing four main findings that answer the four research questions outlined above. First, MNCs shifted over \$850 billion in profits to tax havens in 2017, which in turn implies \$200--300 billion in revenue losses for other countries. Our total estimates of profit shifting are broadly comparable to existing estimates such as \textcite{torslovMissingProfitsNations2023} and \textcite{wierGlobalProfitShifting2022}, who estimate profit shifting to be \$616 billion in 2015 and \$969 billion in 2019---see Table \ref{fig:torslov} for a more detailed comparison with \textcite{torslovMissingProfitsNations2023} and Table \ref{tab:trl_comparison} for a comparison with additional studies. 
By combining our modelling approach with the extreme non-linearity and exceptionally high coverage of our dataset (214 countries), we arrive at semi-elasticity estimates that are consistent with higher shares of profits in tax havens.
%Our estimates reconcile so called micro and macro estimates (e.g. \citealp{dharmapalaProfitShiftingGlobalized2019, wierDominantRoleLarge2022}) or the differences between the relatively low estimates of tax semi-elasticity in the earlier literature \citep{beerInternationalCorporateTax2020,heckemeyerMultinationalsProfitResponse2017} and the high shares of profits reported by MNCs in tax havens \citep{zucmanHiddenWealthNations2015}.

Second, we proceed to estimating which tax havens are the largest. The large majority of shifted profits are shifted to a small group of countries with extremely low effective tax rates (ETRs), defined as the ratio of accrued taxes over profits. 
We find that the Cayman Islands, Luxembourg, Singapore, Canada, the Netherlands, Switzerland, Hong Kong, Bermuda, Puerto Rico and Ireland are the largest destination of profits. In contrast with the consistent estimates across models of the overall scale, the largest tax havens---as well as the countries affected by them---differ substantially between models. The United Kingdom is seen as a source of profits in the misalignment model, while a recipient of profits (given its low tax rate) in other models. The misalignment and logarithmic model agree that the vast majority of profits in small tax havens (e.g., the Cayman Islands or Luxembourg) are shifted there, while the quadratic and linear models are not able to capture the extent of profit shifting. Moreover, high-income countries capture most of the tax revenue gains due to being destinations of profit shifting.

Third, among headquarter countries reporting on over 20 jurisdictions, MNCs headquartered in the United States, Brazil and Singapore are the most aggressive in terms of profit shifting. In contrast, we find no evidence of profit shifting towards tax havens by MNCs headquartered in South Africa and Malaysia. While a great deal of previous research has been carried out on US-headquartered MNCs due to data availability, our results highlight that they are not necessarily representative of all MNCs and that there are important differences across countries. Consequently, policymakers might negotiate international agreements differently if they know how aggressive their own MNCs are in comparison with other countries' MNCs with respect to profit shifting.

Fourth, we contribute to the ongoing discussion of which countries lose more tax revenue due to profit shifting. CBCR data has much higher country coverage (214 countries) and includes many lower-income countries for the first time. We find that it is precisely these lower-income countries that tend to lose more tax revenue due to profit shifting relative to their total tax revenue, directly contravening one of the goals of the 2030 Agenda for Sustainable Development, namely to: ``Strengthen domestic resource mobilization, including through international support to lower-income countries, to improve domestic capacity for tax and other revenue collection''. In absolute terms, the United States is estimated to suffer the most from profit shifting while other high-income countries such as Germany and France are estimated to lose up to half of their profit base in this manner. 

Overall, our paper's enhanced methodology and data provide a possible resolution to the inconsistency between so-called micro and macro estimates of profit shifting found in existing literature. Specifically, the relatively low (micro) estimates of tax semi-elasticity in earlier studies \citep{beerInternationalCorporateTax2020,heckemeyerMultinationalsProfitResponse2017} could not explain the (macro) estimates of the high shares of profits reported by MNCs in tax havens \citep{zucmanHiddenWealthNations2015, dharmapalaProfitShiftingGlobalized2019}. In this paper, we show that one explanation for the apparent inconsistency is the manner in which the relationship between profits and tax rates has been modelled using mostly linear and, far less frequently, quadratic functions \citep{dowdProfitShiftingMultinationals2017,muttiTaxesLocationUS2019}. When we instead use a logarithmic function to allow for the relationship's extreme non-linearity, we arrive at high estimates of tax semi-elasticity at low levels of ETRs and, consequently, a very high share of profits in a number of tax havens with low ETRs. While the CBCR data may need to be provided at firm level or for several years to facilitate even more nuanced findings, such as on the incidence or industry heterogeneity of profit shifting worldwide, our improved methodology and dataset do help reconcile the micro and macro estimates.

We structure the rest of the paper as follows. In Section~\ref{sec:methodology}, we introduce the new logarithmic model designed to estimate the scale and distribution of profit shifting to tax havens and compare it with existing specifications of the semi-elasticity model. We then describe how we reallocate the shifted profit from tax havens to other countries, as well as including the so-called misalignment model as an alternative to the semi-elasticity model. In Section~\ref{sec:data}, we discuss the available datasets used to estimate profit shifting, mainly the CBCR data. In Section~\ref{sec:results}, we show how our methodology improves profit shifting estimation using the US CBCR data, applies the methodology to the OECD CBCR data to obtain global estimates, and describes how profit shifting differs by countries' per capita income. In Section ~\ref{sec:results}, we also summarise several robustness checks and sensitivity analyses with which we show that our findings are robust to changes in the methodology. In Section~\ref{sec:conclusion}, we conclude the paper.

\section{Methodology for estimating profit shifting}\label{sec:methodology}

In this section we first introduce the traditional methodology for estimating profit shifting using linear and quadratic specifications (Section~\ref{sec:methodology_model}). We then detail our logarithmic specification as this paper's preferred way of estimating the scale of profit shifting to tax havens (Section~\ref{sec:methodology_log}). We proceed to describe how the shifted profit is reallocated from tax havens to other countries on the basis of economic activity (Section~\ref{sec:methodology_real}). Finally, we describe how we apply this logic of shifted profit reallocation to estimate the scale of profit shifting itself using the so-called misalignment model (Section~\ref{sec:methodology_mis}).

\subsection{Semi-elasticity model}\label{sec:methodology_model}

MNCs can, and many of them do, engage in shifting profit to tax havens where they seek lower taxation of their profits---a recent review of existing literature is provided by \textcite{beerInternationalCorporateTax2020}. The profit booked in a jurisdiction $i$ by MNCs ($\pi_i$) can be expressed as a sum of the ''real unobserved profits'' ($p_i$) and profits shifted into the jurisdiction ($S_i$) minus the cost of profit shifting incurred by the MNCs ($c_i$):

\begin{equation}\label{eq:1}
    \pi_i = p_i + S_i - c_i.
\end{equation}

While various methodologies have been used to estimate profit shifting (e.g. \cite{huizingaInternationalProfitShifting2008, weichenriederProfitShiftingEU2009, alvarez-martinezHowLargeCorporate2021, dharmapalaEarningsShocksTaxmotivated2013, crivelliBaseErosionProfit2016, auerbachInternationalTaxPlanning2017}), profit shifting is most frequently modelled using the method proposed by \textcite{hinesFiscalParadiseForeign1994}. This method assumes that the cost of profit shifting increases quadratically with the fraction of profit shifted (this assumption and other theory-related issues are discussed in more detail in section \ref{sec:appendix_theory}). The booked profits ($\pi$) are maximised subject to the existence of profit shifting. Subsequently, theoretical profits are identified with the Cobb-Douglas production function, yielding equation \ref{eq:3} for the first-order Taylor expansion and equation \ref{eq:4} for the first-order Taylor expansion around the two values of the Lagrange multiplier where profit shifting becomes zero (it is worth noting that the points where profit shifting becomes zero are far from the most interesting cases: tax havens, see section \ref{sec:appendix_theory}):

\begin{equation} \label{eq:3}
    \log{(\pi_i)} = \beta_0 + \beta_1
    \log{(K_i)} + \beta_2\log{(L_i)} +  \beta_3(\tau_i) + \beta_\chi \chi + \epsilon,
\end{equation}

%If instead a second order Taylor expansion is employed, we reach the following equation:
\begin{equation} \label{eq:4}
    \log{(\pi_i)} = \beta_0 + \beta_1\log{(K_i)} + \beta_2\log{(L_i)} +  \beta_3(\tau_i) + \beta_4 (\tau_i)^2 + \beta_\chi \chi + \epsilon,
\end{equation}

where $\pi_i$ represents profits booked in country $i$, including both real profit and profit shifted, and $K_i$ and $L_i$ are the capital and labour components of the Cobb-Douglas production function, usually operationalised with total tangible assets and wages. $\tau_i$ is either the tax rate of the subsidiary, the difference of tax rates between the subsidiary and the parent, or, less frequently (due to lacking data), between the subsidiary and other subsidiaries, and $\chi$ are controls including e.g. GDP per capita and population.

Both equations are currently viewed as traditional methods. However, follow-up studies use equation \ref{eq:3} and its modifications much more than equation \ref{eq:4}, even though \citealp{hinesFiscalParadiseForeign1994} noted that the results of \ref{eq:4} suggest that the effect is strongest at low tax rates. Recent research has revisited the possibility of significant curvature in the relationship between tax rates and reported profits. In particular, \textcite{dowdProfitShiftingMultinationals2017} apply equation \ref{eq:4} to a panel dataset of US tax returns spanning the 2002--2012 period and find that the effect of tax on profit shifting is not linear, namely that incentive to shift profits from a country with a tax rate of 20\% to one with a tax rate of 0\% is more than double compared to incentive to shift profits from a country with a tax rate of 20\% to one with a tax rate of 10\%. \textcite{dowdProfitShiftingMultinationals2017} account for this non-linearity by including a quadratic term.

\subsection{Addressing extreme non-linearity: A logarithmic model}\label{sec:methodology_log}

In this paper, we argue that the non-linearity of tax semi-elasticity is too extreme to be adequately accounted for using linear or quadratic models. We argue that the assumption of the quadratic relationship between the fraction of profit shifted and the cost of profit shifting, while suitable for the transfer pricing of physical goods where arm-length prices are more readily available, is not suitable for profit shifting strategies based on financial assets such as intellectual property or intra-group lending. In these strategies, the costs of profit shifting are largely fixed and do not differ significantly with profits. As such, the cost as a fraction of profits shifted is high for low fractions of profit shifted and subsequently decreases \citep{dischingerCorporateTaxesLocation2011}. Once a tax avoidance structure is in place (e.g. intellectual property located in a tax haven), we assume that the costs do not increase much with each additional dollar of profit shifted through it. Consequently, since they constitute a small share of their overall costs and are typically much lower than the tax avoided through profit shifting, these costs are minor for large MNCs (and only those are included in the CBCR data). By contrast, smaller companies may not find it viable to set up such tax avoidance structures at all and this dichotomy has been observed previously \citep{johannesenAreLessDeveloped2020, daviesKnockingTaxHaven2018}; indeed, large MNCs tend to be responsible for the bulk of profit shifting \citep{wierDominantRoleLarge2022}. Moreover, these costs are comparable regardless of which tax haven profits are shifted to. Firms thus have an incentive to shift profits to tax havens with the lowest effective taxation available, not merely to countries with lower ETRs, and, therefore, models including a logarithmic semi-elasticity would more effectively model profit shifting.

\begin{figure}[ht!]
 \caption{Profit per employee as a function of the ETR using the OECD data}
  \includegraphics[width=0.7\textwidth]{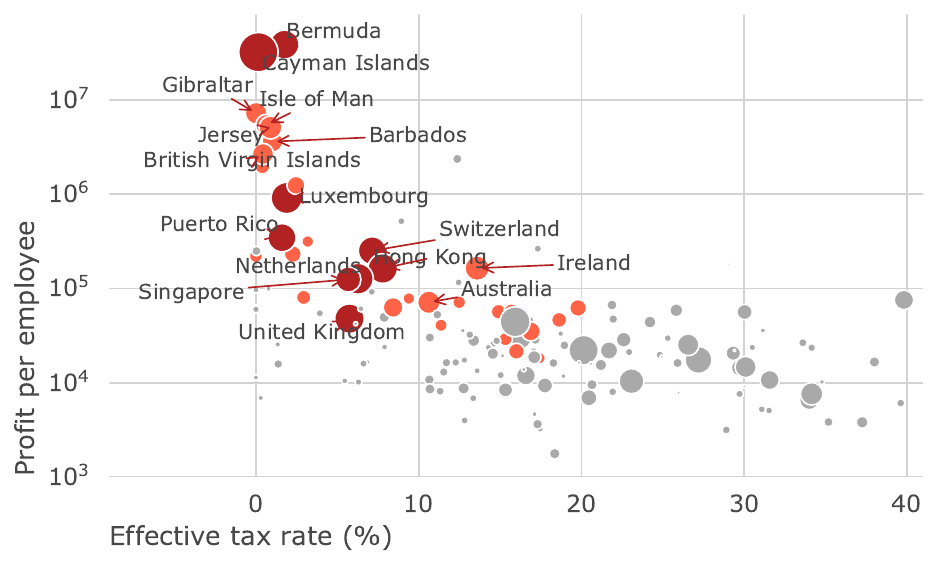}
  \label{fig:profit_employee_etr}
\footnotesize
\newline
Notes: Profit per employee as a function of the ETR using the OECD data. Colour indicates the ETR (below or above 10\%). Note that the horizontal axis is logarithmic, and as such the effect of effective tax rate on the profitability per employee, which is driven in a large part by profit shifting, is extremely non-linear. The horizontal axis is cut at 40\% to increase readability. Only nine countries have a tax rate above 40\% (Angola, Guyana, Malawi, Niger, Nigeria, Pakistan, Reunion, Rwanda, Tonga).
\end{figure}

This theoretical prediction is empirically backed by three observations. The first takes into account the extreme non-linearity of the profitability of firms in tax havens. The reported profit per employee is relatively constant around \$30,000 to \$50,000 per employee in all countries with an ETR over 10\%, and exponentially increases as the ETR falls below that level (Fig. \ref{fig:profit_employee_etr}). The second observation focuses on the empirical results by \textcite{dowdProfitShiftingMultinationals2017}. Their discontinuity model yields semi-elasticities for ETRs below 10\% which are twice as large as those yielded by their quadratic model, indicating that the extreme non-linearity is not fully captured using a quadratic term. The third observation is derived from our data-driven exploration of the data (see Appendix \ref{sec:appendix_eureqa}), where we use symbolic regression to obtain models which best fit the data. All of these models include a term which allows for extreme non-linearity semi-elasticities: a logarithmic term.

We argue that including a logarithmic term enables us to capture the extreme non-linearity better than the inclusion of a quadratic term, and we show this empirically in the results section. In order to model the extreme non-linearity, we propose to modify the equation as follows:

\begin{equation}\label{eq:5}
    \log{(\pi_i)} = \beta_0 + \beta_1\log{(K_i)} + \beta_2\log{(L_i)} + \beta_3(\tau_i) + \beta_4\log{(t+\tau_i)} + \beta_\chi \chi + \epsilon.
\end{equation}

where $\tau$ is the tax rate faced by the subsidiary which we proxy by ETRs (and $t$ is an offset parameter, which we discuss below). Likewise, three recent influential profit shifting studies all use ETRs to estimate profit shifting in one way or another \citep{clausingProfitShiftingTax2020,guvenenOffshoreProfitShifting2022, torslovMissingProfitsNations2023}. ETRs are superior to statutory rates because they are capable of better capturing the actual tax rate faced by MNCs and are more likely to be used by MNCs for profit shifting decision-making. A great deal of existing literature (e.g. \citealp{dharmapalaWhatWeKnow2014}) uses statutory rates, arguing that they are determined by governments and are thus generally exogenous to firms' choices (and mostly for comparability, we present estimates using statutory rates for both data sets in Tables \ref{tab:cit_etr_us} and \ref{tab:cit_etr_oecd}, which show that these models generally perform worse than those with ETRs: for the US data, they have lower R-squared in every case and the BIC values are higher with the exception of the linear model). Such endogeneity might be of importance for the research question of whether MNCs shift profits to countries with low tax rates, of which there is now abundant evidence \citep{beerInternationalCorporateTax2020}. However, once we move on to the scale, destinations and origins of profit shifting, the informativeness about the actual rates MNCs face becomes more important. In particular, statutory rates are not very informative about the taxes MNCs face (e.g. what \citealp{dharmapalaWhatWeKnow2014} highlighted as some possibility of mismeasurement of actual tax rates), correlate only weakly with various measures of ETRs \citep{garcia-bernardoMultinationalCorporationsTax2021a} and do not seem to sufficiently explain the location of profits (e.g. Section \ref{sec:appendix_eureqa}). For example, Luxembourg had a statutory rate of 29.22\% in 2017, whereas our ETR estimate for the same year is 1\% (Table \ref{tab:shifted}) and is in line with ETR estimates from other data sources \citep{garcia-bernardoEffectiveTaxRates2023}. 

%Overall, ETRs are a better alternative than statutory rates, since ETRs are more realistic and more likely to be used by MNCs for making profit shifting decisions (despite the downside of ETRs being potentially endogenous to profit shifting, since, for example, governments might lower ETRs to attract shifted profits). 

In equation \ref{eq:5}, $t$ is an offset parameter, included in order to avoid obtaining extremely high differences in the tax semi-elasticity for countries with similar but extremely low tax rates. We obtain the optimal value of the offset (0.0011 for US data and 0.0023 for OECD data) numerically by iterating over the range 0--0.2 and keeping the value that minimises the Bayesian Information Criterion. In section \ref{sec:robustness} we show that our results are highly robust to the choice of the offset, and that including this parameter in the linear and quadratic models does not increase their predictive power. We further include headquarter-country fixed effects to account for differences in profitability and data reporting methods between MNCs headquartered in different countries, and interaction terms between the country fixed effects and $log(t+\tau_i)$, which capture differences in the profit shifting aggressiveness of the MNCs of different reporting countries.

Consistently with literature, we operationalise capital ($K$) using tangible assets, and labour ($L$) using wages. A limitation of the operationalisation of the capital component through tangible assets is that tangible assets are affected by profit shifting strategies. For example, US MNCs in Luxembourg report the second highest value of tangible assets in Europe according to the CBCR data, with a combined value of \$223 billion (equal to values reported by US MNCs in the rest of the European Union, excluding tax havens). As a consequence, the use of tangible assets yields conservative estimates of the tax semi-elasticity. Since the data does not include wages, we model them using the product of employees and the average salary in each country, obtained from the International Labour Organization. Missing values in the average salary are estimated using a linear model containing log-GDP and log-Population ($R^2 = 0.91$).

After estimating the tax semi-elasticities using equation \ref{eq:5}, we calculate for each pair of countries (headquarter country and jurisdiction of operation) the underlying profits without profit shifting, $\hat{p_i}$. To do this we remove the effect of tax rates by comparing the profits reported in country \textit{i} with profits in a hypothetical scenario where the country's ETR is 25\%:

\begin{equation}\label{eq:pi}
    \hat{p_i} = \pi_i\cdot\frac{ e^{\left(\beta_3(0.25) + \beta_4\log{(t+0.25)}\right)}}{ e^{\left(\beta_3(\tau_i) + \beta_4\log{(t+\tau_i)}\right)}}.
\end{equation}

This ETR threshold of 25\% corresponds roughly to a zero marginal effect of the ETR on profits in the quadratic and logarithmic models, as explored further in the results section. Since MNCs do not appear to shift profits to countries with ETRs above 15\%, and our threshold is 25\%, $\hat{p_i}$ is almost always larger than $\pi_i$. The results are robust to changes in this threshold. A threshold of 20\% reduces the estimate of profit shifting using the logarithmic model by 8\%, the estimate using the quadratic model by 11\% and the estimate using the linear model by 19\%. This is expected, since the vast majority of profits are shifted towards countries with extremely low tax rates, which the logarithmic models can account for. In addition, we observe in Orbis that most MNCs indeed have opportunities to shift profits to tax havens. Specifically, 3,804 out of 5,391 (70\%) corporate groups, accounting for 89.7\% of profits in Orbis, have a subsidiary in one of the following tax havens: Bermuda, Ireland, Luxembourg, Netherlands, Switzerland, Cayman Islands, British Virgin Islands, Hong Kong, and Singapore.

\subsection{Reallocating shifted profits}\label{sec:methodology_real}

We now proceed to describe how the shifted profit is reallocated from tax havens to other countries and start by discussing how it is linked with the equation \ref{eq:1} above. Profit shifting is calculated as the difference of the booked profits and the estimated profits, assuming that the cost of profit shifting is negligible (as discussed in section \ref{sec:methodology_log} above):

\begin{equation}\label{eq:diff}
    \hat{S_i} = \pi_i - \hat{p_i}
\end{equation}

Since $\hat{p_i}$ is almost always larger than $\pi_i$, $S_i$ does not correspond to profit shifted in or out of the country (i.e. $\sum_i{\hat{S_i}})>0$). For this to happen, we need to redistribute shifted profits to where real economic activity takes place:

\begin{equation}
    \Delta P_i = -\hat{S_i} + (\sum_i{\hat{S_i}}) \cdot R_i,
\end{equation}	

where the change in profits due to profit shifting, $\Delta P_i$ , is defined as the profits shifted out of the country, $-\hat{S_i}$ (we reverse its sign since $\hat{S_i}$ measures profits shifted into a country), plus the share of total profit shifted redistributed back to the country.

We define the redistribution formula, $R_i$, operationalising real economic activity, as
\begin{equation}\label{eq:red}
    R_i =   1/4 \frac{L_i}{\sum_i{L_i}} + 1/4 \frac{W_i}{\sum_i{W_i}} + 1/2 \frac{Rev_i}{\sum_i{Rev_i}},
\end{equation}

where 25\% of the weight is given to employees ($L_i$), 25\% to wages ($W_i$) and 50\% to unrelated party revenues ($Rev_i$). We use unrelated party revenues, which are less affected by tax-planning strategies than, for example, tangible assets. This is the same formula used by the misalignment model described in section \ref{sec:methodology_mis}. The reallocation of shifted profits to the jurisdictions where economic activity takes place is also used in the impact assessment of the \citep{oecdTaxChallengesArising2020} BEPS plan in both pillar one (excess profit allocation) and pillar two (operationalization of the undertaxed payments rule), as well as by \textcite{beerExploringResidualProfit2020}---i.e., it is common to use a formulary approach to identify where the economic activity takes place. In the sensitivity analysis, we test that our results are robust to changes in the redistribution formula. The alternative redistribution of using bilateral balance of payments data \citep{torslovMissingProfitsNations2023} is not feasible due to the poor coverage of that data for many countries worldwide \citep{cobhamStatisticalMeasurementIllicit2021}.

After the redistribution, the sum of the change in profits due to profit shifting, $\sum{\Delta P_i}$, sums to zero.

Finally, tax revenue loss, $TRL_i$, is the product of the change in the profit base and the ETR (and we use the statutory rate as a robustness check):

\begin{equation}
    TRL_i = \Delta P_i\cdot ETR_i
\end{equation}

\subsection{Misalignment model}\label{sec:methodology_mis}

In addition to various semi-elasticity model specifications, we estimate the scale of profit shifting based on profit misalignment. The misalignment model applies basic arithmetic to the data to observe how well the location of reported profits are aligned with the location of economic activity, typically approximated by a combination of labour (measured using wages and employees), capital (often approximated with tangible assets) and revenue. Profit misalignment is then calculated as the difference between reported profits ($\pi$) and estimated theoretical profits (($\hat{p}$)). In our version of this method, and as in equation \ref{eq:red}, we calculate $\hat{p}$ giving 25\% of the weight to employees, 25\% of the weight to wages, and 50\% of the weight to unrelated party revenues (eq. \ref{eq:misaligned}). Since the majority of profits are shifted towards a small number of tax havens, the exact formula has little impact on the aggregated estimation of profit shifting, although can affect the results for individual countries (which we explore in section \ref{sec:robustness}).

\begin{equation}\label{eq:misaligned}
    \hat{p_i} = R_i \cdot \sum_i{\pi_i}.
\end{equation}

Profit shifting is again calculated as the difference between booked profits and the estimated profits (eq. \ref{eq:diff}). In a pure misalignment model, the sum of profit shifting is equal to zero ($\sum{\hat{S_i}} = 0$ and $\Delta P_i = \hat{S_i}$). We, however, add one extra constraint, similarly to \textcite{oecdTaxChallengesArising2020}. We set the profit misalignment of all foreign observations (pairs of reporting and investment countries where the reporting and investment countries are different) with a tax rate higher than 25\% to zero, since we assumed that an MNC would not shift profits to a country with a tax rate over 25\%. This corrects for extreme outliers, such as the high profits of of MNCs in resource-rich countries compared with the economic activity in the countries. In order to ensure that $\sum{\Delta P_i} = 0$, we redistribute the profits as in section \ref{sec:methodology_real}.

The logarithmic and misalignment models have different advantages. The logarithmic model explicitly models the observed extremely non-linear relationship between profits and tax rates (Fig. \ref{fig:profit_employee_etr}), and as such provides an estimate on the tax semi-elasticities (which we later visualize in Figures ~\ref{fig:elasticity_us} and ~\ref{fig:elasticity_oecd}). In contrast, the misalignment model provides a better estimate of the origin and destination countries of profit shifting because it takes into consideration the current distribution of profits. The logarithmic model is agnostic to this fact, and redistributes the profits only as a function of the location of economic activity.

The redistribution of shifted profits works differently in the two methods, as we illustrate in the following example. Assume that \$9 million profits are located in the US, \$0 million are located in India and \$1 million are located in the Cayman Islands. In contrast, the wages and sales in the United States add up to \$9 million, \$1 million in India and in \$0 million in the Cayman Islands. Both the logarithmic and the misalignment models would find that the shifted profit or the total misalignment is approximately \$1 million (located in the Cayman Islands), but the redistribution would differ. The logarithmic model would redistribute 90\% of those shifted profits to the US and 10\% to India. Since the profits in the US are comparable to the economic activity, the misalignment model would redistribute 0\% of those profits to the US and 100\% to India. Since the misalignment model takes into consideration the degree of profit shifting out of a country---as our example illustrates---the redistribution of profits is more accurate and realistic under the misalignment model than under the logarithmic specification.

\section{Data}\label{sec:data}
Our paper exploits the CBCR dataset that became available in July 2020, and is of unprecedented quality. The dataset was created thanks to a CBCR regulation that stems from OECD Base Erosion and Profit Shifting (BEPS) Action 13, and requires all large MNCs to report how much tax they pay in individual countries, including tax havens. The regulation impacts MNCs with consolidated annual group revenue of €750 million and above, headquartered in any country that has adopted the CBCR regulation. The firm-level data is collected by the headquarter country (a template is depicted in Fig. \ref{fig:cbcr_template}), aggregated by country of operations, and published by the OECD. The published data, which we use in this paper, is thus aggregated at the country level for each reporting country --- for example, India publishes data on the operations of India-headquartered MNCs in Ghana, Switzerland and many other countries.

To our knowledge, there are now several concurrent research papers using CBCR data from the US \citep{demooijAssessmentGlobalFormula2019, cobhamGlobalInequalitiesTaxing2019, clausingProfitShiftingTax2020, garcia-bernardoMultinationalCorporationsTax2021a,garcia-bernardoDidTaxCuts2022,nessaEffectUSCountrybyCountry2022}, Italy \citep{brattaAssessingProfitShifting2021}, Germany \citep{fuestCorporateProfitShifting2022a} and, most recently, Germany and other countries \citep{fuestGlobalProfitShifting2022}, and this has been the first paper using the OECD CBCR data.

We use the 2017 OECD CBCR data, which contains data for 38 headquarter countries (see Table~\ref{tab:n_jur_country}). The US IRS has been publishing CBCR data approximately one year before it is published by the OECD, which has allowed previous researchers to compare US CBCR with other sources \citep{clausingProfitShiftingTax2020,garcia-bernardoMultinationalCorporationsTax2021a}, and established a good correlation between various types of data sources. Moreover, the CBCR data is outstanding in at least three dimensions.
%Since reporting by US MNCs was voluntary in 2016, we replaced the 2016 data on US MNCs provided by the OECD with the 2017 data published by the US Internal Revenue Service (IRS). 

First, one of the most obvious advantages of CBCR data over other data sources is its much more substantial country coverage. This is especially relevant for lower-income countries and for selected parts of the world, for which coverage from other data sources is notoriously limited \citep{garcia-bernardoMultinationalCorporationsTax2021a}. For example, US CBCR data includes information on taxes and profits for US MNCs in 25 African countries, while the frequently used data from the Bureau of Economic Analysis of the United States Department of Commerce only covers 3. CBCR data includes data on large MNCs' profits and tax payments in, for example, up to 145 (United States) and 198 (Japan) jurisdictions in the full dataset. The exceptional data coverage of up to 214 countries enables us to estimate the scale of profit shifting for lower-income countries. This country coverage is one reason why \textcite{unodcConceptualFrameworkStatistical2020} propose to use this CBCR data for the Sustainable Development Goals indicator of illicit financial flows, likely in a similar way that we implement the profit misalignment method outlined in Section \ref{sec:methodology_mis}.

Second, CBCR ensures that profits and taxes are defined consistently with the concepts of corporate profits and taxes. By contrast, this is not the case with, for example, Bureau of Economic Analysis data, where profits are imputed from a combination of net profits, intra-group dividends, interest paid and other variables, as recently discussed by \textcite{clausingProfitShiftingTax2020, clausingFiveLessonsProfit2020, blouinDoubleCountingAccounting2020, garcia-bernardoMultinationalCorporationsTax2021a}. Consequently, CBCR data excludes double-counting in revenue.

However, a certain extent of double counting in profit due to intercompany dividends, for which we correct, and stateless entities, which we drop from our analysis, is inevitable and confirmed by existing evidence on which we build. Specifically, we correct for double counting in profit in five ways that we briefly summarise here and discuss in more detail in the Appendix \ref{sec:methodology_extrapolation}. First by excluding stateless entities. Second, some countries such as the Netherlands investigated the extent of double counting in domestic profits and we use their corrections. Third, we remove double counting of US profits, as estimated by \citealp{garcia-bernardoDidTaxCuts2022}. Fourth, we remove 10\% of profits in tax havens for non-US MNCs (a similar ratio that the one found by \citealp{garcia-bernardoDidTaxCuts2022} for US MNCs). Fifth, we remove certain domestic profits in other countries depending on the relative difference between ETR on domestic and foreign profits. When corrected for double-counting in profits, CBCR data offers the best available information on MNCs' tax payments for many countries, it thus provides us with the first suitable dataset for a high-quality cross-country comparison---until now various proxies for profits were used, for example, by \textcite{haberlyTaxHavensProduction2015}, \textcite{bolwijnEstablishingBaselineEstimating2018} or \textcite{damgaardWhatRealWhat2019}.

Third, CBCR data is provided in two separate datasets, for all subsidiaries (``All Sub-Groups''), as well as for those subsidiaries that had positive profits and so not losses (``Sub-Groups with Positive Profit''). While the data on affiliates with positive profits has lower coverage (Table~\ref{tab:n_jur_country}), it allows for more accurate estimates of the ETRs. %The following section details how we use the data in this paper.

\subsection{Use of data in the logarithmic and misalignment models}
We use different subsets of the data for different parts of the methodology. We estimate ETR as the ratio of accrued taxes over profits, using the data on ``sub-groups with positive profits''. By using the data with positive profits only, we avoid offsetting firms with losses and firms with profits, and we can thus estimate ETRs more precisely. Since taxes are typically paid by companies earning profits, including companies making losses would overstate ETRs. We use ETRs in two parts of the paper: to calculate profit shifting in the semi-elasticity models, and to calculate tax revenue losses. For the semi-elasticity models we remove outliers---country dyads with tax rates above 50\%. This eliminates outliers and allows for a more efficient estimation of the semi-elasticities. To calculate tax revenue losses we use the average ETR in the country, using the average ETR paid by foreign MNCs and the statutory tax rate as robustness checks. The average ETR is weighted by profits booked: $\frac{ETR_i \pi_i}{\sum{\pi_i}}$. For countries that are only available in the data on all sub-groups but not in the data on sub-groups with positive profit, we used the statutory corporate income tax rate (which was the case for Anguilla, Antigua and Barbuda, Cuba, Djibouti, French Guiana, Guadeloupe, Haiti, Kiribati, Kosovo, Kyrgyz Republic, Sao Tome and Principe, St. Lucia, St. Vincent \& Grenadines, Syria, Turkmenistan, Turks and Caicos Islands). The ETRs are reported in Table \ref{tab:n_ETR_country}.

First, we estimate the semi-elasticity model (detailed in section \ref{sec:methodology_model}) using data on ``sub-groups with positive profits''. In addition, we run a robustness check and find no significant differences with the full sample, see section \ref{sec:robustness} for more information. Table~\ref{tab:stats} shows the summary statistics of the CBCR data for the countries in this sample, distinguishing between domestic and foreign activities of MNCs---domestic ones are those in the reporting (i.e. headquarter) countries, while foreign ones are those in all other countries (i.e. except for the domestic one). For most countries domestic profits and activities are higher than foreign ones. The observed balance between domestic and foreign activities provides useful guidance for when we estimate missing data in Section \ref{sec:methodology_extrapolation}.

Second, we reallocate profits shifted (equation \ref{eq:red}) using the dataset including all sub-groups for the 38 countries that reported some information. Using the complete dataset allow us to more accurately measure information on real economic activities of MNCs regardless of whether the affiliates are profit- or loss-making. Since MNCs prefer to report losses in countries with high taxes while locating their profits in countries with low taxes, excluding loss-making affiliates would exclude an important component of profit shifting (see Figure ~\ref{fig:loss_making} for a visualisation of this behaviour in the CBCR data, and \textcite{desimoneUnprofitableAffiliatesIncome2017} for an empirical confirmation using tax returns data in the United Kingdom). The dataset on all sub-groups is also more suitable for comparison with other datasets (e.g. from the Bureau of Economic Analysis).

Finally, and for the same reasons explained in the previous paragraph, we used data on all sub-groups for the misalignment model. The ETRs (used to calculate tax revenue losses) are still calculated from the data on sub-groups with positive profit. Since the misalignment method is not affected by outliers we keep all observations in the sample.

While we make use of the substantial country coverage and other advantages of CBCR data, we carefully deal with several remaining challenges associated with the new data source, particularly missing data and double counting of profits. We describe how we address the data limitations in detail in Appendix \ref{sec:methodology_extrapolation}. 

\section{Results}\label{sec:results}

The results section is composed of four parts. In the first part we demonstrate the advantage of our methodology using the US CBCR data. In the second part we apply our methodology to the OECD CBCR data, and compare it to estimates generated using other methodologies. In the third part we test whether the scales of profit shifting and associated tax revenue loss are higher or lower in some country groups. In the fourth part, we present a series of robustness tests and sensitivity analyses.

\subsection{Estimation of profit shifting (US data): the logarithmic model versus other models}\label{sec:us_results}
\begin{figure}[h]
        \begin{center}
 \caption{Comparison of tax semi-elasticity estimates using the 2017 US data}
  \includegraphics[width=1\textwidth]{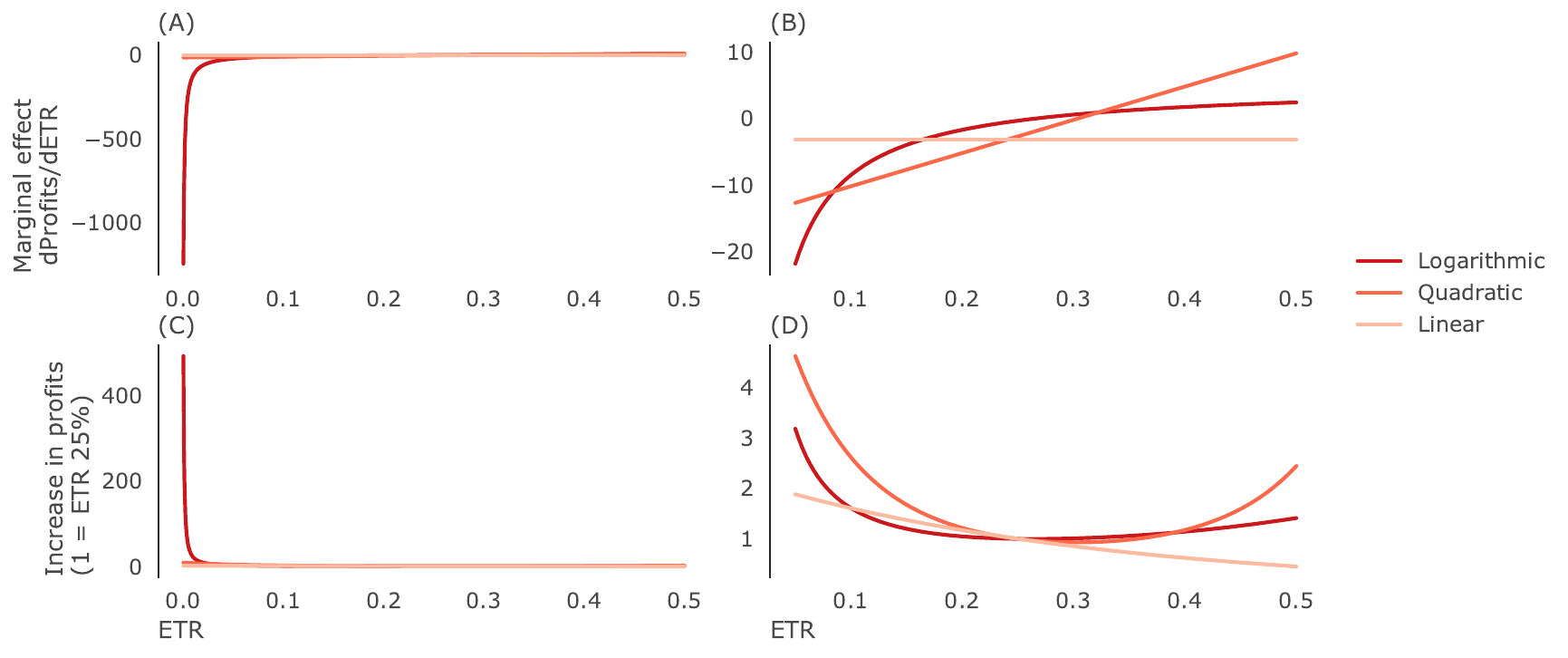}
  \label{fig:elasticity_us}
\end{center}
        \footnotesize
Notes: Graphical representation of Table \ref{tab:comparison} for the logarithmic, quadratic, linear models. (A, B) Marginal effect of ETR on profits. (C,D) Relative increase in profits due to profit shifting, compared with a country with an ETR of 25\%. Plots B and D are close-ups of plots A and C respectively, constraining ETRs between 5 and 50\%. Note that the marginal effects for the logarithmic model decreases (becomes more negative) faster than other models as the ETR approaches 0\%.
%Comparison of semi-elasticities for the logarithmic (Log), quadratic (Quad) and linear (Linear) models. 
\end{figure}

We first test our methodology using only US CBCR data. Restricting our analysis to US data allows us to compare with previous analysis, including one of the best-regarded papers on profit shifting using tax semi-elasticities, \cite{dowdProfitShiftingMultinationals2017}. The results of our regressions (Table~\ref{tab:comparison}) shows that the logarithmic model fits the data better than any other model. This not only involves a higher R-square and lower Bayesian information criteria, but also a better disaggregation of the origin and destination of profits shifted. Fig.~\ref{fig:elasticity_us} shows a graphical interpretation of the coefficients. The logarithmic model is capable of accounting for extreme ratios of profit shifted in small countries with low ETRs, while at the same time avoiding the overestimation of profit shifted in countries with tax rates above 15\%. For a country with a tax rate of 15\% (e.g. Australia), the logarithmic model estimates that 16\% of the profits are shifted in, while the quadratic and linear models estimate that number to be 40 and 26\% respectively. Importantly, the logarithmic model and the misalignment model clearly identify that the majority of profits in small countries with extremely low tax rates are shifted there. This effect is less pronounced for the quadratic model, and especially so for the linear model (Table \ref{tab:shifted}). For an ETR of 1.7\% (e.g. Bermuda), the logarithmic and quadratic model estimates that 91\% and 86\% of the booked profits have been shifted into the country (and 98.5\% with the misalignment method below), while the linear model estimates that only 52\% have.

\begin{table}[h]
\caption{Comparison of tax semi-elasticity estimates using the 2017 US data}
\label{tab:comparison}
\begin{center}
\begin{tabular}{lllll}
\toprule
{} &         Log &         Quad &    Log+Quad &         Linear \\
\midrule
ETR                    &   5.1998*** &  -15.1545*** &      8.4752 &  -3.1654*** \\
                       &    (1.4324) &     (3.0356) &    (5.2472) &    (1.0062) \\
ETR\textsuperscript{2} &             &   24.9745*** &     -5.0609 &             \\
                       &             &     (6.0246) &    (7.7973) &             \\
log(0.0011 + ETR)      &  -1.3813*** &              &  -1.5210*** &             \\
                       &    (0.1959) &              &    (0.2915) &             \\
log(Tangible assets)   &   0.5456*** &    0.6745*** &   0.5412*** &   0.7595*** \\
                       &    (0.0751) &     (0.0815) &    (0.0756) &    (0.0861) \\
log(Wages)             &   0.2481*** &       0.1163 &   0.2541*** &      0.0442 \\
                       &    (0.0885) &     (0.0977) &    (0.0893) &    (0.1049) \\
log(Population)        &     0.1638* &       0.1094 &     0.1602* &      0.0200 \\
                       &    (0.0859) &     (0.0983) &    (0.0863) &    (0.1046) \\
log(GDP per capita)   &      0.1664 &       0.2132 &      0.1638 &      0.2347 \\
                       &    (0.1316) &     (0.1510) &    (0.1322) &    (0.1646) \\
N                      &          91 &           91 &          91 &          91 \\
R-squared              &      0.8850 &       0.8481 &      0.8856 &      0.8170 \\
R-squared Adj.         &      0.8768 &       0.8372 &      0.8759 &      0.8062 \\
BIC                    &      230.48 &       255.83 &      234.53 &      268.26 \\
\bottomrule
\end{tabular}
\end{center}
\footnotesize
Notes: Comparison of semi-elasticities for the logarithmic (Log), quadratic (Quad), the combination of the two (Log+Quad) and linear (Linear) models using the 2017 US CBCR data. The dependent variable for all models is profits booked in country in logarithm. (We include the combination of the two (Log+Quad) models in this and other regression tables as a robustness check to test whether the Log model was not better because it was more complex; the results show that adding a Quad term to the Log model does not improve the model.)
\end{table}

\begin{table}[ht!]
\small
\caption{Percentage of profits shifted into countries with at least \$10 bn reported using the 2017 US data}
\begin{center}
\begin{tabular}{llllp{1cm}p{1cm}p{1cm}p{1cm}}
\toprule
Country &    ETR & Profits (+) & Profits (all) & Misal. & Log & Quad & Linear  \\
\midrule
Jersey         &   0.1\% &   \$12.8 bn &   \$10.5 bn &              97.3\% &                   99.4\% &                    89.0\% &                   54.5\% \\
Cayman Islands &   0.6\% &   \$56.1 bn &   \$52.7 bn &              98.8\% &                   97.6\% &                    88.3\% &                   53.9\% \\
Other Europe   &   0.8\% &   \$13.6 bn &    \$0.0 bn &                  - &                   96.5\% &                    87.9\% &                   53.6\% \\
Luxembourg     &   1.0\% &   \$54.4 bn &   \$22.4 bn &              92.0\% &                   95.2\% &                    87.5\% &                   53.2\% \\
Puerto Rico    &   1.6\% &   \$31.7 bn &   \$30.9 bn &              94.9\% &                   91.8\% &                    86.4\% &                   52.3\% \\
Bermuda        &   1.7\% &   \$31.9 bn &   \$29.2 bn &              98.5\% &                   91.4\% &                    86.2\% &                   52.2\% \\
Other America  &   2.4\% &   \$12.2 bn &   \$-0.1 bn &                  - &                   86.4\% &                    84.7\% &                   51.1\% \\
\bottomrule
\end{tabular}
\label{tab:shifted}
\end{center}
\footnotesize

Notes: Profit shifted into countries estimated using a variety of models and the US CBCR data. The table shows the percentage of profits shifted for the misalignment (Misal.) model using data on all sub-groups, and---using data on affiliates with positive profits---logarithmic (Log), quadratic (Quad), and linear (Linear) models.  The column ``profits (+)'' indicates the profits of affiliates with positive profits, the column ``profits (all)'' indicates the profits of all affiliates. 
\end{table}

Figure \ref{fig:elasticity_us} shows a U-shaped relationship in the effect of ETRs on profits. The semi-elasticity is negative until the ETR reaches approximately 25\%, thereafter becoming positive. This is due to high profits in countries rich in natural resources, such as Angola, the United Arab Emirates, Qatar, Norway and Nigeria. These countries levy resource taxes while carrying out activities that produce vast amounts of profit in relation to the labour and capital costs. In order to correct for this in our estimates of profit shifting, we assume a tax semi-elasticity of zero if the ETR is higher than 25\%. This approach is also used in the Impact Assessment of the BEPS plan \cite{oecdTaxChallengesArising2020}.

\begin{figure}[h!]
\begin{center}
 \caption{Profits shifted in and out of countries using the US data}
  \includegraphics[width=1.05\textwidth]{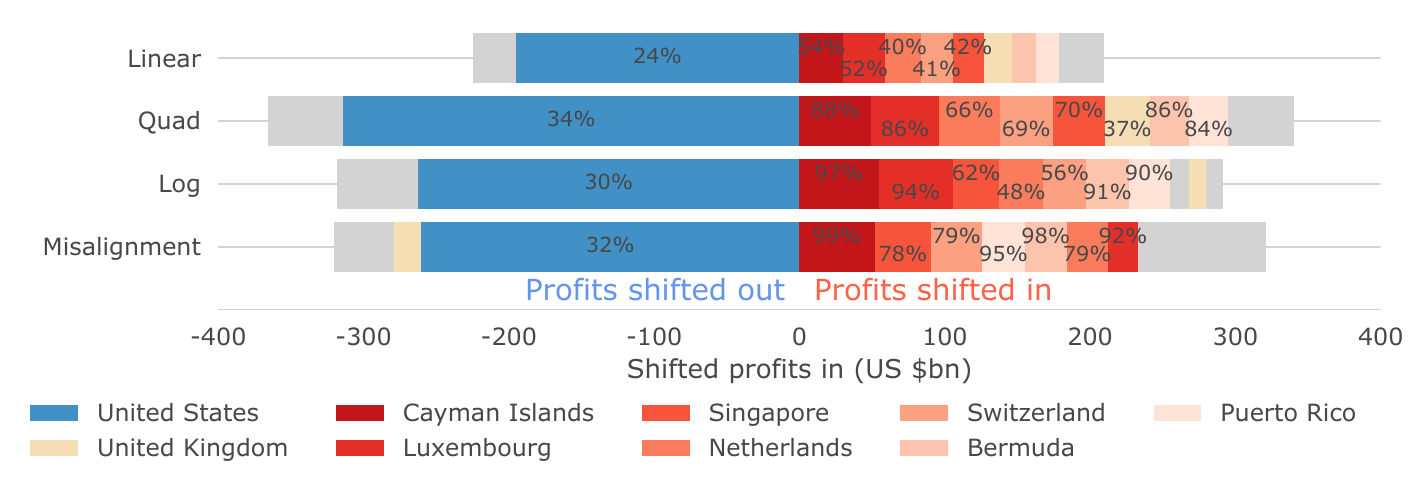}
  \label{fig:breakdown_us}
  %Note: Labels adjusted with inkscape to improve visibility
\end{center}
\small
Notes: Profits shifted in and out of countries using the US data, estimated with with the linear, quadratic (Quad), logarithmic (Log) and misalignment models. MNCs shift profits from countries with negative shifted profits to countries with positive shifted profits. The largest origins of the profits are visualised in blue, and the largest destinations in red. All other countries are visualised together in grey. The annotations indicate the percentage of profit shifted out of the country (compared to estimated profits) or into the country (compared to booked profits).
\end{figure}

Next, we redistribute the profits shifted according to equation \ref{eq:red} to calculate global profit shifting. The logarithmic model yields an estimate of \$364 billion of profit shifted, comparable to the \$323 billion of profit shifted found by the misalignment strategy (Figure~\ref{fig:breakdown_us}). Since our objective in this section is to compare the different methodologies and not to present the scale of profit shifted for individual countries, we do not try to disaggregate categories, such as ``Other Europe'', into individual countries, as detailed in section \ref{sec:methodology_extrapolation} and as applied in section \ref{sec:oecd_results}. The destination of shifted profits is similar between models (Figure~\ref{fig:breakdown_us} and \ref{fig:compare_log_mis_US}). The majority of these profits are in a small group of tax havens. The large majority of profits shifted, are shifted to the top 10 countries shown in Table~\ref{tab:shifted}. Moreover, over 75\% of the profits booked in those 10 countries are artificially shifted there. The Cayman Islands, Luxembourg, the Netherlands, Switzerland, Singapore, Bermuda and Puerto Rico are the largest destinations. However, several differences may be observed. Profit shifted to Luxembourg is more than two times larger in the logarithmic rather than misalignment model, because of the presence of many companies with losses and different data used for the two models. Compared with the \$22 billion of profits found in Luxembourg in the data on ``All Sub-Groups'' (used for the misalignment model), there is \$54 billion of profits in the data ``Sub-Groups with positive profits''. Similarly, while the ``Other Europe'', ``Other Asia\&Oceania'' and ``Other America'' groups appear as profit destinations in the logarithmic model, they appear as places of profit origin in the misalignment. This is due to the higher granularity of the data with ``All Sub-Groups'' used for the misalignment model. There, ``Isle of Man'', ``Barbados'', ``Gibraltar'', ``Macao'' and ``the British Virgin Islands'' appear as standalone countries, with \$22 billion booked into those countries.

The origin of shifted profits is, however, considerably different across the models. We observe that 72\% of the employees, wages and sales of US MNCs are located in the United States and the ETR of the United States is 43\%. This together implies that no profits are shifted into the US, and approximately 72\% of the global shifted profits are redistributed back to the United States in the models relying on semi-elasticity. In the misalignment model a similar level of profits are shifted out of the USAsince 48\% of the profits are already reported there. In addition to the US, the UK also seems to lose out in the misalignment model, but not in the other models (Fig. ~\ref{fig:breakdown_us}). In the tax semi-elasticity models, the low ETR of the UK (12\%) implies that profits are expected to be shifted into the country. In the misalignment model profits are found to be shifted out of the country, since the profits reported in the UK (1.5\% of the total) are lower than their share of the economy (3.3\% of the total). The low aggregated profits recorded in the UK are a consequence of MNCs reporting zero or negative profits \citep{bilickaComparingUKTax2019}.

Overall, while we consider the logarithmic specification to be more accurate with respect to estimating the global scale of profit shifting, the misalignment method might provide more accurate estimates of the redistribution of these shifted profits. The misalignment method takes into consideration the current distribution of profits, and in this respect provides a more accurate way of redistributing profits, as we have discussed at the end of the methodology section \ref{sec:methodology}. The location of profits and economic activity is often more balanced (i.e. less misaligned) in countries with high per capita income (Fig. \ref{fig:compare_log_mis_US} in the Appendix). The logarithmic model is agnostic to this fact, and redistributes the profits only as a function of the location of economic activity. As a consequence, the misalignment model typically redistributes more profits back to lower-income countries.

\subsection{Estimation of profit shifting (OECD data)}  \label{sec:oecd_results}

Having shown that the logarithmic model is superior to both the quadratic and linear models, we apply it to the OECD CBCR data. We use the same methodology as in the previous section, but add fixed effects for the reporting countries to correct for differences in profitability due to the location of headquarters, and add an interaction between the reporting country and either $\log(t+ETR)$, $ETR^2$ or $ETR$ in our logarithmic, quadratic and linear models.

Table~\ref{tab:oecd_reg_table} shows the estimates of tax semi-elasticity using the OECD data (with an interpretation of the coefficients given visually in Figure \ref{fig:elasticity_oecd}). The US is used as the reference group for country comparisons. We again observe that the logarithmic model fits the data better than the quadratic and linear specifications. The logarithmic model estimates that over 40\% of profit shifting takes places towards countries with an ETR below 1\% (Table \ref{tab:tax_havens}). The quadratic and linear models are not able to capture this fully (Table \ref{tab:tax_havens}). The misalignment model yields similar results to the logarithmic model, reinforcing the accuracy of the logarithmic model.

\begin{table}
\caption{Comparison of tax semi-elasticity estimates using the OECD data}
\label{tab:oecd_reg_table}
\scriptsize
\begin{center}
\begin{tabular}{llllll}
\toprule
{} & Logarithmic &   Quadratic & Log*FE + Quad & Log + Quad*FE &      Linear \\
\midrule
ETR                    &      0.9313 &  -8.5950*** &        1.5638 &        0.6666 &  -3.7698*** \\
                       &    (0.7316) &    (2.0412) &      (1.8934) &      (2.0052) &    (0.3623) \\
ETR\textsuperscript{2} &             &  13.0086*** &       -1.1865 &       -1.2082 &             \\
                       &             &    (3.9734) &      (3.2164) &      (3.5384) &             \\
log(0.0023 + ETR)      &  -0.7626*** &             &    -0.7817*** &    -0.4012*** &             \\
                       &    (0.0788) &             &      (0.0848) &      (0.0930) &             \\
Australia:tax          &   0.3816*** &   4.2074*** &     0.3807*** &      2.7669** &   1.6438*** \\
\bottomrule
\end{tabular}
\end{center}
\footnotesize
Notes: Regression table for the OECD data. Clustered standard errors are shown in parenthesis. Country:tax represents the interaction effect between the country and $\log(0.0023+ETR)$, $ETR^2$ and $ETR$ for our three specifications (logarithmic, quadratic and linear). The intercept and country fixed effects are not shown and are generally negative and significant at the 0.1\% significance level; since the treatment group is the US, this indicates the higher profitability of US MNCs. Only countries reporting in over 20 partner jurisdictions are shown.
\end{table}

\begin{table}
\begin{center}
\caption{Share of profit shifted into countries, grouped by the effective tax rates}
\label{tab:tax_havens}
\small
\begin{tabular}{lrrrr}
\toprule
ETR &      Misalignment &      Logarithmic &     Quadratic &      Linear \\
\midrule
Below 5\%     &  40.0\% &  40.6\% &  33.5\% &  31.6\% \\
5-10\%   &  30.0\% &  43.1\% &  40.6\% &  39.8\% \\
10-15\%  &  15.4\% &  11.8\% &  16.4\% &  17.2\% \\
15-25\%  &   9.7\% &   2.7\% &   4.1\% &   6.1\% \\
Above 25\%    &   4.9\% &   1.7\% &   5.4\% &   5.3\% \\
\bottomrule
\end{tabular}
\\
\end{center}
\footnotesize
Notes: Share of profit shifted into countries, grouped by the average foreign ETR in the country. The quadratic and linear models are not able to account for the large share of profits shifted into countries with ETRs below 1\%. Since profit shifting is estimated at the bilateral level (reporting:partner) for the semi-elastictiy methods, a country can (rarely) have an average ETR above 25\% and an ETR below 25\% for some of those reporting:partner relationships.
\end{table}

The location of MNCs' headquarters have been shown both theoretically and empirically to be an important consideration in the profit shifting carried out by MNCs \citep{dischingerRoleHeadquartersMultinational2014, bilickaComparingUKTax2019}. A significant number of existing studies observed profit shifting in the case of US headquartered MNCs \citep{guvenenOffshoreProfitShifting2022, dowdProfitShiftingMultinationals2017,clausingProfitShiftingTax2020}. For example, the ETR paid on foreign profits by US MNCs in sectors other than oil has fallen by half since the late 1990s and nearly half of this decline is estimated to be the outcome of the rise of profit shifting to tax havens \citep{wrightExorbitantTaxPrivilege2018}.
%Moreover, previous research has suggested, though, to the best of our knowledge, not empirically confirmed, that US MNCs are more aggressive than other MNCs with respect to their tax planning strategies.

The introduction of the interaction term between the country fixed-effect and $\log(ETR)$, $ETR^2$ or $ETR$ allows us to understand the aggressiveness of each country's MNCs with respect to profit shifting. We find that MNCs from Singapore, Brazil and the United States are the most aggressive (the magnitude of the interaction is more negative). This is in contrast with a recent paper by \textcite{bilickaComparingUKTax2019} that studies headquarter location heterogeneity for MNCs active in the United Kingdom and finds US MNCs to have a similar size of the estimated profit ratio gap as French and German MNCs. Furthermore, we find that the difference is statistically significant for all countries (Table \ref{tab:oecd_reg_table}). %Specifically, using the logarithmic the semi-elasticity of ETR around 0\% is highest for US MNCs, i.e. the magnitude of $log(0.0007+ETR)$ is more negative for the United States than for other countries.
In fact, the relationship completely disappears for South Africa and Malaysia (Fig. \ref{fig:elasticity_oecd_by_country}).
%The latter three countries have a credit system for the relief of foreign taxes paid on dividends, which encourages the repatriation of profits to the headquarter country (instead of a tax haven). Every other country in the sample exempt dividend tax on related party transactions \citep{tjnView2020Results2020}.
Japan is an interesting case to study due to its historically perceived distinct attitude towards tax planning \citep{izawaWhoMeTax2019}. 
Our results show that profit shifting by Japanese MNCs are similar to other MNCs. 

\begin{figure}[h]
 \caption{Profits shifted in and out of countries using the OECD data}
  \includegraphics[width=1\textwidth]{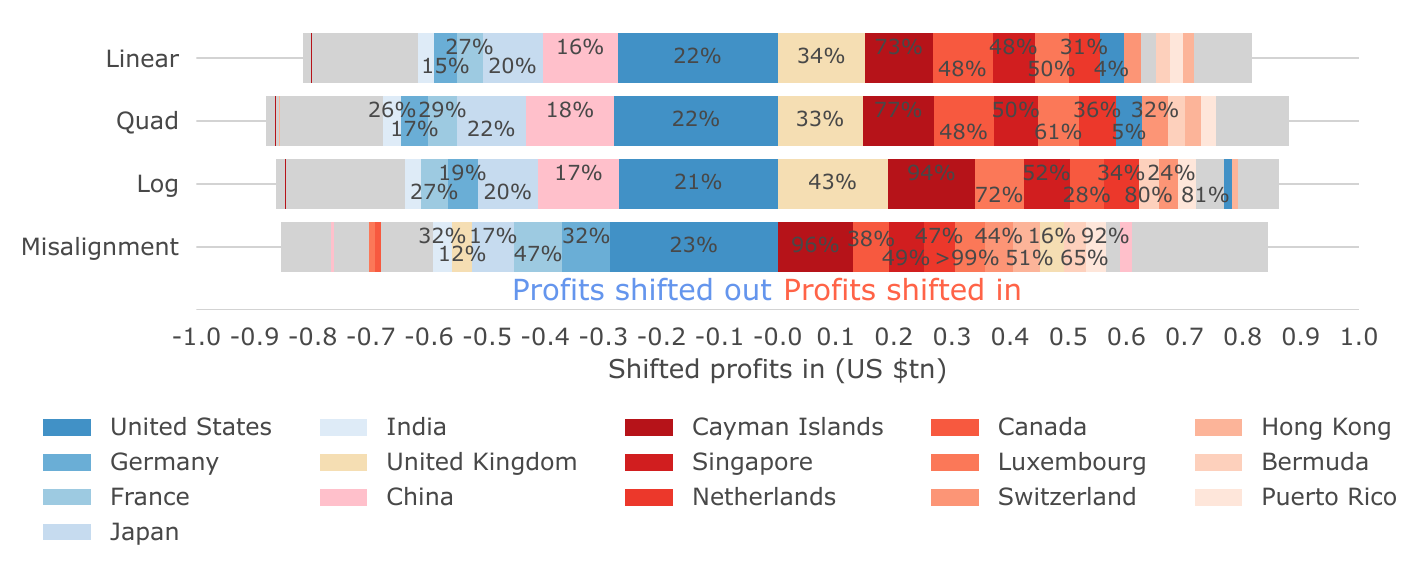}
  \label{fig:breakdown_oecd}
\footnotesize
Notes: Profits shifted in and out of countries using the OECD CBCR data, estimated with the linear quadratic (Quad), logarithmic (Log) and misalignment models. MNCs shift profits from countries with negative shifted profits to countries with positive shifted profits. The largest origins of the profits are visualised in blue, and the largest destinations in red. All other countries are visualised together in grey. The annotations indicate the percentage of profit shifted out of the country (compared to estimated profits) or into the country (compared to booked profits). Tables \ref{tab:top_dest} and \ref{tab:top_sources} show the top sources and destinations.
\end{figure}

Next, we calculate the extent of profit shifting for all models. We reach an estimate of \$862 billion shifted for the logarithmic model and a 95\% confidence interval of \$838--1,022 billion for the misalignment model, of which we use the median, \$867 billion (Table \ref{tab:trl} and Fig.~\ref{fig:breakdown_oecd}).\footnote{Note that the sum of the median misalignment for each country is different from the median of the total misalignment. We use the median misalignment for each country which adds up to a positive misalignment of \$841 billion and negative of 854B.}
We compare the results obtained using both methodologies in Figure \ref{fig:compare_log_mis_US} in the Appendix. In general, there is a good correlation between the origin and the destination of profit shifted, albeit with some outliers (the United Kingdom, Brazil, Malaysia, China), with the United Kingdom previously discussed in section \ref{sec:us_results}. Our total estimates are comparable to existing estimates such as \textcite{torslovMissingProfitsNations2023} and \textcite{wierGlobalProfitShifting2022}, who estimate profit shifting to be \$616 billion in 2015 and \$969 billion in 2019, albeit using a smaller sample of countries ((see Figure \ref{fig:torslov} for a detailed comparison)). Our findings imply that revenue losses total approximately \$200–300 billion. This is comparable with recent leading estimates of revenue losses which range from \$100 to \$300 billion, as compared in Table \ref{tab:trl_comparison}. Furthermore, it is important to keep in mind that some aspects of our methodological approach are conservative. For example, we aggregate at the country level and as such offset profits shifted in with profits shifted out. The estimated scale of profit shifting and tax revenue loss would be higher if we would not net gains and loses.

In addition to outlining the overall scale of the practice, Figure ~\ref{fig:breakdown_oecd} provides an overview of the origins and destinations of profit shifting. Using both the logarithmic model and misalignment methods, we estimate that the Cayman Islands and the Netherlands have 94--96\% and 34--47\% of their respective booked profits shifted in from other countries, while also ranking among countries most benefiting from profit shifting in absolute terms. The US is estimated to suffer the most from profit shifting in absolute terms according to most of the methods, while  France is affected to a substantially greater degree relative to estimated profits, with an estimate ranging from 47\% (misalignment method) to 27\% (logarithmic model). Table \ref{tab:top_dest} and \ref{tab:top_sources} shows the largest destinations and sources of profit shifting respectively.\footnote{Perhaps surprising is the inclusion of Canada as one of the top destinations. This is mostly driven by Canadian MNCs shifting profits towards the headquarters. However, this may be explained by the relatively low ETR of 12.9\% observed in the data, and the characteristics of its tax system \citep{vantrietPotentialBenefitsTax2014,capursoBurgersDoughnutsExpatriations2016}.} Figure \ref{fig:compare_log_mis_OECD} in the Appendix shows profit shifting at the country level for all countries, 214 countries for the log and 210 for the misalignment model, and for the misalignment model we show results in Table \ref{tab:winners_mis} and we visualise the uncertainty in Figure \ref{fig:ci_misalignment}.%--\ref{tab:lossers_log}

\begin{table}
\small
\caption{Top destinations of profit shifting}% and \ref{tab:winners_log}.
\label{tab:top_dest}
\begin{tabular}{l|l|lll|lll}
                &  ETR  & \multicolumn{3}{l}{Misalignment}  &  \multicolumn{3}{l}{Logarithmic}  \\ 
Country       & (\%)  & P (all groups) & PS (B) & PS (\%) & P (groups>0) & PS (B) & PS (\%) \\ \hline
United Kingdom         &      7.3 &        260,194 &    6,041 &     2.3 &      441,372 &   190,160 &    43.1 \\
Cayman Islands         &      0.2 &        136,292 &  130,276 &    95.6 &      159,811 &   149,000 &    93.2 \\
Luxembourg             &      1.8 &         48,845 &   41,520 &    85.0 &      116,346 &    83,360 &    71.6 \\
Singapore              &      5.3 &        121,237 &   59,602 &    49.2 &      151,568 &    79,058 &    52.2 \\
Canada                 &     12.9 &        162,820 &   51,011 &    31.3 &      215,831 &    59,342 &    27.5 \\
Netherlands            &      6.0 &        114,845 &   54,029 &    47.0 &      172,608 &    59,256 &    34.3 \\
\bottomrule
\end{tabular}
\begin{flushleft}
\footnotesize
Notes: The table shows the top destinations of profit shifting (PS (B)) for misalignment and logarithmic models and as a percentage of the total profits booked in the jurisdiction (PS (\%)). All countries with at least \$10 bn shifted are included. The total profits for all groups ((P (all groups)) and groups with positive profits (P (groups>0) are shown for comparison.  The full table can be found in Tables \ref{tab:winners_mis} and \ref{tab:winners_log} for misalignment and logarithmic models, respectively. 
\end{flushleft}
\end{table}

\begin{table}
\small
\caption{Top sources of profit shifting}
\label{tab:top_sources}
\begin{tabular}{l|l|lll|lll}
                &  ETR  & \multicolumn{3}{l}{Misalignment}  &  \multicolumn{3}{l}{Logarithmic}  \\ 
Country       & (\%)  & P (all groups) & PS (B) & PS (\%) & P (groups>0) & PS (B) & PS (\%) \\ \hline
\multicolumn{8}{l}{Sorted by total profit shifted (average of the two models)} \\
United States &     29.6 &        954,310 &  287,971 &    23.2 &      997,467 &  259,263 &    20.6 \\
Japan         &     27.4 &        346,261 &   71,961 &    17.2 &      414,662 &  104,218 &    20.1 \\
Germany       &     18.3 &        175,971 &   82,717 &    32.0 &      223,988 &   51,277 &    18.6 \\
France        &     25.0 &         92,626 &   82,377 &    47.1 &      122,962 &   45,784 &    27.1 \\
China         &     25.9 &        583,000 &  -15,511 &    -2.7 &      683,371 &  139,404 &    16.9 \\
\bottomrule
\end{tabular}
\begin{flushleft}
\footnotesize
Notes: The table shows the top sources of profit shifting (PS (B)) for misalignment and logarithmic models and as a percentage of the total profits in the jurisdiction (PS (\%)). Note that the total profits is the sum of the profits booked and the profits shifted. All countries with at least \$10 bn shifted out (top table) or \$1bn and 20\% of the total profits (bottom table) are included. The total profits for all groups ((P (all groups)) and groups with positive profits (P (groups>0) are shown for comparison.  The full table can be found in Tables \ref{tab:lossers_mis} and \ref{tab:lossers_log} for misalignment and logarithmic models, respectively. 
\end{flushleft}
\end{table}

\subsection{Profit shifting and tax revenue loss by income groups} \label{sec:income_results}

The analysis presented above compares different methodology approaches and establishes the largest origins and destinations of profits in absolute terms. In this section, we focus on the distribution effects of profit shifting, and find that lower-income countries tend to lose more tax revenue relative to their total tax revenue (i.e. total tax revenue collected by government, from across all types of taxes), which is in line with some of the earlier estimates (e.g., \textcite{fuestInternationalDebtShifting2011, crivelliBaseErosionProfit2016}.

We first focus on profit shifting. While countries from all income groups lose similarly relative to their GDP, profit shifting takes place predominantly to high-income countries (Figure \ref{fig:diffs_aggs_income_rev_prof_shift}). This is expected, since the majority of tax havens are included in this group (see, e.g., \textcite{cobhamFinancialSecrecyIndex2015, torslovMissingProfitsNations2023}). Although we present results for both the misalignment and the logarithmic model, we argue that the results of the misalignment model might be more accurate for two reasons. First, we use all available data in the misalignment model, imputing missing data. As previously mentioned, countries often do not report on small countries, but group them together into categories (e.g. ``Other Africa''). For example, only Germany, Japan and India report operations on Gambia (three, two and two MNCs respectively), while the remaining reporting countries with operations in the country group Gambia with other African countries. For the logarithmic model this leads to an underestimation of the losses of lower-income countries. For the misalignment model, we estimate the expected employees and revenue of all country pairs, and use this information to correct the amount of profit shifted more accurately (Section \ref{sec:methodology_mis}). While this only increases total profit shifted by 30\%, it is key to estimating profit shifting in lower-income countries accurately. A second, closely related, reason in favour of the misalignment model is its observation that profits are less aligned with economic activity in lower-income countries (discussed in Section \ref{sec:us_results}).

\begin{figure}[h!]
\caption{Tax revenue loss as a percentage of total tax revenue}
  \includegraphics[width=1\textwidth]{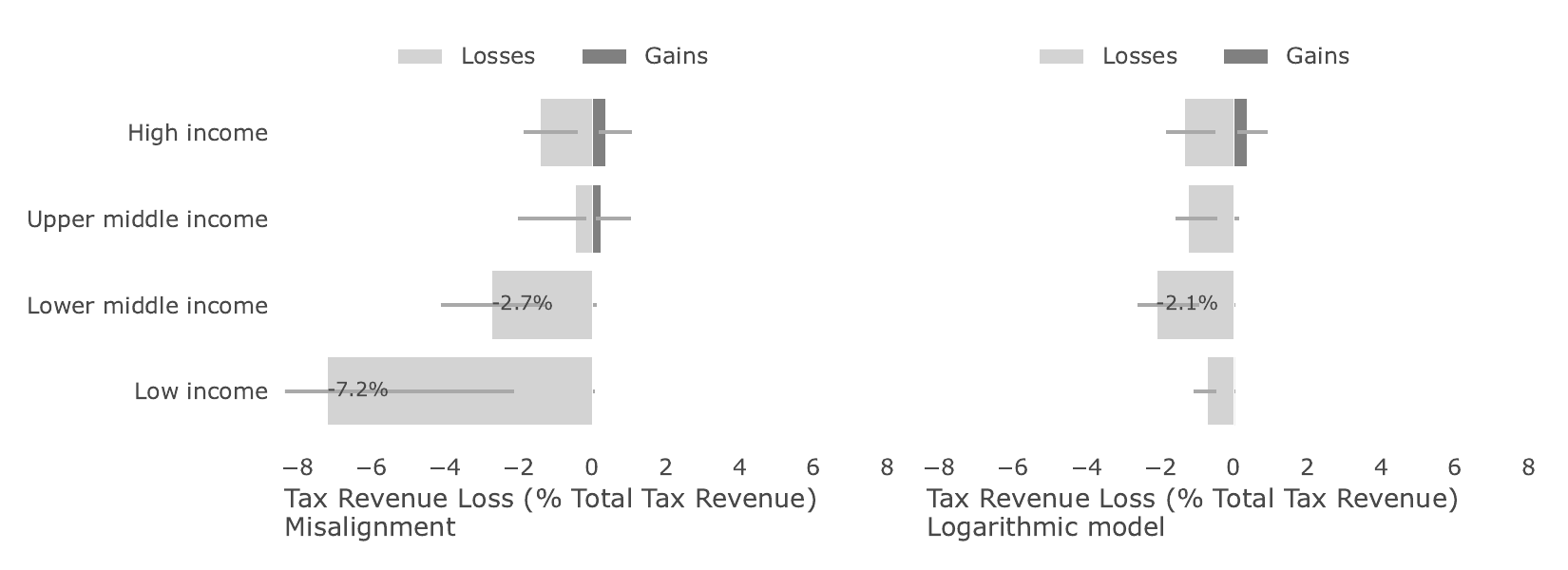}
  \includegraphics[width=1\textwidth]{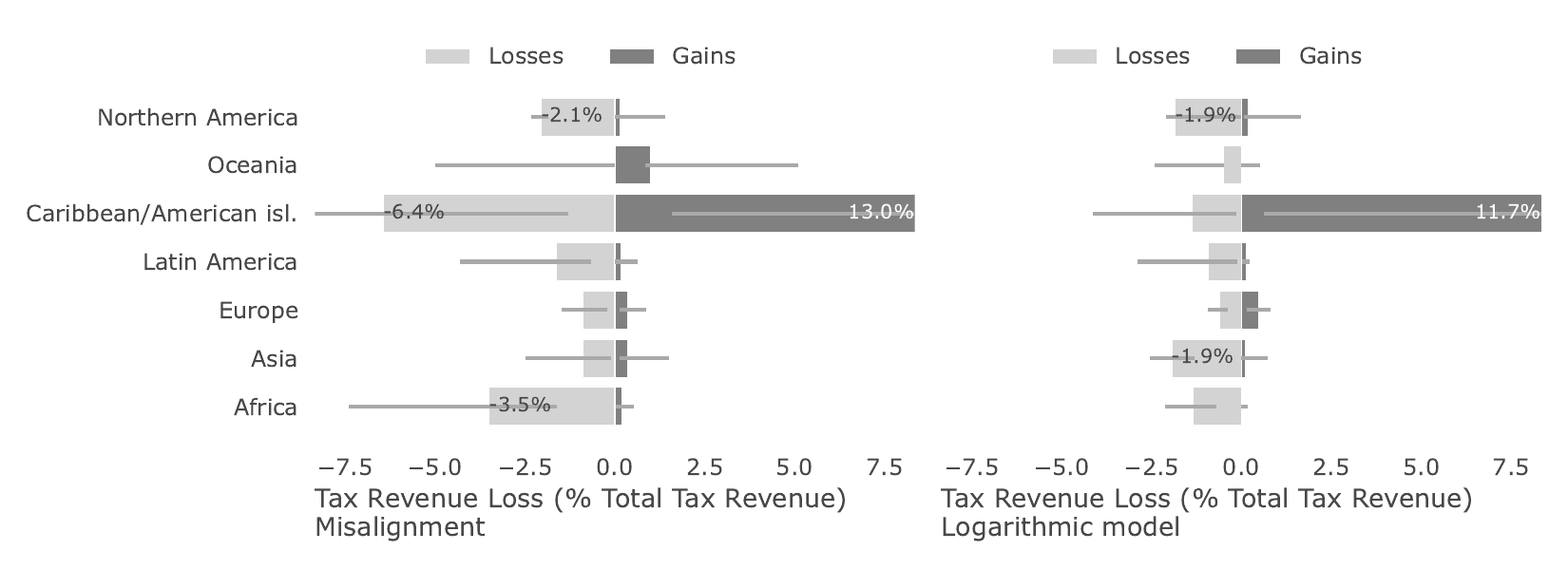}
  \label{fig:diffs_aggs_region_rev}\label{fig:diffs_aggs_income_rev}
\footnotesize
\noindent Notes: Tax revenue loss as a percentage of total tax revenue for countries in different income groups (top row) and different geographical regions (bottom row), as estimated by the misalignment (left side of graph) and logarithmic (right side of graph) models. Confidence intervals show 95\% intervals, calculated via bootstrapping.
\end{figure}

We continue by looking at tax revenue loss (the product of profits shifted and the ETR) as a function of the total tax revenue in each income group and region (Figure ~\ref{fig:diffs_aggs_income_rev}). In general, we find that lower-income-countries – those in Africa and Latin America---tend to lose more tax revenue relative to their total tax revenue. Countries with low and middle per capita incomes (Figure ~\ref{fig:diffs_aggs_income_rev}), are thus the largest profit-shifting losers. MNCs shift an equivalent of 7.19\% (95\% CI; 2.12--17.42) of their total tax revenue out of low-income countries, while receiving influxes equivalent to 0.02\% (0.01--0.07). On the other hand, high-income countries lose an equivalent of 1.43\% (0.40--1.87), while gaining 0.37\% (0.18--1.07). These low tax revenue gains for high-income countries contrast with the high volume of profits shifted in those countries (Figure \ref{fig:diffs_aggs_income_rev_prof_shift}). Furthermore, losses for lower-middle-income countries (1.29--4.11) are also significantly higher than those of higher-income countries, while upper-middle-income countries exhibit only moderate loses (0.15-2.01).

When analysing each country separately (Figure \ref{fig:diffs_income_TRL_rev}), we once again find that lower-income countries lose significantly more tax revenue than high- and upper-middle-income countries. Similar results are found for comparisons of tax revenue losses with corporate income tax revenue (Figure \ref{fig:diffs_income_TRL_crev}) and GDP (Figure \ref{fig:diffs_income_TRL_gdp}). There are, however, differences within lower-income countries. In general, African countries tend to lose the higher share of their tax revenue to profit shifting (Figure \ref{fig:lower_boxen}). 
Overall, our analysis shows that only a small number of countries gain any tax revenue. Profit shifting is thus a phenomenon where the majority of countries lose, and especially so lower-income countries. 
%The magnitude of the tax revenue loss in those countries can be better appreciated by comparing it with public expenditure. Lower-income countries loss the equivalent of 28.2--82.4\% of their government health expenditure (Figure \ref{fig:diffs_income_TRL_health}), or 8.65--31.9\% of their government education expenditure (Figure \ref{fig:diffs_income_TRL_educ}).

With this finding on lower-income countries we contribute to the ongoing discussion of which countries lose more to profit shifting. Few existing studies identify how countries in various income groups are distinctly affected by profit shifting, and the nature of these differences varies across the studies. On the one hand, the theoretical case for such countries' higher vulnerability is strong \citep{hearsonWhenDevelopingCountries2018}, and several studies indicate that low- and lower-middle-income countries (which we label as lower-income countries) are more vulnerable to profit shifting by MNCs than countries at higher levels of income \citep{fuestInternationalDebtShifting2011, johannesenAreLessDeveloped2020}. On the other hand, \textcite{janskyEstimatingScaleProfit2019} compare five sets of country-level estimates---\textcite{torslovMissingProfitsNations2023,clausingEffectProfitShifting2016,cobhamGlobalDistributionRevenue2018, cobhamMeasuringMisalignmentLocation2019} and their own estimates---and four of the five do not suggest that lower-income countries are disproportionately affected by profit shifting. These five studies rely on data from a different number of countries (25, 102, 34, 37, 79), which are all small in comparison with our results for up to 214 countries.

%Our results confirm the theoretical arguments that lower-income countries are more vulnerable to profit shifting \citep{hearsonWhenDevelopingCountries2018,fuestInternationalDebtShifting2011,johannesenAreLessDeveloped2020a}, and indicate that the previous lack of empirical confirmation may have driven by data limitations.

\subsection{Robustness checks and sensitivity analyses}\label{sec:robustness}

We address the data and methodology limitations of this paper by testing the consistency of the main results to our methodological choices. In total, we carry out 10 robustness checks and sensitivity analyses. We briefly summarise them here. Overall, these robustness checks and sensitivity analyses show how our results are robust to changes in the methodology and data.

First of all, we discuss the main sensitivity analyses of our methodology that are not related to the data.
(i) We use a variety of models based on linear, quadratic and logarithmic semi-elasticities, as well as the misalignment method. The scale of profit shifting is estimated to be similar across all models---\$817 (linear), \$880 (quadratic), \$862 (logarithmic), and \$867 (misalignment) billions (Sections \ref{sec:us_results} and \ref{sec:oecd_results}). 
(ii) We test the robustness of the 25\% ETR threshold in equation  \ref{eq:pi}. Reducing the threshold to 20\% would reduce our estimate of profit shifting by the logarithmic model by 6\%. Increasing the threshold to 30\% would increase the estimate of profit shifting by 8\%. 
(iii) We compare our results to those of \textcite{torslovMissingProfitsNations2023} and \textcite{wierGlobalProfitShifting2022}, observing a high correlation with a much increased country sample (Figure \ref{fig:torslov}). 
(iv) We compare the tax revenue loss with other benchmarks, corporate tax revenue (Figure \ref{fig:diffs_income_TRL_crev}) and GDP (Figure \ref{fig:diffs_income_TRL_gdp}). We find that lower-income countries lose comparatively the most in all specifications, as discussed above in section \ref{sec:income_results}. 

We further test the consistency of the main results to other methodological choices in three additional ways:
(v)	We analyse the sensitivity of our results to the offset in the logarithmic model, showing a robust estimation of the coefficients for a wide range of offsets (Figure \ref{fig:offset_robust}).
%[TODO potentially for poitn vii] Mention that future work could investigate a theoretical model of why this extreme non-linear relationship? 
(vi) We compare the logarithmic specification with other specifications that can accommodate extreme non-linearities, including $1/(\tau+ETR)^1$, $1/(\tau+ETR)^2$, $1/(\tau+ETR)^3$ and $coth(\tau+ETR))$. The logarithmic specification allows for higher non-linearities, and exhibits a higher R2 and lowest Bayesian Information Criteria (see Table \ref{tab:app:robustness_log} and Figs. \ref{fig:elasticity_robustness} and \ref{fig:elasticity_robustness_dummies}).
(vii) We test a different redistribution formula. For this, we first regressed the share of profits booked in a country against the shares of employees, capital, sales and wages (Table \ref{tab:red_robust}). We then used the coefficients as our new redistribution formula, after normalising them to sum to one. Profit shifting is reduced by 9\%, with a similar distribution of the origin and destination of profits (Figure \ref{fig:Origin_profits_OECD_ut_weights}).
%ix. We show that the results do not change when we use the original full OECD sample rather than the baseline one we created on the basis of the data for 10 reporting headquarter countries with good coverage of their subsidiaries in tax havens (regression results in Table \ref{tab:oecd_reg_table_full_sample}).

Additional robustness checks and sensitivity analyses focus on the data itself and, in particular, on missing data imputation. In accordance with the design of the individual methods, this missing data imputation does not affect our preferred semi-elasticity methods of estimating the scale of profit shifting, but only influences the measures of misalignment and the subsequent redistribution of the shifted profit for all methods: 
(viii) We estimate missing data using 1,000 bootstrapped data samples (Section \ref{sec:methodology_extrapolation}) to show the consistency of our results in relation to variations in data coverage. In the main results we use the median of the samples. The confidence intervals are included in Figure \ref{fig:ci_misalignment}. 
(ix) We compare the location of employees and revenue according to our missing data model with the information in the original data as well as GDP, showing how our method addresses the limitations of these two alternatives (Figure \ref{fig:share_economies}). 
(x)	Finally, we compare our missing data imputation method with other models, including with penalised linear regression (Appendix \ref{sec:appendix_missing}), showing that our method has higher predictive power.
%xiii. Finally, we run a robustness test in which the data of China was not adjusted. This decreases profit shifted by 9\%, especially reducing profit shifted towards China.

\section{Conclusion}\label{sec:conclusion}

Exploiting the combination of a new methodology and a new dataset, we establish that MNCs shifted around \$900 billion in profits to tax havens in 2017. Our results show that existing linear and quadratic models underestimate profit shifting to countries with extremely low tax rates while simultaneously overestimating it for countries with moderate rates. However, the new logarithmic model as well as the misalignment model are able to capture this behaviour accurately. Using these two preferred models, we show that 40--41\% and 70--83\% of profit shifted is shifted to countries with an ETR below 1\% and 10\%, respectively. Overall, our findings are consistent with the hypothesis that MNCs exploit the combination of globalisation and the sovereignty of individual countries, in particular tax havens, to avoid paying taxes at the expense of countries worldwide regardless of income level.
%Finally, we show that while profit shifting affects both lower- and higher-income countries, lower-income countries lose a higher share of their tax revenue due to profit shifting.

Our findings provide two key insights. First, the extremely non-linear relationship between the location of profits and tax rates has implications for both research and policy. In research, we show that accurately accounting for this relationship affects the estimated scale and distribution of profit shifting. In policy, this modelling choice can significantly influence the assessment of international tax reform, as may be the case with the global minimum tax rate agreement reached in 2021, so-called Pillar Two. While this assessment assumes that profit shifting incentives decrease in linear fashion as the minimum ETR increases, we show in this paper that this linearity assumption is unlikely to hold. Our findings indicate the importance of the specific value of the minimum ETR, which is 15\% in the case of Pillar Two. 
%will determine whether MNCs will continue shifting a similar share of their profits to countries offering the minimum ETR, or, on the other hand, that the minimum ETR will eliminate any profit-shifting incentives.
%Accounting for the extreme non-linearity shows that less than 10\% of profit shifting takes place to countries with ETRs above 10\%, versus 30\% of profit shifting if the non-linearity is not addressed.

Our second key insight is based on our finding that lower-income countries tend to lose more tax revenue relative to total tax revenue due to profit shifting. This could nudge their governments into using confidential tax return data for more detailed analyses---as South Africa recently did, thus learning that profit shifting is highly concentrated among a few large MNCs \citep{wierDominantRoleLarge2022}. In policy, our results indicate that the current international tax system may be hindering the achievement of one of the goals of the 2030 Agenda for Sustainable Development: to strengthen domestic resource mobilisation \citep{unTransformingOurWorld2015, factiFACTIPanelReport2021}. This supports the arguments of lower-income countries that they should be represented on an equal footing at reform discussions and that such reforms should be geared towards creating a level playing field in the corporate taxation of MNCs.
%Our results show that the aggregation of small countries in the data increases the uncertainty about the full effect of profit shifting in these countries.
%when international corporate tax system reform is being debated and decided at the OECD or G20 forums

The findings presented in this paper open up additional avenues for future research. We see two such research directions as especially fruitful. The first is obtaining more accurate estimates as new CBCR data become available in the future. For example, the guidelines to report intra-company dividends have been updated and the 2020 data, which should be published no later than in 2024, will thus contain no double counting. The second research avenue is obtaining more accurate estimates of tax semi-elasticity using firm-level data. An increasing number of MNCs (e.g. Vodafone and Shell) are voluntarily publishing their own CBCR data, and more are likely to do so in the future, either of their own accord or due to government pressure. For example, in 2021 the European Union approved a regulation whereby companies will be required to publish parts of their CBCR data, to be available from around 2025. If firm-level CBCR data become available for a large number of MNCs, it would enable us to understand even more accurately the extent of profit shifting, as well as which MNCs are responsible for the bulk of profit shifting worldwide.
%has a number of limitations that 

\printbibliography

\newpage

\appendix
\renewcommand{\thefigure}{A\arabic{figure}}
\setcounter{figure}{0}
\renewcommand{\thetable}{A\arabic{table}}
\setcounter{table}{0}

% Uncomment to include the appendix
\section{Appendix}

\subsection{Additional data corrections}\label{sec:methodology_extrapolation}

\subsubsection{Data limitations and corrections: imputing missing data}

\begin{figure}[h!]
    \centering
    \includegraphics[width=1\textwidth]{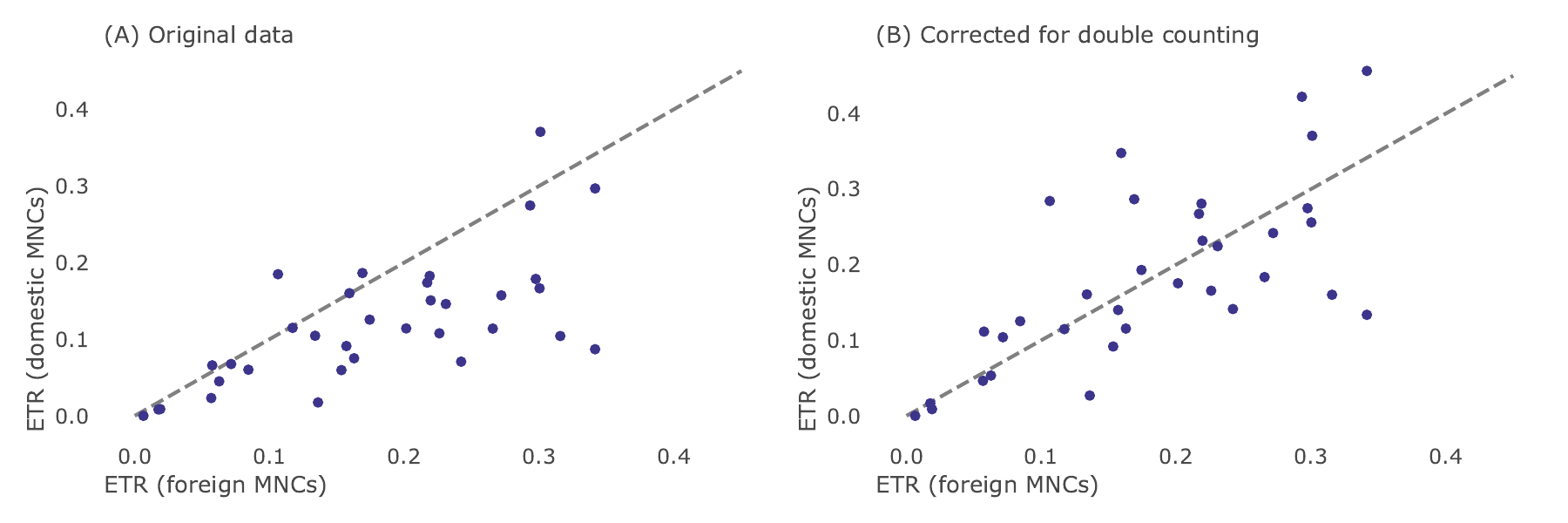}
    \caption{Domestic and foreign ETRs before and after correction for double counting}
    \label{fig:correction_etr}
\end{figure}

While the substantial country coverage, as well as the other advantages of CBCR data, open new avenues for research, several challenges associated with the new data source remain (we summarise them and the discussed advantages in Table~\ref{tab:advantages}). First, a certain extent of double counting in profit due to intercompany dividends is inevitable---MNCs are instructed not to double count intercompany dividends in revenue, but not instructed to do so explicitly in profit. Some countries (e.g. the Netherlands and Sweden) have published associated notes together with their data, showing that domestic operations of MNCs may be considerably affected by this. We correct for double counting in the following ways. First by excluding stateless entities.\footnote{Stateless entities include not only entities whose stateless status results from a mismatch between the legislation of two jurisdictions (e.g. the case of Apple Sales International in Ireland), but includes also flow-through entity (tax-transparent entities). The latter are not considered separate legal entities from their owners, and whose profits are taxed at the level of the owner.}. Second, some countries such as Sweden, the Netherlands, Italy and the United Kingdom investigated the extent of double counting in domestic profits. These range from 16\% in the Netherlands to 51\% in Sweden and the United Kingdom. We include such corrections. Third, by exploiting the fact that multiple data sets are available for US MNCs and remove double counting of US profits---54\% of US domestic profits (as estimated by \citealp{garcia-bernardoDidTaxCuts2022} using the methodology of \citep{horstAssessingDoubleCount2020}) and 10\% of profits in tax havens. Fourth, by removing 10\% of profits in tax havens for non-US MNCs. Fifth, by removing 35\% of domestic profits in all other countries, except Slovenia, Latvia, and Luxembourg, where the ETR on domestic profits is higher than foreign profits, and Belgium, Singapore, Isle of Man and Bermuda, where we remove 50\% since the ETR on domestic profits is much lower than of foreign profits. As a result, we see a higher correlation between the ETRs paid by domestic and foreign companies (Fig. \ref{fig:correction_etr}).

\begin{table}[ht!]
    \caption{Summary of selected advantages and disadvantages of CBCR data}
    \label{tab:advantages}
    \footnotesize
    \setlength{\tabcolsep}{4pt}
\begin{tabular}{p{17cm}}
\toprule
Selected advantages \\
\midrule
Includes data on large MNCs’ profits and tax payments in over 150 jurisdictions for at least 5 headquarter countries. \\
Does not include double counting in revenue.  \\
Enables to use data on large MNCs and those with positive profit only (the latter estimates ETRs more precisely).  \\
\toprule
Selected disadvantages    \\
\midrule
Might include some double counting in profit due to intercompany dividends or stateless entities (which we drop). \\
Includes a sample of large MNCs for 2017 for some countries in aggregated and anonymised form (which we address). \\
\bottomrule
\end{tabular}
\footnotesize
Notes: The table summarises some of the most important advantages and disadvantages of CBCR data from the point of view of using them to estimate profit shifting of multinational corporations worldwide.
\end{table}

\begin{figure}[ht!]
        \begin{center}
 \caption{Country availability in the 2017 OECD data}
  \includegraphics[width=.8\textwidth]{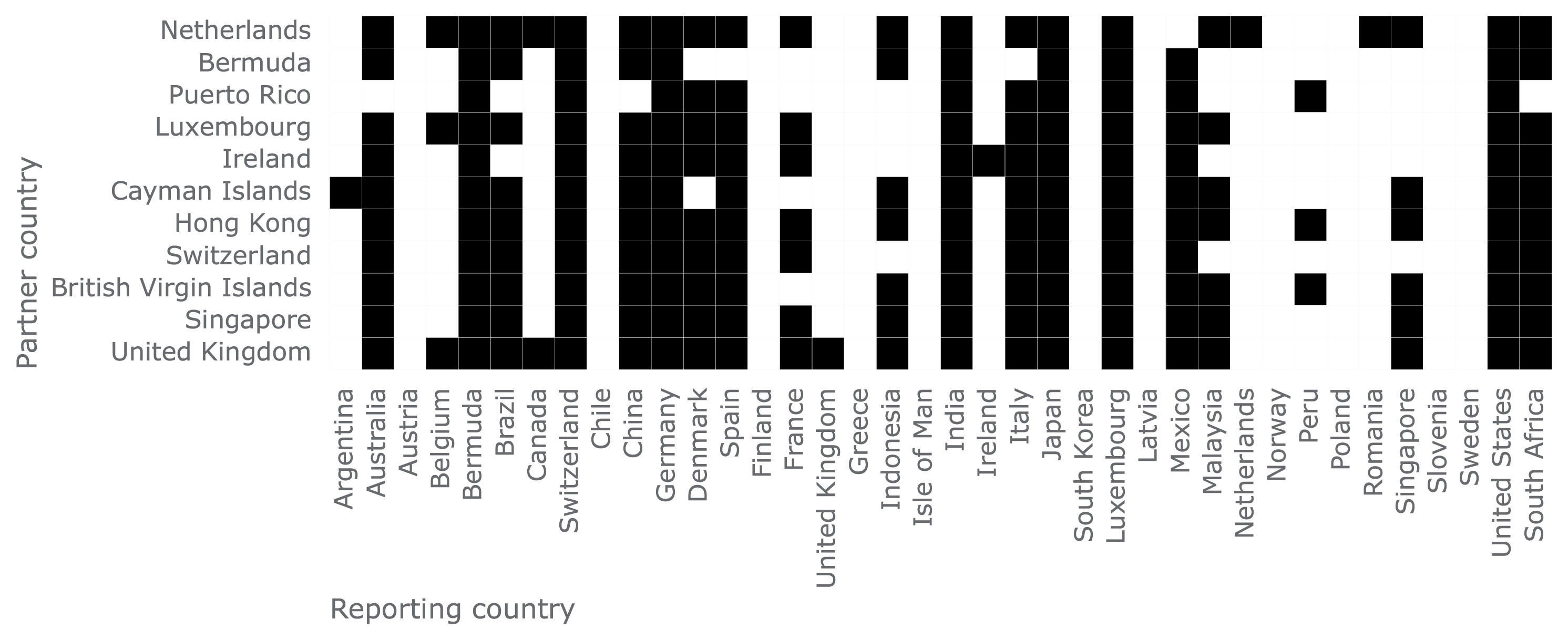}
  \label{fig:sample}
        \end{center}
        \footnotesize
    Notes: Country availability in the 2017 OECD CBCR data. Reporting countries (horizontal axis) reporting on selected countries often considered tax havens (vertical axes) are depicted with black squares. While every MNC reports on its activities in every jurisdiction, it is decided by each country whether it enables publication of its CBCR data through the OECD and how aggregated the published data are. This figure depicts the heterogeneity in countries' decisions. Only countries that decided to share some data and report on at lest one of the major tax havens are on the horizontal axis (reporting country). The figure depicts whether the reporting country allows the OECD to publish the country-level data for each of the selected partner countries (major tax havens) on the vertical axis. For example, the United States enables publication of information for all of those tax havens, while Austria or Norway or Poland for none of them (and therefore not shown in the table).
\end{figure}

Second, while the availability of CBCR data constitutes a significant step forward, and partially corrects this issue, the data is still not complete and is not systematically disaggregated by jurisdiction. The CBCR regulation has been implemented by approximately 100 countries so far---only 38 of them agreed to share their data publicly in aggregated and anonymised form; moreover, some have chosen to aggregate data to a far greater extent than others (Table~\ref{tab:ratio}), with the US and Japan (145 and 198 jurisdictions, respectively) leading the way (Table \ref{tab:n_jur_country}). Figure~\ref{fig:sample} shows how many reporting countries report on selected countries often considered tax havens and it captures well the heterogeneity in countries' decisions to aggregate the data.

In the remainder of this section, we deal empirically with three issues related to data completeness: the lack of completeness in the data of reporting countries; the varying combinations of countries in the aggregated country categories; and the lack of reporting by some countries. Other limitations of the CBCR data (e.g. revenue unavailable according to the location of the final customer) are discussed by the OECD, which published the data with an ``Important disclaimer regarding the limitations of the country-by-country report statistics'', and by \textcite{garcia-bernardoMultinationalCorporationsTax2021a}, \textcite{clausingProfitShiftingTax2020} or \textcite{sullivanFiveYearsCbC2023}.

The first limitation concerns the lack of completeness in the data of reporting countries. We address this limitation by comparing the number of companies in Orbis, a frequently used database covering over 300 million public and private firms worldwide, with the number of companies observed in CBCR (Table~\ref{tab:ratio}). Orbis has good coverage regarding the number and consolidated revenue of large MNCs \citep{garcia-bernardoEffectsDataQuality2018}, but poor information at the subsidiary level \citep{bajgarUseNotUse2018, garcia-bernardoMultinationalCorporationsTax2021a, phillipsGroupSubsidiariesTax2020}. 
%While the number of companies observed and expected are similar for most countries, we observe large differences in the case of some countries. We therefore multiplied all reported financial information by a ratio listed in Table~\ref{tab:ratio} in case that ratio was above one, with the exception of two countries – the US and China. In the case of the US, we expected 1,501 companies according to Orbis. Instead, we find 1,101 companies in the 2016 data. This is due to a lack of completeness of 2016 data in the US \citep{garcia-bernardoMultinationalCorporationsTax2021a}. US IRS data for 2017 indicates that we should observe approximately 1,575 companies---1,548 with profits in at least one jurisdiction. In order to correct for this disparity we use US data for 2017 (the US is the only country that has published data for 2017). In China, instead of the expected 583, only 82 companies reported satisfactory data to the OECD. However, those 82 companies reported US\$2.9 trillion of sales domestically, and US\$0.45 trillion abroad; for comparison, the numbers for the US in 2016 were US\$7.8 trillion domestically and US\$3.4 trillion abroad. This indicates that the data is not as erratic as it may appear. Lacking a better heuristic, we multiply the financial information for China by a conservative factor of two, and run robustness tests to assess the impact of our correction (Section \ref{sec:robustness}).
We find a good correlation between the Orbis and CBCR data Table~\ref{tab:ratio}, which indicates that CBCR data is complete in 2017. 

The second limitation concerns the combination of countries in aggregated categories---for example, Chile and the British Virgin Islands may be grouped together in ``Other Americas''. The aggregation criterion is different for different countries. While India and South Africa do not seem to aggregate data, the US aggregates countries with a low number of reporting MNCs. This is problematic, as aggregation affects particularly lower-income countries and low tax jurisdictions.  For instance, only three countries report information on Gambia, and only seven countries report on the Isle of Man. The other countries aggregate information on Gambia and the Isle of Man in larger categories such as Other Africa and Other Europe. If we decided to ignore this grouped data, we would be missing a significant part of the operations in those countries, leading to an underestimation of the extent of profit shifting. This is acknowledged by the Economic Analysis and Impact Assessment of the OECD \cite{oecdTaxChallengesArising2020}, who impute missing sales by extrapolating using a gravity model using data available in the CBCR, Orbis, and the OECD’s Activity of Multinational Enterprises database, as well as foreign direct investment and GDP data.

We address these biases by modelling the location of employees and sales for each pair of countries using a Histogram-based Gradient Boosting Regression Tree, a type of gradient boosting based on decision trees that frequently outperforms other machine learning algorithms, while offering some interpretability on the most relevant features \citep{keLightgbmHighlyEfficient2017,friedmanGreedyFunctionApproximation2001}. Specifically, we use the Python implementation in scikit-learn \citep{pedregosaScikitLearnMachineLearning2011}. Another of its advantages is that it offers native support for missing values, and as such is able to use a large range of features without data imputation. We train the location of profits, employees, sales and tangible assets using variables from the gravity dataset of CEPII, imports and exports from UN Comtrade, and foreign direct investment from the World Bank, as well as from other sources detailed in Table E1 in Appendix \ref{sec:appendix_missing}. We obtain a mean out-of-sample R-square of 0.74, 0.56, 0.57 and 0.44 respectively for employees, sales, profits and tangible assets.

\begin{figure}[ht!]
 \caption{Available information on CBCR}
  \includegraphics[width=1\textwidth]{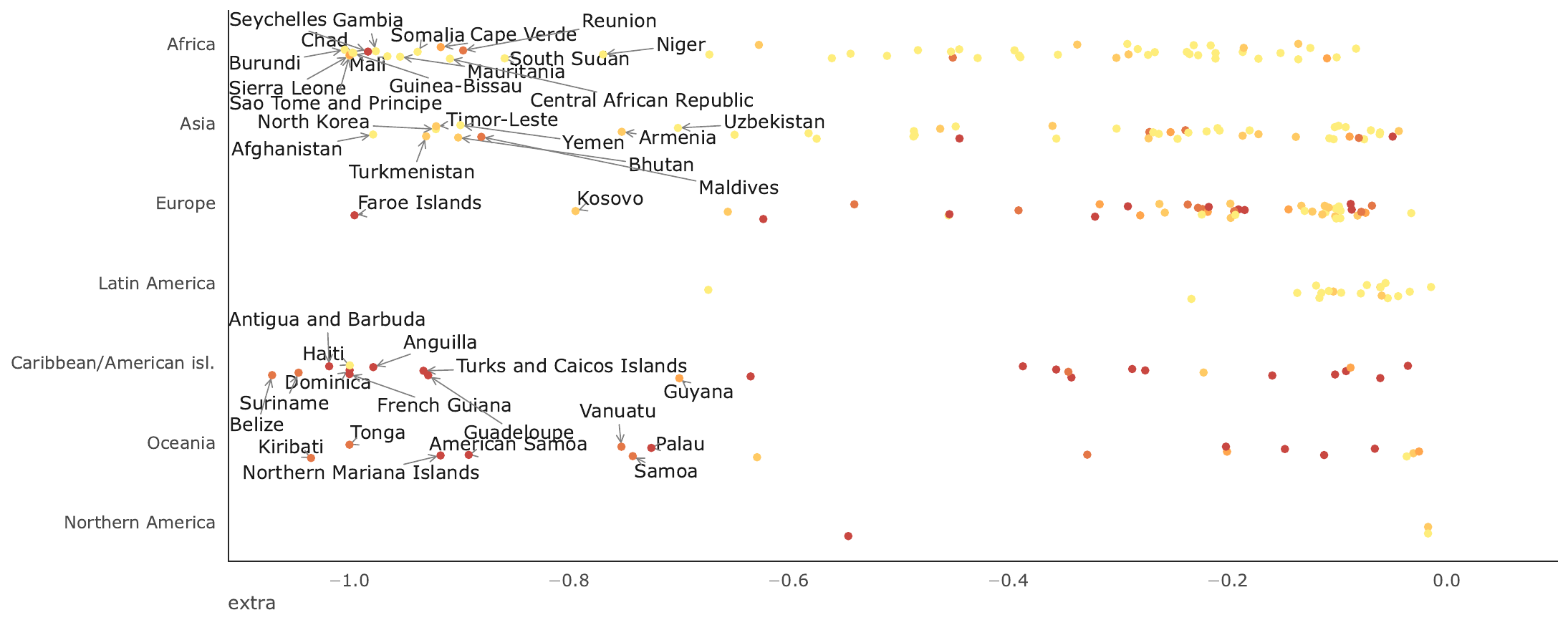}
  \label{fig:available_cbcr}
\footnotesize
Notes: Colour denotes increasing GDP per capita. Countries with availability below 20\% are annotated. All countries with availability below 10\% are placed to the left of the 10\% horizontal line.
\end{figure}

We use the model to estimate the total number of employees, unrelated party sales and tangible assets for each pair of countries in the world. For reporting countries, we then adjust the estimated values so their sum corresponds to the aggregated sum in CBCR. We demonstrate our approach using the following model scenario: French MNCs have 10,000 employees in Other America, and Other America comprises Paraguay and Suriname---we can establish this by checking which countries are missing from the CBCR data of France. If our model estimates 6,000 employees in Paraguay and 5,000 employees in Suriname, we multiply the employees of those countries by 10,000 and divide by 11,000. In the next step, we compare for each country the sums of those estimated values with the sums of the values observed in the CBCR data. We then use the lowest of the two ratios (estimated vs. reported employees and sales) to adjust the profits shifted in order to correct for the combination of small countries in aggregated groups. We cap this ratio at 10---that is, if the model expects that the OECD data is less than 10 per cent complete, we consider it to be 10 per cent complete. While the estimation of missing economic activity increases total shifted profits by approximately 30 per cent, it is key with respect to accounting for missing data in countries underrepresented in the sample---typically lower-income countries. Without this step, we would redistribute too few profits to those countries. Figure~\ref{fig:available_cbcr} shows the available information on CBCR, displaying how data coverage is especially worrisome in the case of lower-income countries.

\begin{figure}[ht!]
 \caption{Relationship between GDP and activity for countries in the OECD data}
  \includegraphics[width=0.5\textwidth]{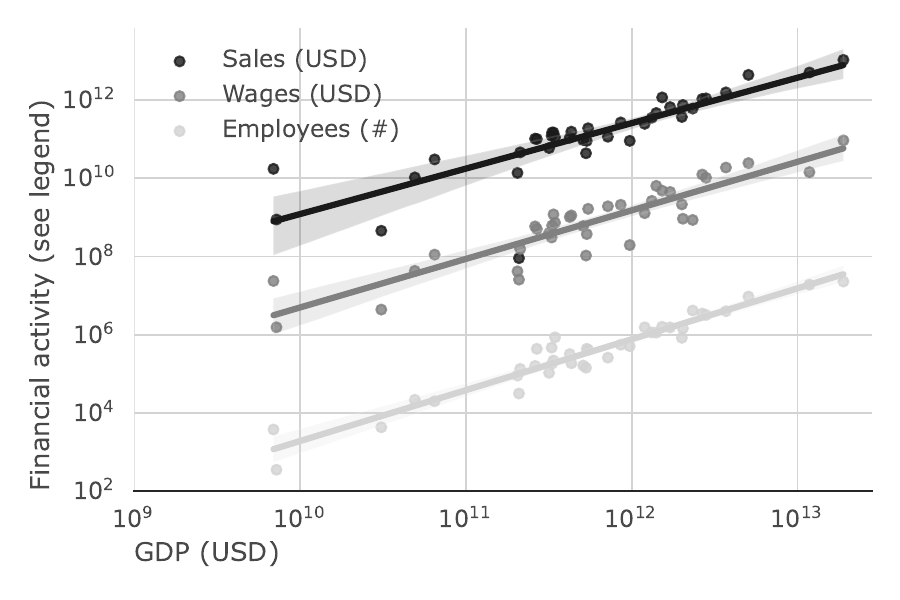}
  \label{fig:domestic_gdp}
\newline
\footnotesize
Notes: Relationship between GDP and domestic employees, sales and tangible assets for countries in the 2016 OECD data. Each dot corresponds to one country in the data.
\end{figure}

\begin{figure}[ht!]
 \caption{Relationship between the number of large MNCs and the total profits reported domestically in the country}
  \includegraphics[width=0.5\textwidth]{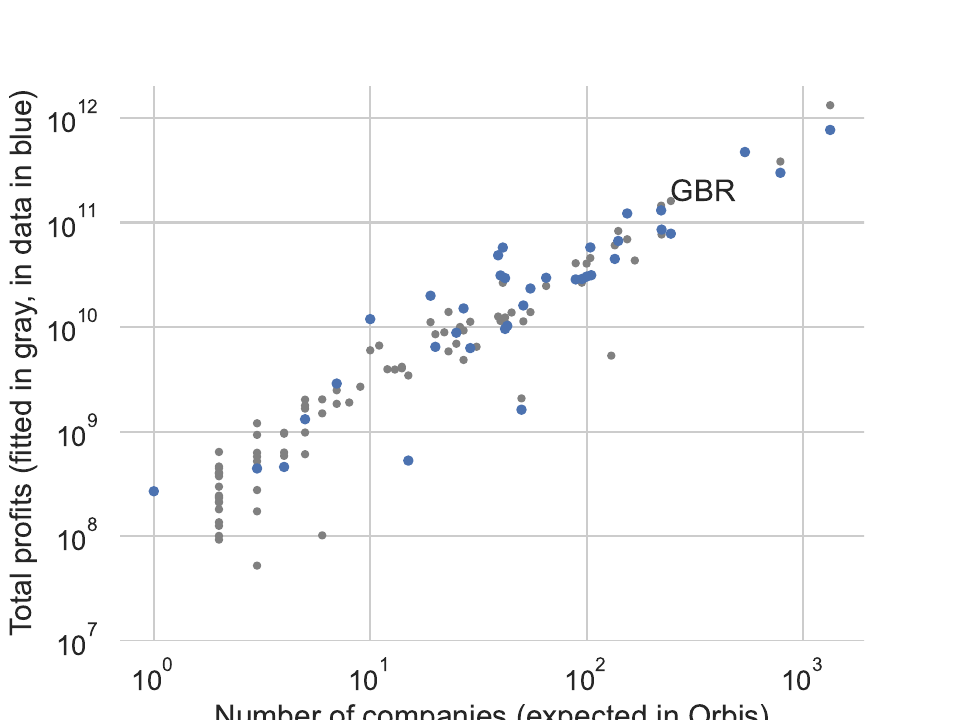}
  \label{fig:orbis_vs_domestic}
\newline
\footnotesize
Notes: Relationship between the number of large MNCs (extracted from Orbis), and the total profits reported domestically in the country. Estimated values using a regression with GDP, population, the average ETRs and the total consolidated banking claims on an immediate counterparty basis (Table B4 of the Bank for International Settlements (BIS) data) are visualised in grey. Empirical values are visualised in blue.
\end{figure}

The third limitation concerns the lack of reporting by some countries, including Russia, which excludes MNCs headquartered in those countries from the sample---but we do have the operations in those countries of MNCs headquartered in reporting countries. This limitation is partially addressed in the previous step, where financial information for all pairs of countries is estimated, even for non-reporting countries. However, the information on domestic activities of MNCs is important, especially for large countries. This is addressed by estimating the number of domestic employees, revenue and tangible assets for all non-reporting countries. We do so by using a linear model based on the number of expected companies in each country, its GDP, population, the average ETRs and the total consolidated banking claims on an immediate counterparty basis (Table B4 of the BIS data) (R-square 0.97, 0.93 and 0.88 respectively; see also Figures \ref{fig:domestic_gdp} and \ref{fig:orbis_vs_domestic}).

Importantly, in the logarithmic model, we only use the fixes to the second and third limitations to redistribute profits back to the home countries, but not to calculate profit shifted. We do this since we do not have accurate estimates of ETR for MNCs of non-reporting countries. Instead, we divide the total profit shifted by the share of GDP of the countries of the sample (83\%). This is a similar figure to the available economic activity estimated in the CBCR data using the fixes to the second and third limitation (94\%). Using the GDP in the logarithmic model may be a conservative strategy, since we are assuming that the MNCs of non-reporting countries (e.g., the Cayman Islands, the British Virgin Islands) are similar to those of reporting countries.

Finally, we assess our results’ sensitivity to the estimation of missing information. To do so, we train the models 1,000 times using bootstrapped samples of the data (i.e. the gradient boosting ensemble to address the second limitation and the linear regression to address the third limitation) and record the impact in our results. Since the sampling randomly removes information, samples without important dyads (e.g. USA-Netherlands, or China–Hong Kong) will be more affected. This thus offers a conservative strategy that allows us to partially understand how our results depend on methodological choices. In the end, we use median values as our preferred point estimates.

\begin{figure}[ht!]
 \caption{Comparison between the redistribution formula}
  \includegraphics[width=1\textwidth]{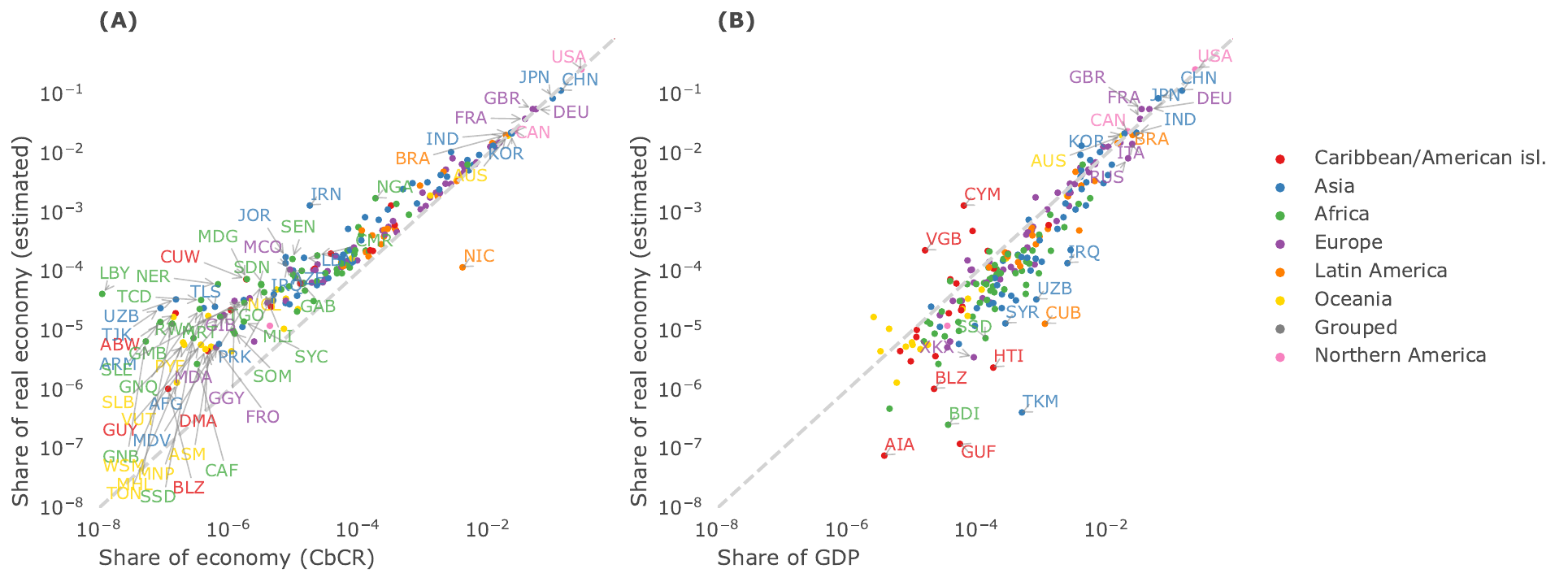}
  \label{fig:share_economies}
\footnotesize
Notes: Comparison between the redistribution formula (eq. \ref{eq:red}) (A) imputing missing data vs using raw data of firms with positive profits; and (B) imputing missing data vs using the share of GDP. Note that the estimated shares of the economy for African countries are higher than the shares of the economy for those countries in the raw data, but lower than the share of GDP of those countries.
\end{figure}

The difference between the observed and estimated location of employees and sales is visualised in Figure~\ref{fig:share_economies}. In comparison with the observed location of the economy, the estimated location is more balanced, giving less weight to reporting countries, and affecting especially Asian and African countries---see the largest outliers in Figure~\ref{fig:share_economies}. Our estimated location of the economy matches closely the share of GDP for richer countries, while departs for developing countries (Figure~\ref{fig:share_economies}B). This is expected given the lower presence of large MNCs in developing countries.

In addition to those limitations discussed above, we briefly refer to other limitations identified in the use of the data and discussed by the OECD, which published the data with an ``Important disclaimer regarding the limitations of the country-by-country report statistics'', an updated version of which has been published in July 2021.

\subsubsection{Modelling missing employees and revenues} \label{sec:appendix_missing}
We train the location of profits, employees and sales using variables from the gravity dataset of CEPII, the World Bank data (WBD), the United Nations data (UN), the International Monetary Fund (IMF), the UN COMTRADE database (COMTRADE), Linkedin, the Tax Justice Network financial secrecy and corporate tax haven indexes (TJN), the Bank of International Settlements (BIS), the International Labour Organization (ILO), the World Health Organization (WHO), and the Government Revenue Dataset (UNU-WIDER GRD). A complete list of variables can be found at the end of this Annex.

\begin{figure}[ht!]
 \caption{The prediction power of the algorithm, in-sample and out-of sample estimations}
  \includegraphics[width=1\textwidth]{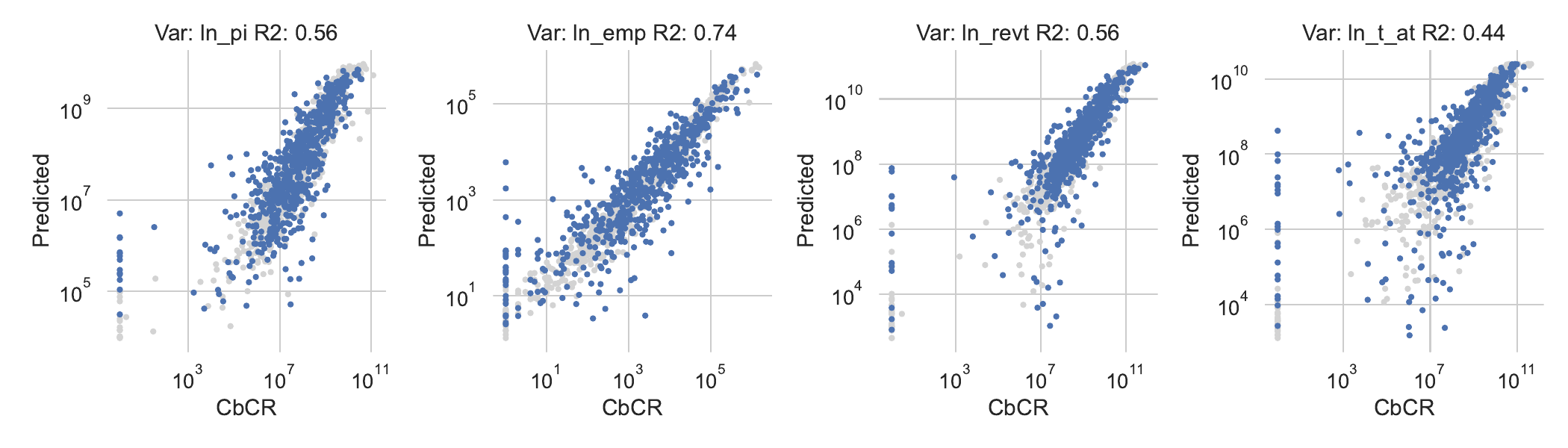}
  \label{fig:predicted_emp_revt}
\footnotesize
Notes: The boosting model was fit on 60\% of the sample and tested in the other 40\%. In-sample estimations are depicted in light grey, while out-of sample estimations of employees, unrelated party sales and tangible assets are visualised in blue. Note that this is one split of the data. The cross-validated r-squares are displayed in the main text. The parameters (learning\_rate=0.105; 0.168; 0.073; 0.042 respectively, l2\_regularization=158.5; 100; 63.1; 39.8 respectively and min\_samples\_leaf=20 for all) were set using cross-validation.
\end{figure}

\begin{figure}[ht!]
 \caption{A comparison with a Lasso regression}
  \includegraphics[width=1\textwidth]{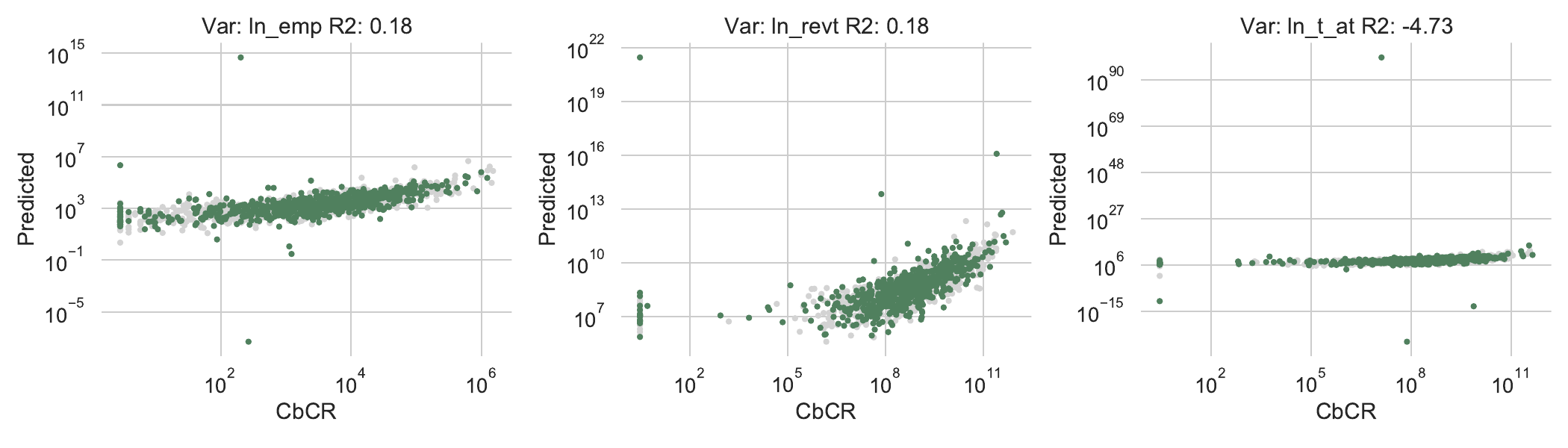}
  \label{fig:predicted_emp_revt_lasso}
\footnotesize
Notes: A penalised linear regression model was fit on 60\% of the sample and tested in the other 40\%. In-sample estimations are depicted in light grey, while out-of sample estimations of employees, unrelated party sales and tangible assets are visualised in blue. Note that this is one split of the data. The penalty parameter (alpha=1.0 for all) was set using cross-validation.
\end{figure}

We used these variables to predict the location of employees and revenues. We tested several models, and Histogram-based Gradient Boosting Regression Tree---a type of gradient boosting based on decision trees which frequently outperforms other machine learning algorithms while offering some interpretability on the most relevant variables \citep{keLightgbmHighlyEfficient2017,friedmanGreedyFunctionApproximation2001}---was shown to perform the best. A sample of the prediction power of the algorithm is visualized in Figure~\ref{fig:predicted_emp_revt}, and a comparison with Lasso regression in Figure~\ref{fig:predicted_emp_revt_lasso}. Since Lasso does not have native support for missing values, these are imputed using the sklearn function IterativeImputer, which provides with an strategy for imputing missing values by modeling each variable as a function of other variables in a round-robin fashion.

\begin{figure}[ht!]
 \caption{Permutation Importance of each variable for the prediction of profits and sales}
  \includegraphics[width=1\textwidth]{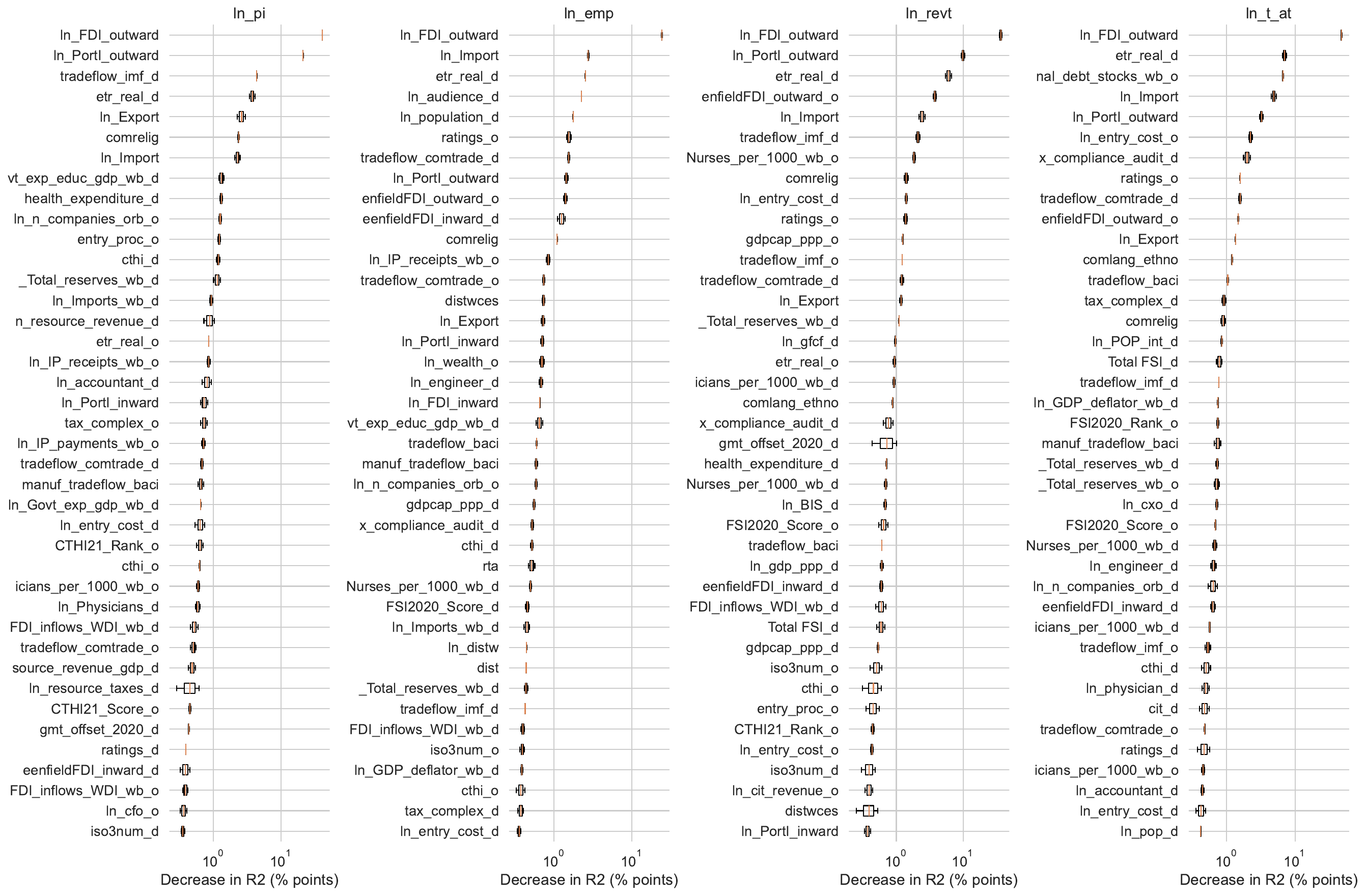}
  \label{fig:permutation_estimation}
\footnotesize
Notes: Permutation Importance of each variable for the prediction of profits (ln\_pi, not used in the paper),  (ln\_emp), and sales (ln\_revt).
\end{figure}

Finally, we investigated which variables were more important in the estimation using permutation feature importance \citep{breimanRandomForests2001}. The permutation feature importance is defined as the decrease in the R-square of the model when the values of a variable are randomly shuffled. We permutated the values of the original data (i.e., not of the 1000 bootstrap samples) 100 times to get a confidence interval, which is visualized in Figure \ref{fig:permutation_estimation}. For the estimation of profits (ln\_pi, not used in the paper), we find that outward FDI and Portfolio Investment are the most important variables. For the estimation of employees (ln\_emp) and sales (ln\_revt), outward FDI and Exports are the most important predictors. Importantly, we are not trying to rationalise the variables chosen by the algorithm, just to create an accurate model to impute missing values.

\clearpage

\subsubsection{Modelling missing employees and revenues: A complete list of variables}

\footnotesize
\begin{longtable}{p{4cm}p{8cm}p{4cm}}
\caption{Variables used in the imputation of missing data}\\
\textbf{Variable}                       & \textbf{Description}                                                                                                      & \textbf{Source}                                                                                          \\
iso3                                    & ISO-3 code                                                                                                                &                                                                                                       \\
revt, emp, pi, txc                      & Revenue, employees, profits and cash taxes                                                                                & CBCR                                                                                                     \\
                                        &                                                                                                                           &                                                                                                          \\
\textbf{Bilateral Variables}            &                                                                                                                           &                                                                                                          \\
ln\_Import                              & Log of total imports from origin to destination                                                                           & COMTRADE                                                                                                 \\
ln\_Export                              & Log of  total exports from origin   to destination                                                                        & COMTRADE                                                                                                 \\
ln\_FDI\_inward                         & Log of total FDI from orogin to destination                                                                               & IMF CDIS                                                                                                 \\
ln\_FDDI\_outward                       & Log of total FDI from destination to origin                                                                               & IMF CDIS                                                                                                 \\
ln\_PortI\_inward                       & Log of total portfolio investment from orogin to destination                                                              & IMF CPIS                                                                                                 \\
ln\_PortI\_outward                      & Log of total portfolio investment from destination to origin                                                              & IMF CPIS                                                                                                 \\
ln\_dClaims                             & Log of total banking claims (derived from partners)                                                                       & BIS Table A6.2                                                                                           \\
ln\_dLiabilities                        & Log of total banking liabilities (derived from partners)                                                                  & BIS Table A6.2                                                                                           \\
ln\_distw                               & Ln of distance between countries                                                                                          & CEPII GravData                                                                                           \\
tdiff                                   & Time zones difference (hours)                                                                                             & CEPII GravData                                                                                           \\
transition\_legalchange                 & Dummy, 1 if common legal origin changed since transition                                                                  & CEPII GravData                                                                                           \\
eu\_to\_acp                             & Dummy,  EU/member exporting to an   ACP country (a preferential trade agreement on imports)                               & CEPII GravData                                                                                           \\
acp\_to\_eu                             & Dummy, ACP country exporting to an EU/member (a preferential trade   agreement on imports)                                & CEPII GravData                                                                                           \\
col45                                   & Dummy, colonial relationship post 1945                                                                                    & CEPII GravData                                                                                           \\
col\_fr                                 & Dummy, origin of colonial relationship post 1945                                                                          & CEPII GravData                                                                                           \\
col\_to                                 & Dummy, destination of colonial relationship post 1945                                                                     & CEPII GravData                                                                                           \\
colony                                  & Dummy, colonial relationship (ever)                                                                                       & CEPII GravData                                                                                           \\
comcol                                  & Dummy, common colonizer post 1945                                                                                         & CEPII GravData                                                                                           \\
comcur                                  & Dummy, common currency                                                                                                    & CEPII GravData                                                                                           \\
comlang\_ethno                          & Dummy, common language (\textgreater{}9\% population)                                                                     & CEPII GravData                                                                                           \\
comlang\_off                            & Dummy, common official language                                                                                           & CEPII GravData                                                                                           \\
comleg\_posttrans                       & Dummy, common legal origins after transition                                                                              & CEPII GravData                                                                                           \\
comleg\_pretrans                        & Dummy, common legal origins before transition                                                                             & CEPII GravData                                                                                           \\
comrelig                                & Religious proximity index                                                                                                 & CEPII GravData                                                                                           \\
contig                                  & Dummy, contiguity                                                                                                         & CEPII GravData                                                                                           \\
curcol                                  & Dummy, current colonial relationship                                                                                      & CEPII GravData                                                                                           \\
cursib                                  & Dummy, current sibling relationship (common colonizer)                                                                    & CEPII GravData                                                                                           \\
sibling                                 & Dummy, ever sibling relationship (common colonizer)                                                                       & CEPII GravData                                                                                           \\
fta\_wto                                & Dummy, regional trade agreement (WTO)                                                                                     & CEPII GravData                                                                                           \\
gsp                                     & Dummy if donator in Generalized System of Preferences                                                                     & CEPII GravData                                                                                           \\
heg\_o                                  & Dummy, 1 if origin is current of former hegemon of destination                                                            & CEPII GravData                                                                                           \\
heg\_d                                  & Dummy, 1 if destination is current of former hegemon of origin                                                            & CEPII GravData                                                                                           \\
gsp\_d\_d                               & Dummy, 1 if origin is donator in Generalized System of Preferences                                                        & CEPII GravData                                                                                           \\
gsp\_o\_d                               & Dummy, 1 if destination is donator in Generalized System of Preferences                                                   & CEPII GravData                                                                                           \\
                                        &                                                                                                                           &                                                                                                          \\
                                        &                                                                                                                           &                                                                                                          \\
\multicolumn{3}{l}{\textbf{Unilateral  variables: Included for the reporting and partner countries}}                                                                                                                                                                           \\
\multicolumn{2}{l}{\textbf{Legislative/historical/Geographical}}                                                                                                    &                                                                                                          \\
entry\_proc                             & Start-up procedures to register a business (Number)                                                                       & CEPII GravData                                                                                           \\
entry\_time                             & Time required to start a business (days)                                                                                  & CEPII GravData                                                                                           \\
entry\_tp                               & Days+Procedures to start a business                                                                                       & CEPII GravData                                                                                           \\
gatt                                    & GATT member                                                                                                               & CEPII GravData                                                                                           \\
EU28                                    & Dummy, country belonging go the EU-28                                                                                     &                                                                                                       \\
OECD                                    & Dummy, country belonging go the OECD                                                                                      &                                                                                                       \\
Ukt                                     & Dummy, UK-territory                                                                                                       &                                                                                                       \\
region\_tjn                             & Region                                                                                                                    & TJN                                                                                                      \\
ln\_area                                & Log of area in sq. kms                                                                                                    & CEPII GravData                                                                                           \\
ln\_entry\_cost                         & Log of cost of business start-up procedures (log of \% GNI per capita)                                                    & CEPII GravData                                                                                           \\
english                                 & Official language 1 in the CEPII GeoDist dataset                                                                          & CEPII GeoDist                                                                                            \\
governance                              & First PCA component of the six dimensions of the Worldwide Governance   Indicators project                                & WBD                                                                                                      \\
                                        &                                                                                                                           &                                                                                                          \\
\textbf{Socio-economic} & \textbf{}                                                                                                                 &                                                                                                          \\
Nurses\_per\_1000                       & Nurses per 1000 inhabitants                                                                                               & WBD                                                                                                      \\
Physicians\_per\_1000                   & Doctors per 1000 inhabitants                                                                                              & WBD                                                                                                      \\
ln\_Nurses                              & Log of number of nurses                                                                                                   &                                                                                                          \\
ln\_physician                           & Log of number of doctos                                                                                                   &                                                                                                          \\
ln\_pop                                 & Population (source CEPII)                                                                                                 & CEPII GravData                                                                                           \\
ln\_population                          & Population (source WBD)                                                                                                   & WBD                                                                                                      \\
ln\_POP\_int                            & Population (manually completed)                                                                                           & WBD, UN, CIA                                          \\
ln\_GDP\_int                            & GDP (manually completed)                                                                                                  & WBD, UN, CIA                                          \\
ln\_gdp\_d                              & GDP                                                                                                                       & CEPII GravData                                                                                           \\
ln\_gdpcap\_d                           & GDP per capita                                                                                                            & CEPII GravData                                                                                           \\
ln\_gdppc\_d                            & GDP per capita                                                                                                            & ln\_GDP\_int - ln\_POP\_int                                                                              \\
ln\_Health\_expenditure\_gdp            & Log of health expenditure (\% of gdp)                                                                                     & WBD                                                                                                      \\
ln\_uhnwi                               & Log10 of the number of high net worth individuals (adults with   wealth above 50 millions)                                & Global Wealth Report 2018 by Credit Suisse                                                               \\
ratings                                 & Trading Economic credit rating, composed from the credit ratings by   Moody’s, S\&P, Fitch and DBRS                       & Feb. 2019 tradingeconomics.com                                                                           \\
ln\_n\_companies\_orb                   & Log of number of MNCs with a turnover higher than 750M in Orbis                                                           & Orbis                                                                                                    \\
ln\_GreenfieldFDI\_inward               & Total greenfield FDI into the country                                                                                     & UNCTAD                                                                                                   \\
ln\_GreenfieldFDI\_outward              & Total greenfield FDI out of the country                                                                                   & UNCTAD                                                                                                   \\
ln\_BIS                                 & Log of total consolidated banking claims on an immediate counterparty   basis                                             & BIS (Table B4)                                                                                           \\
ln\_ExternalDebtStocks                  & Log of External Debt Stock                                                                                                & WBD                                                                                                      \\
ln\_consumption                         & Log10 of final consumption expenditure by households and non-profit   institutions serving households (constant 2010 USD) & \begin{tabular}[c]{@{}l@{}}Mean 2014-2018\\      NE.CON.PRVT.KD\end{tabular}                             \\
ln\_gfcf                                & Log10 of gross fixed capital formation (constant 2010 USD)                                                                & \begin{tabular}[c]{@{}l@{}}Mean 2014-2018\\      NE.GDI.FTOT.KD\end{tabular}                             \\
ln\_FDI\_Inflows\_WDI\_d                & Log of total FDI inflows                                                                                                  & WBD                                                                                                      \\
ln\_imports\_wbd                        & Log of total imports in the country                                                                                       & WBD                                                                                                      \\
ln\_ip\_payments\_wbd                   & Log of IP payments in the country                                                                                         & WBD                                                                                                      \\
ln\_ip\_receipts\_wbd                   & Lof of IP receipts in the country                                                                                         & WBD                                                                                                      \\
ln\_exports\_wbd                        & Log of total exports in the ocuntry                                                                                       & WBD                                                                                                      \\
ln\_month\_wage                         & Log of monthly wage                                                                                                       & ILO                                                                                                      \\
ln\_govt\_exp\_educ\_sgdp\_wb           & Log of government expenditure in education (\%GDP)                                                                        & WBD                                                                                                      \\
ln\_who\_gvt\_health\_expenditure       & Log og public expenditure in health care                                                                                  & WHO                                                                                                      \\
ln\_cit\_revenue                        & Log of government revenue from corporate income tax                                                                       & UNU - WIDER GRD                                                                                          \\
ln\_resource\_revenue                   & Log of government revnue from resource taxes and fees                                                                     & UNU - WIDER GRD                                                                                          \\
ln\_resource\_taxes                     & Log of government revnue from resource taxes                                                                              & UNU - WIDER GRD                                                                                          \\
ln\_resource\_revenue\_gdp              & Log of government revnue from resource taxes and fees (\% GDP)                                                            & UNU - WIDER GRD                                                                                          \\
ln\_total\_taxes\_revenue               & Log of total government revenue from taxes                                                                                & UNU - WIDER GRD                                                                                          \\
                                        &                                                                                                                           &                                                                                                          \\
\textbf{Secrecy and tax}                & \textbf{}                                                                                                                 &                                                                                                          \\
Total FSI                               & Financial Secrecy Score                                                                                                   & TJN                                                                                                      \\
cit                                     & Statutory Corporate Income tax rates                                                                                      & \begin{tabular}[c]{@{}l@{}}Mean 2014 - 2018,\\      \citep{janskyEstimatingScaleProfit2019}\end{tabular} \\
cthi                                    & Corporate Tax Haven Score                                                                                                 & TJN                                                                                                      \\
etr\_real                               & Effective tax rate, capped at 0.6 and using CIT for missing and negative   values.                                        & CBCR weighted average                                                                                    \\
ln\_cit                                 & Log of cit                                                                                                                & KPMG, EY, PwC                                                                                            \\
ln\_etr\_real                           & Log of etr\_real                                                                                                          & CBCR weighted average                                                                                    \\
tax\_complex                            & Time to prepare and pay taxes (hours)                                                                                     & Mean 2014 - 2018, IC.TAX.DURS (WBD)                                                                      \\
ln\_accountant\_d                       & Log of number of accountants                                                                                               & Linkedin (Garcia-Bernardo and Stausholm, Forthcoming)                                                    \\
ln\_all\_tax                            & Log of number of all tax professionals                                                                                    & Linkedin (GB\&S)                                                    \\
ln\_audience                            & Log of number of linkedin users                                                                                           & Linkedin (GB\&S)                                                    \\
ln\_banker                              & Log of number of bankers                                                                                                  & Linkedin (GB\&S)                                                    \\
ln\_ceo                                 & Log of number of CEOs                                                                                                     & Linkedin (GB\&S)                                                    \\
ln\_cfo                                 & Log of number of CFOs                                                                                                     & Linkedin (GB\&S)                                                    \\
ln\_coo                                 & Log of number of COOs                                                                                                     & Linkedin (GB\&S)                                                    \\
ln\_cxo                                 & Log of number of Chief Executives                                                                                         & Linkedin (GB\&S)                                                    \\
ln\_engineer                            & Log of number of engineers in country                                                                                     & Linkedin (GB\&S)                                                    \\
ln\_finance                             & Log of number of finance workers                                                                                          & Linkedin (GB\&S)                                                    \\
ln\_other\_corporate                    & Log of number of corporate tax professionals                                                                              & Linkedin (GB\&S)                                                    \\
ln\_wealth                              & Log of number of wealth managers                                                                                          & Linkedin (GB\&S)                                                    \\
ln\_transfer\_pricing                   & Log of number of transfer pricing specialists                                                                             & Linkedin (GB\&S)                                                    \\
ln\_tax\_compliance\_audit              & Log of number of tax compliance experts and auditors                                                                      & Linkedin (GB\&S)                               \end{longtable}
\normalsize
\clearpage

\subsection{Additional theory discussion} \label{sec:appendix_theory}

Profit shifting is most frequently modelled using the method proposed by \textcite{hinesFiscalParadiseForeign1994}. In this section we show why the assumptions of that model are not consistent with our empirical observation of modern profit shifting behaviour. 

The \textcite{hinesFiscalParadiseForeign1994} model method assumes that the cost of profit shifting in country \textit{i}, $Cost_i$, increases quadratically with the profit shifted ($S_i$). This assumption implies that low levels of profit shifting are practically cost-free, while the cost quadratically increases as the intensity of profit shifting increases. The relationship is normalized by $\alpha/2$.

\begin{equation} \label{eq:cost}
    Cost_i = \frac{\alpha}{2}\frac{S_i^2}{p_i}
\end{equation}

The profits booked in a country ($\pi_i$) are defined by the sum of real profits ($p_i$) and profits shifted minus the cost of profit shifting:
\begin{equation} \label{eq:main}
    \pi_i = p_i + S_i - Cost_i
\end{equation}

\textcite{hinesFiscalParadiseForeign1994} maximize the after-tax profits ($(1-\tau_i)\cdot \pi_i$) subject to the existence of profit shifting, finding that the relationship between profits shifted and tax rates is:

\begin{equation} \label{eq:lag}
S_i = p_i \left( \frac{1-\tau_i-\lambda}{\alpha(1-\tau_i)}\right),
\end{equation}

Substituting equation \ref{eq:lag} in equation \ref{eq:main} and \ref{eq:cost}, we obtain:
\begin{equation} \label{eq:main}
    \pi_i = p_i + p_i \left( \frac{1-\tau_i-\lambda}{\alpha(1-\tau_i)}\right) - p_i \frac{\alpha}{2}\left( \frac{1-\tau_i-\lambda}{\alpha(1-\tau_i)}\right)^2
        = p_i\cdot\left(1 + \frac{\alpha}{2\alpha} - \frac{\lambda^2}{2\alpha(1-\tau_i)^2}\right)
\end{equation}

where $\lambda$ is the Lagrange multiplier and $\tau_i$ is the tax rate in the country. 

While this method has been widely used \citep{beerExploringResidualProfit2020}, the assumptions of the model are inadequate to study modern profit shifting. First, the Taylor expansion of equation \ref{eq:main} is taken around the point where no profit shifting takes place (i.e., where $\lambda^2 = (1-\tau_i)^2$, which simplifies equation \ref{eq:main} to $\pi_i = p_i$). This simplifies

\begin{equation}\label{eq:log_main}
  \log(\pi_i) = \log(p_i) + \log{\left(1+ \frac{1}{2a}-\frac{\lambda^2}{2a(1-\tau_i)^2}\right)}  
\end{equation}

around $\tau_i  = (1 - \lambda)$ into 

\begin{equation} \label{eq:f_o_st} 
    \log(\pi_i) \sim \log(p_i) + \left(\frac{1-\lambda}{\alpha\lambda}-\frac{\tau_i}{\alpha\lambda}\right),
\end{equation}

and around $(1-\tau_i)^2  = \lambda^2$ into 

\begin{equation} \label{eq:s_o_st}
    \log(\pi_i) \sim \log(p_i) + \left(\frac{1}{1\alpha}-\frac{\lambda^2}{2\alpha(1-\tau)^2}\right).
\end{equation}

Subsequently, theoretical profits ($p_i$) are identified with the Cobb-Douglas production function, yielding equations \ref{eq:f_o} and \ref{eq:s_o} from equations \ref{eq:f_o_st} and \ref{eq:s_o_st}:

\begin{equation} \label{eq:f_o}
    \log{(\pi_i)} = \beta_0 + \beta_1
    \log{(K_i)} + \beta_2\log{(L_i)} +  \beta_3(\tau_i) + \beta_\chi \chi + \epsilon,
\end{equation}

\begin{equation} \label{eq:s_o}
    \log{(\pi_i)} = \beta_0 + \beta_1\log{(K_i)} + \beta_2\log{(L_i)} +  \beta_3(\tau_i) + \beta_4 (\tau_i)^2 + \beta_\chi \chi + \epsilon,
\end{equation}

where $\pi_i$ represents profits booked in country $i$, and $K_i$ and $L_i$ are the capital and labour components of the Cobb-Douglas production function, usually operationalised with total tangible assets and wages. $\tau_i$ is either the tax rate of the subsidiary, the difference of tax rates between the subsidiary and the parent, or, less frequently (due to lacking data), between the subsidiary and other subsidiaries, and $\chi$ are controls including e.g. GDP per capita and population.

\begin{figure}[ht!]
        \begin{center}
 \caption{Residuals of the regressions}
  \includegraphics[width=\textwidth]{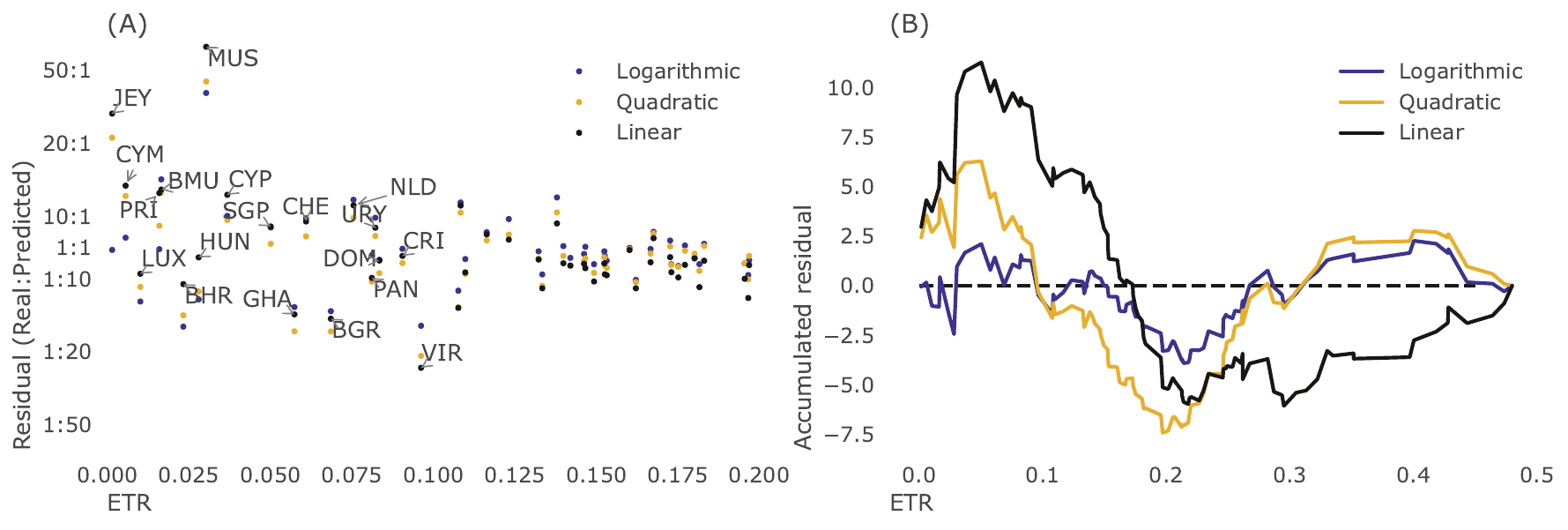}
  \label{fig:non_linear}
        \end{center}
        \footnotesize
    Notes: (A) Residuals ($\log{(\frac{booked profits}{predicted profits})}$) and (B) Cumulative sum of residuals for the linear (green), quadratic (orange) and logarithmic (blue) models, as a function of the ETR in the country.
\end{figure}

These two equations are convenient but are only valid around the point where no profit shifting is present ($\pi_i=p_i$). Extending these approximations far from that point is not adequate.  While, as \textcite{hinesFiscalParadiseForeign1994} put it, ``it  is  helpful  to  transform the  second term  on the  right  side [of eq. \ref{eq:log_main}] into  a linear  function  of tax  rates'', the response of profit shifting on the tax rate is, as we show in the main text and below, highly non-linear. As a result, these simplifications (both in the linear and the quadratic forms) are not able to model profit shifting correctly (Fig. \ref{fig:non_linear}). 

For both the linear (eq. \ref{eq:f_o}) and quadratic (eq. \ref{eq:s_o}) models, profit shifting is undervalued for countries with a tax rate below 10\% and overvalued for other countries (Fig. \ref{fig:non_linear})---the residuals are typically positive for countries with an ETR below 10\% for both the linear and quadratic model (Fig. \ref{fig:non_linear}). The residuals are only corrected once the logarithmic term is reintroduced, with the cumulative residual fluctuating only slightly (Fig. \ref{fig:non_linear}).

The second assumption that requires careful consideration is the choice of the cost function (Eq.~\ref{eq:cost}). Profit shifting strategies based on intangibles require high costs to set up, and low functioning costs---the opposite of what it is assumed in equation \ref{eq:cost}.  

The inadequacy of equation \ref{eq:cost} to model the cost of profit shifting in the 21\textsuperscript{st} century can also be shown by estimating $\alpha$ (normalization constant) and $\lambda$ (the Lagrange multiplier). We start with the estimates in the literature of the semi-elasticity of the tax rate (which is equivalent to $-\frac{1}{\alpha\lambda}$), at around 1 \citep{beerInternationalCorporateTax2020} or around 4, as implied in this paper. Furthermore, we can estimate (using the logarithmic or the misalignment models) that the share of profits shifted ($S_i/(S_i + p_i)$) in countries with no corporate income tax rate is around 90--99\%, and 30--50\% in countries with moderate tax rates. This allows us to plug $S_i$/$p_i$ in equation \ref{eq:lag}, keeping only $\tau_i$, $\alpha$ and $\lambda$. Using the relationship of our parameters of interest ($\alpha$ and $\lambda$) and the semi-elasticity on the one hand and the share of profit shifting on the other, we can calculate the value of $\alpha$ for different values of the profits shifted ($S_i/(S_i+p_i$), the semi-elasticity and $\tau_i$ (Table \ref{tab:alpha}). This exercise allows us to estimate $\alpha$ to be around 0.14--4.19.

\begin{table}[]
\centering
\begin{tabular}{ll|lll|lll}
\toprule
                                 &      & \multicolumn{3}{l|}{Semi-elasticity: 1} & \multicolumn{3}{l}{Semi-elasticity: 4} \\ \hline
                                 &      & \multicolumn{3}{l|}{Tax rate}           & \multicolumn{3}{l}{Tax rate}           \\
                                 &      & 0           & 0.2         & 0.4        & 0           & 0.2         & 0.4        \\ \midrule
\multirow{3}{*}{Profits shifted} & 95\% & 0.26        & 0.28        & 0.32       & 0.14        & 0.16        & 0.18       \\
                                 & 50\% & 1.62        & 1.72        & 1.89       & 1.21        & 1.25        & 1.32       \\
                                 & 25\% & 3.79        & 3.95        & 4.19       & 3.23        & 3.29        & 3.37     \\ \bottomrule 
\end{tabular}
\caption{Values of $\alpha$ depending on the tax rate ($\tau_i$), the semi-elasticity, and the fraction of profit shifted ($S_i/(S_i + p_i)$)}
\label{tab:alpha}
\end{table}

For a tax haven with no corporate income tax rate and 95\% of the profits shifted in ($S_i = 0.95 (S_i + p_i)$), equation \ref{eq:cost} becomes
\begin{equation} \label{eq:new_cost}
    Cost_i = \frac{\alpha}{2}\frac{S_i^2}{0.05S_i/0.95} = \frac{\alpha}{2}19  S_i > \frac{0.14}{2}19  S_i > 1.33 S_i
\end{equation}

\begin{equation} 
\label{eq:new_cost_alpha}
    Cost_i = \frac{\alpha}{2}19  S_i > \frac{0.14}{2}19  S_i > 1.33 S_i
\end{equation}

Using the most conservative value of alpha (0.14), as in equation \ref{eq:new_cost_alpha}, we would estimate that the cost of profit shifting is at least 33\% higher than the amount of profits shifted for tax havens. This points to a mismatch between the quadratic cost function proposed by ~\citep{hinesFiscalParadiseForeign1994} and profit shifting in the 21\textsuperscript{st} century.

\clearpage

\subsection{Supplementary Figures} \label{sec:appendix_figures}

\begin{figure}[ht!]
 \caption{Template for the country-by-country report by the OECD}
  \includegraphics[width=1\textwidth]{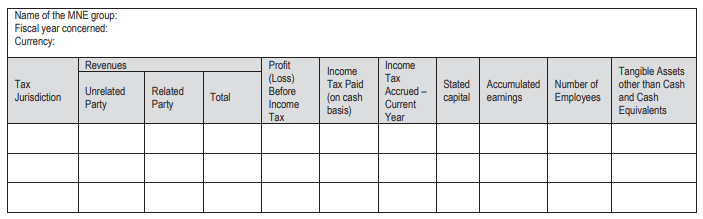}
\newline
  \includegraphics[width=1\textwidth]{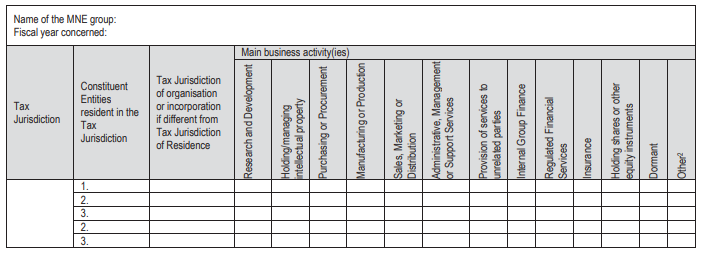}
\newline
  \includegraphics[width=1\textwidth]{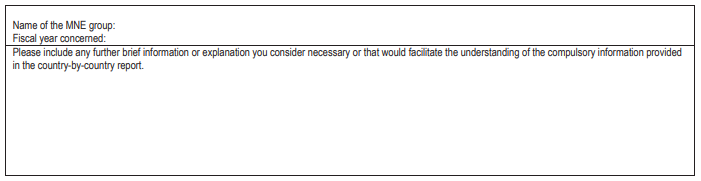}
  \label{fig:cbcr_template}
\newline
\footnotesize
Notes: This figure includes pictures of a three-part template for CBCR by \textcite{oecdCountrybyCountryReportingXML2019}: (i) overview of allocation of income, taxes and business activities by tax jurisdiction; (ii) list of all the constituent entities of the MNC group included in each aggregation per tax jurisdiction; and (iii) additional information. Source: \textcite{oecdCountrybyCountryReportingXML2019}
\end{figure}

%Figure A1
\begin{figure}[ht!]
 \caption{Loss-making affiliates as a profit shifting strategy}
  \includegraphics[width=.6\textwidth]{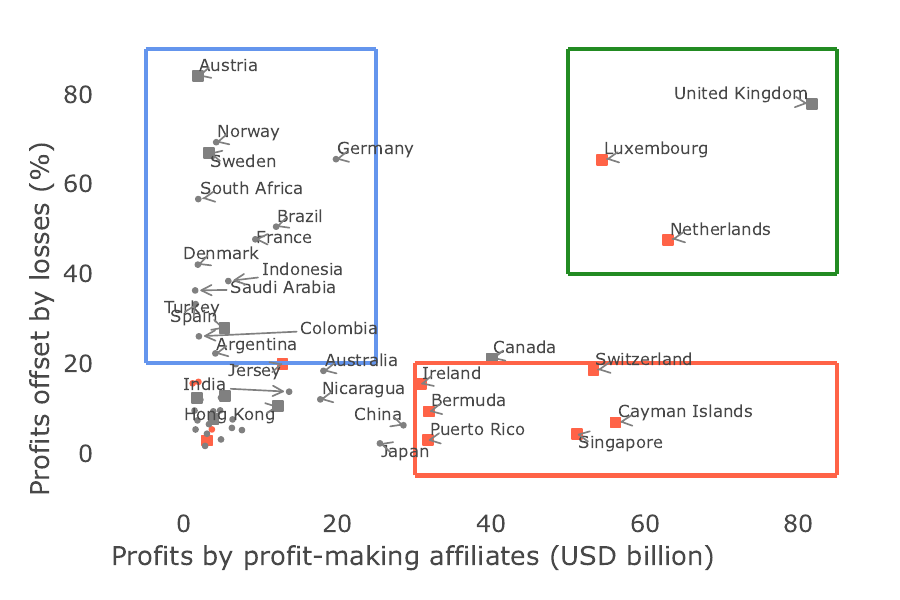}
  \label{fig:loss_making}
\newline
\footnotesize
Notes: Loss-making affiliates as a profit shifting strategy, using US data. The total profits made by profit-making affiliates is plotted against the percentage of profits offset by losses. Three types of countries are highlighted with boxes in line with \textcite{reurinkCompetingCapitalsGreat2020}. In red are ``profit centers'', reporting very high profits not offset by losses. In green are ``coordination centers'' (or conduits), reporting very high profits offset by losses. In blue are origin countries, reporting profits offset by losses. Only countries reporting at least \$10 billion profits are reported, the USA (profits of US 1,310 and offset ratio of 10\% is excluded). Countries in red exhibit profitabilities above \$100,000 per employee.
\end{figure}

%Fig A2
\begin{figure}[ht!]
   \caption{Comparison of profit shifting in and out for the misalignment and the logarithmic models using the US data}  \includegraphics[width=\textwidth]{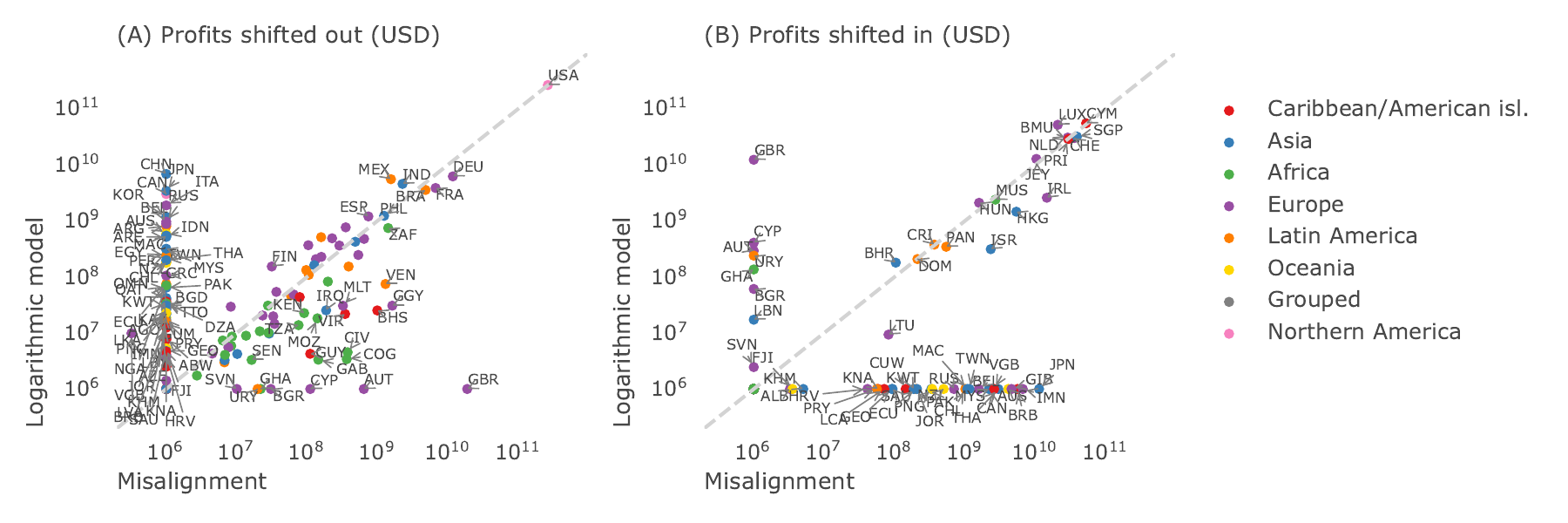}
  \label{fig:compare_log_mis_US}
\footnotesize
Notes: Comparison of profits shifted out (A) and profits shifted in (B) for the misalignment and the logarithmic models for the US data. Each dot represents a country, coloured by region. Note that profit shifting out of African countries is higher in the misalignment model.
\end{figure}

%Fig A2
\begin{figure}[ht!]
   \caption{Comparison of profit shifting in and out for the misalignment and the logarithmic models using the OECD data}  \includegraphics[width=\textwidth]{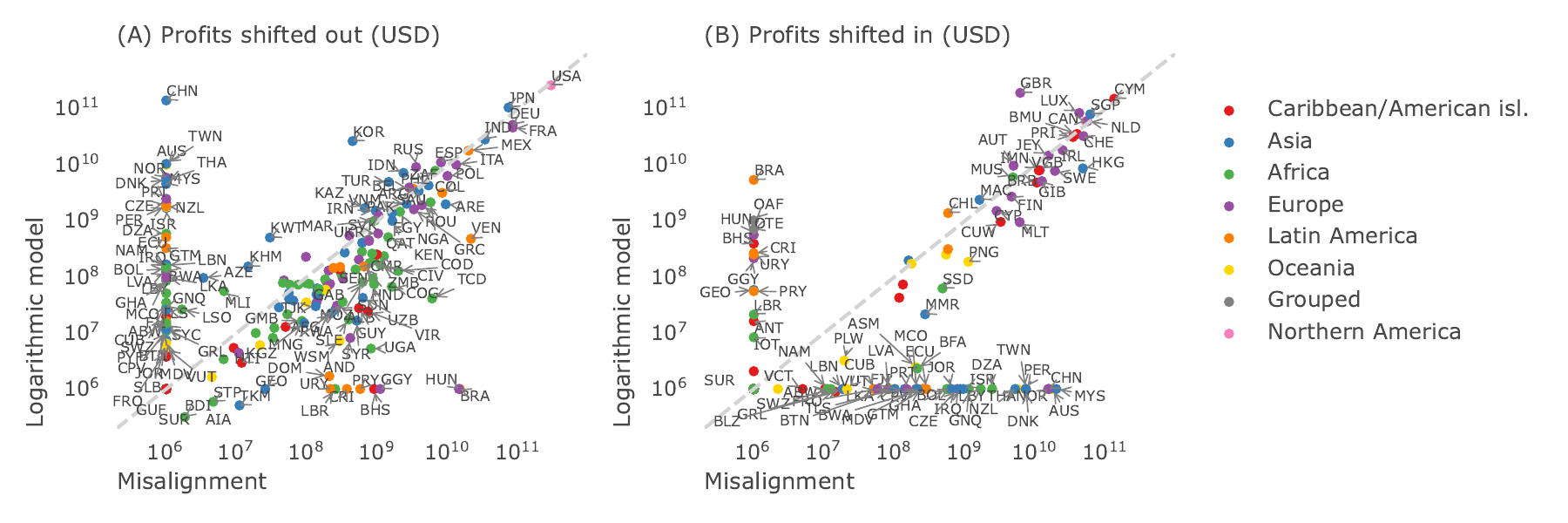}
  \label{fig:compare_log_mis_OECD}
\footnotesize
Notes: Comparison of profits shifted out (A) and profits shifted in (B) for the misalignment and the logarithmic models for the OECD data. Each dot represents a country, coloured by region. Note that profit shifting out of African countries is higher in the misalignment model.
\end{figure}

\begin{figure}[ht!]
 \caption{Distribution of the scale of profit shifted estimated by the misalignment model at the country level}
  \includegraphics[width=1\textwidth]{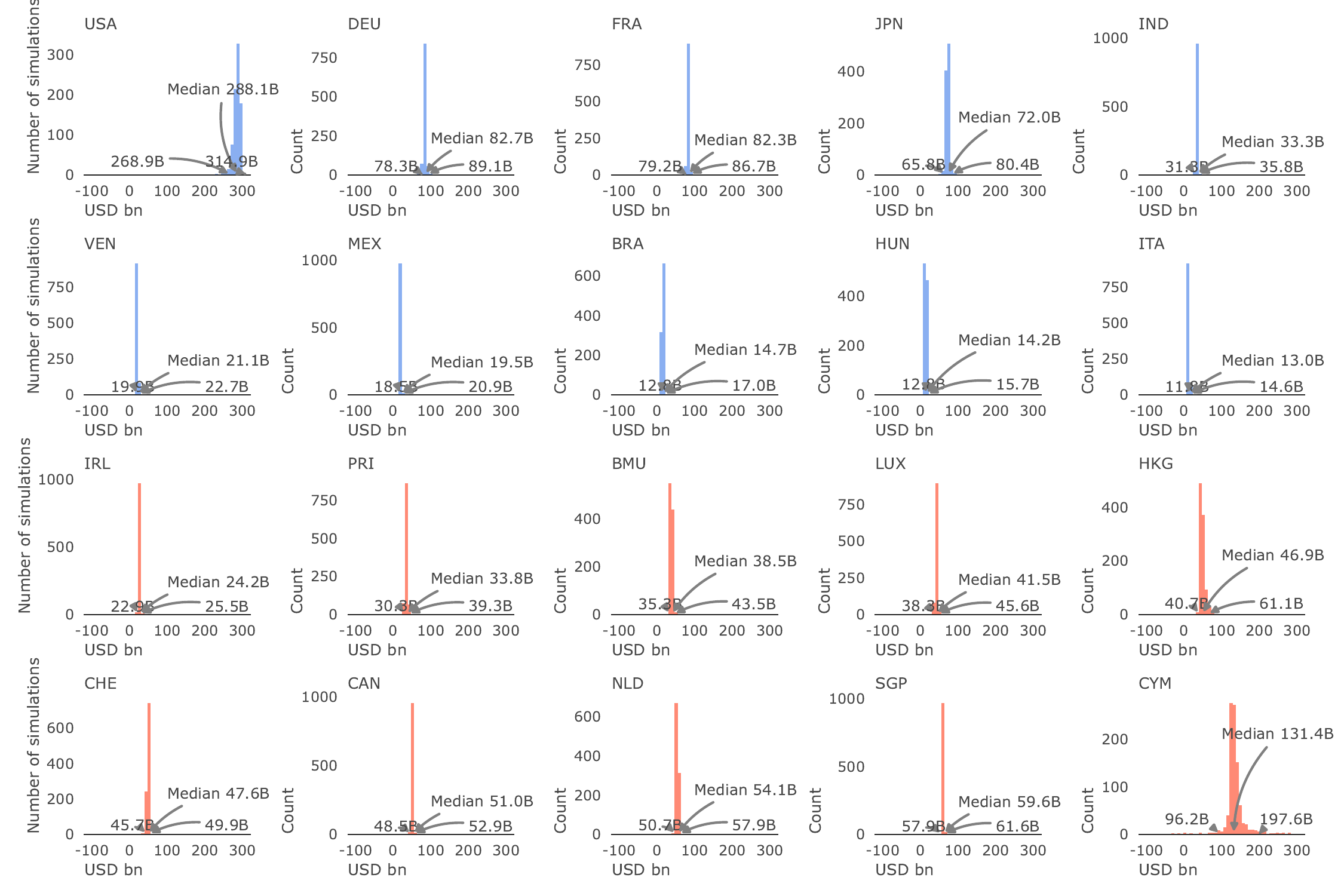}
  \label{fig:ci_misalignment}
\footnotesize
Notes: Distribution of the scale of profit shifted estimated by the misalignment model at the country level. The largest origins (top two rows, in blue) and destinations (bottom two rows, in red) are shown. The variance observed is created by the bootstrapping process detailed in Section \ref{sec:methodology_extrapolation}. Non reporting countries (Germany (DEU), the United Kingdom (GBR), Cayman Islands (CYM) have higher uncertainty than reporting countries such as France (FRA), Italy (ITA) or Bermuda (BMU). The 5\% percentile, the median, and the 95\% percentile are annotated.
\end{figure}

\begin{figure}[ht!]
 \caption{Graphical representation of Table \ref{tab:oecd_reg_table} for the logarithmic, quadratic, and linear models}
  \includegraphics[width=1\textwidth]{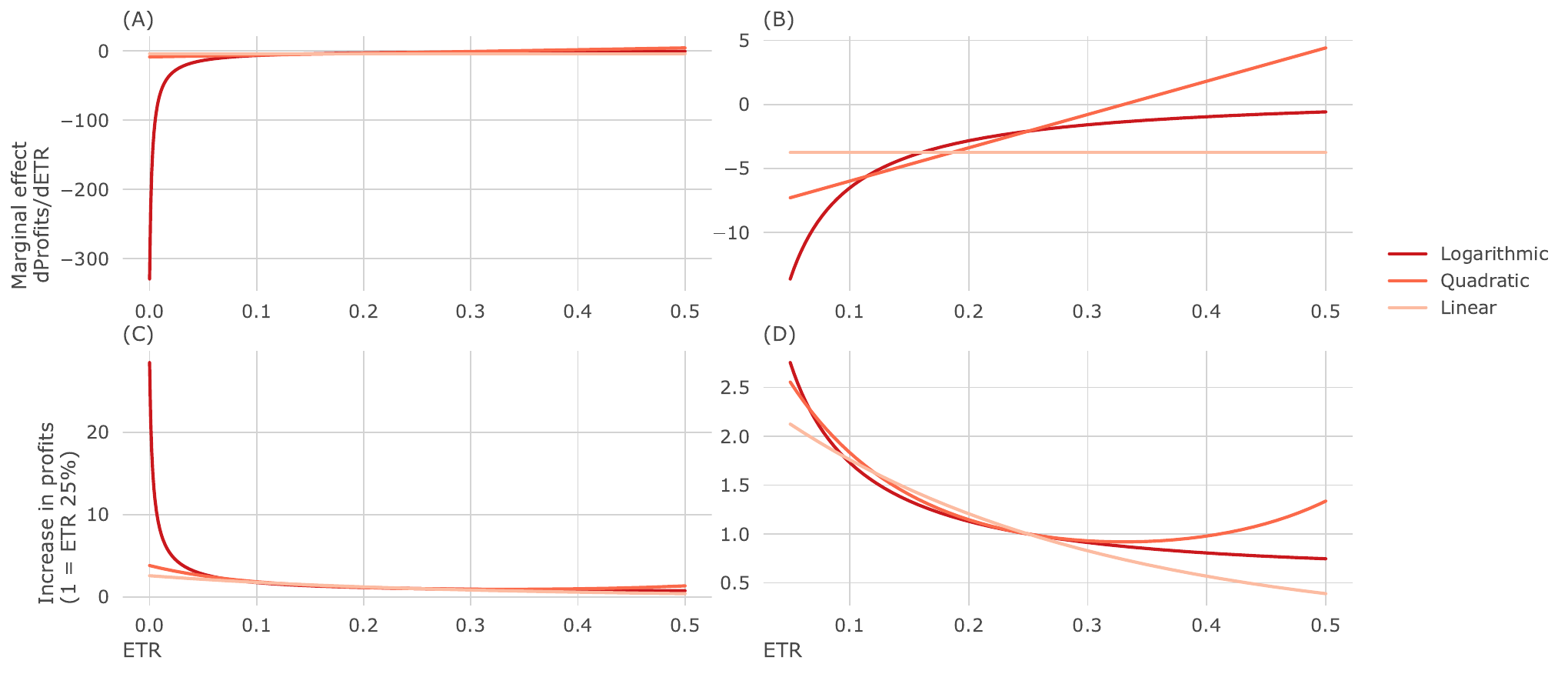}
  \label{fig:elasticity_oecd}
\footnotesize
Notes: Graphical representation of Table \ref{tab:oecd_reg_table} for the logarithmic, quadratic, linear and DLM models. (A, B) Marginal effect of ETR on profits. (C,D) Relative increase in profits due to profit shifting, compared with a country with an ETR of 25\%. Plots B and D are close-ups of plots A and C respectively, constraining ETRs between 5 and 50\%. Note that the marginal effects for the logarithmic model decreases faster than other models as the ETR approaches 0\%.
\end{figure}

\begin{figure}[ht!]
 \caption{Graphical representation of Table \ref{tab:oecd_reg_table} for the logarithmic model, at the country level}
  \includegraphics[width=1\textwidth]{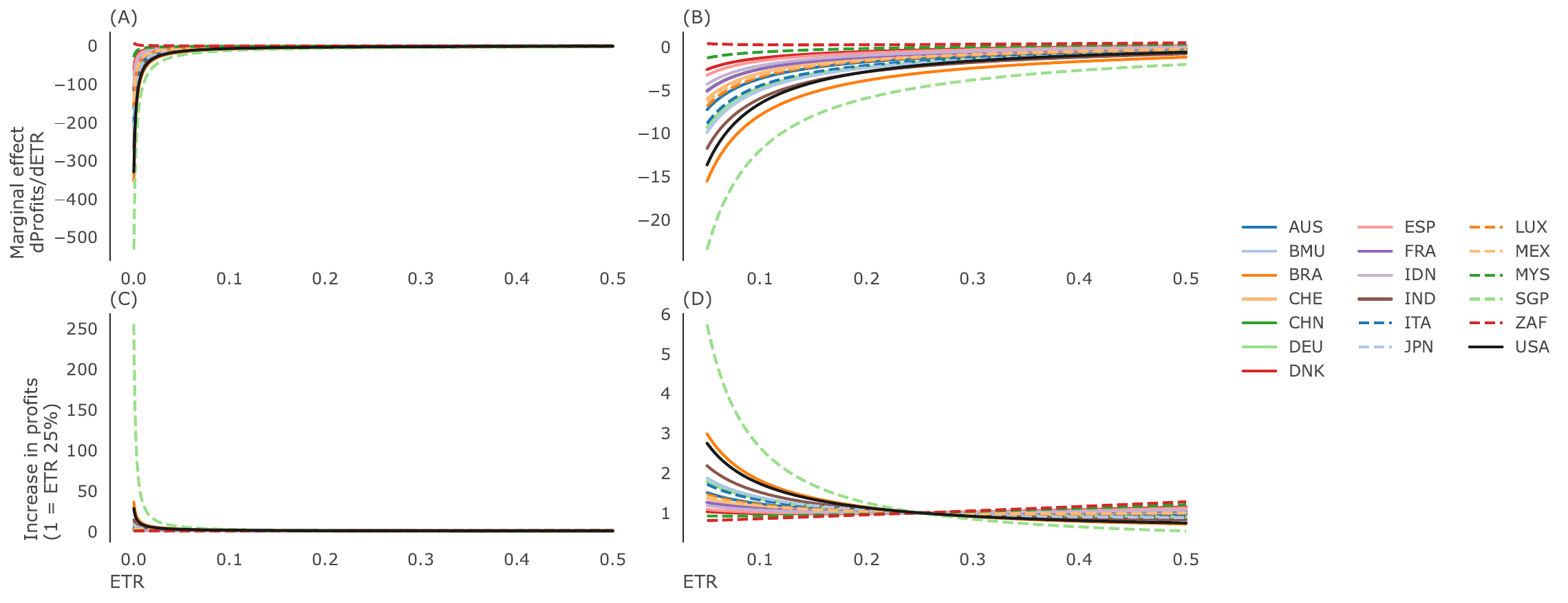}
  \label{fig:elasticity_oecd_by_country}
\footnotesize
Notes: Graphical representation of Table \ref{tab:oecd_reg_table} for the logarithmic model, at the country level. (A, B) Marginal effect of ETR on profits. (C,D) Relative increase in profits due to profit shifting, compared with a country with an ETR of 25\%. Plots B and D are close-ups of plots A and C respectively, constraining ETRs between 5 and 50\%. Only countries reporting on at least 20 other countries are included.
\end{figure}

\begin{figure}[h!]
 \caption{Profits shifted as a percentage of GDP}
  \includegraphics[width=1\textwidth]{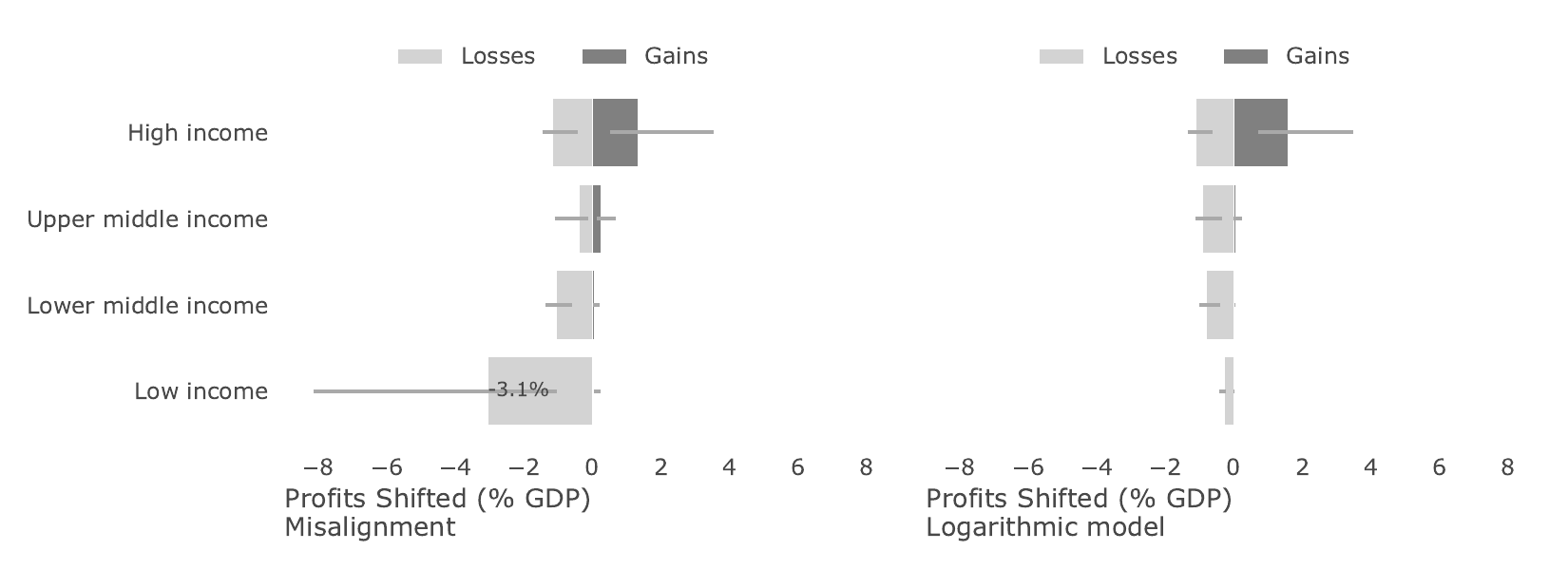}
  \label{fig:diffs_aggs_income_rev_prof_shift}
\footnotesize
Notes: The figure shows the shifted profits as a percentage of GDP for countries in different income groups, as estimated by the misalignment (left graph) and logarithmic (right graph) models. Confidence intervals show 95\% intervals, calculated via bootstrapping.
\end{figure}

\begin{figure}[h!]
 \caption{Tax revenue loss as a percentage of total tax revenue estimated with the misalignment and logarithmic models}
  \includegraphics[width=1\textwidth]{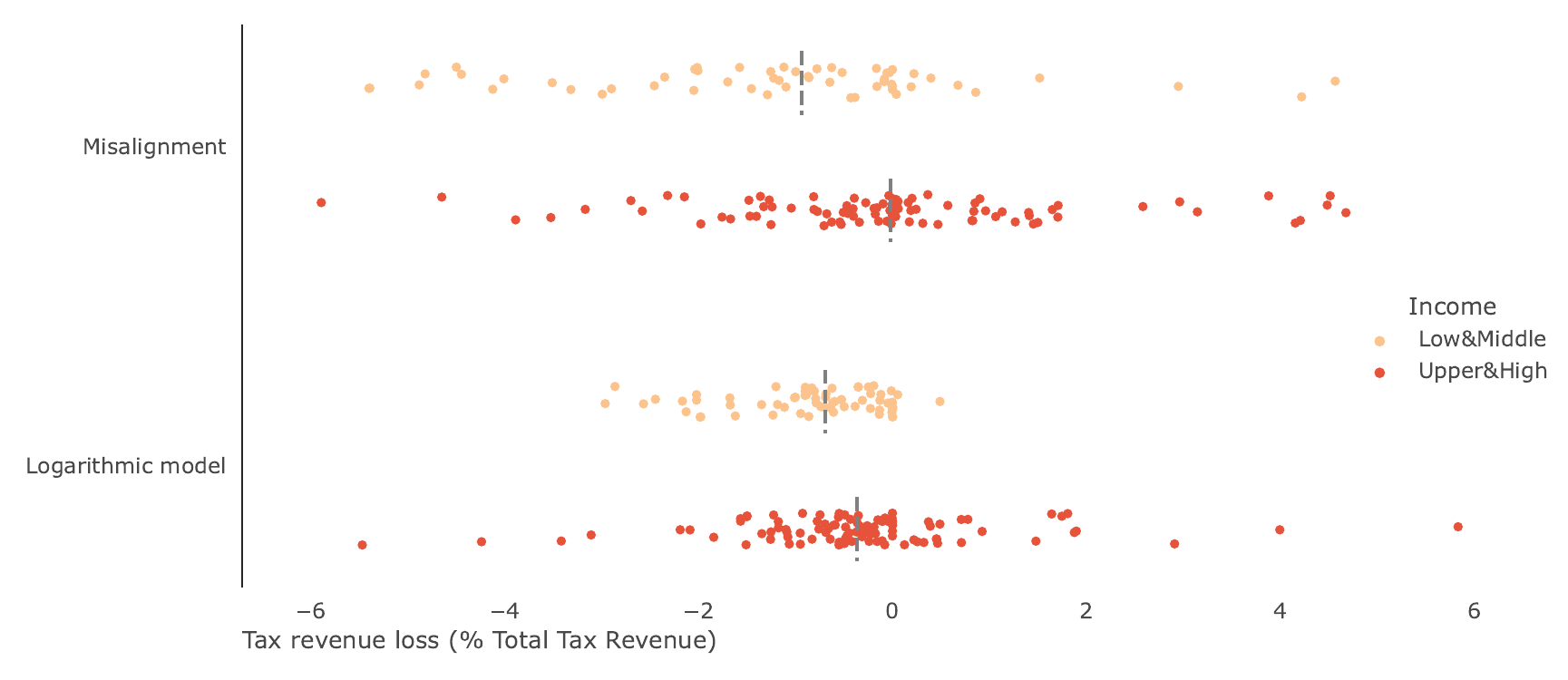}
  \label{fig:diffs_income_TRL_rev}
\footnotesize
Notes: Tax revenue loss as a percentage of total tax revenue estimated with the misalignment and logarithmic models. Each dot represents an individual country, and the median values are visualised with a dashed line. The data is split into low and lower-middle income countries (Low\&Middle, light orange) upper middle and high income countries (Upper\&High, darker red). The statistical differences between the median are assessed with a Mann-Whitney test. Only observations within a distance from the median of 5 interquartile ranges are shown. %Significance level: p<0.05 (*), p<0.01 (**), p<0.001 (***).
\end{figure}

\begin{figure}[ht!]
 \caption{Tax revenue loss as a percentage of corporate income tax revenue}
  \includegraphics[width=1\textwidth]{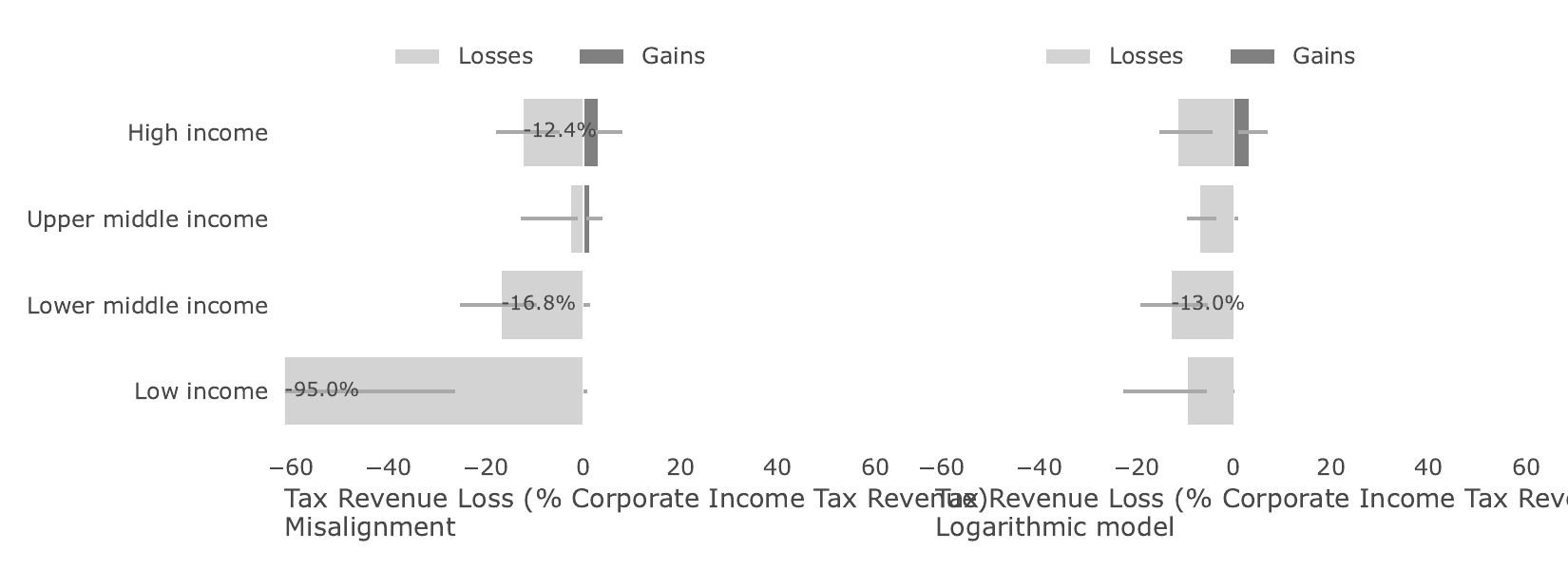}
  \includegraphics[width=1\textwidth]{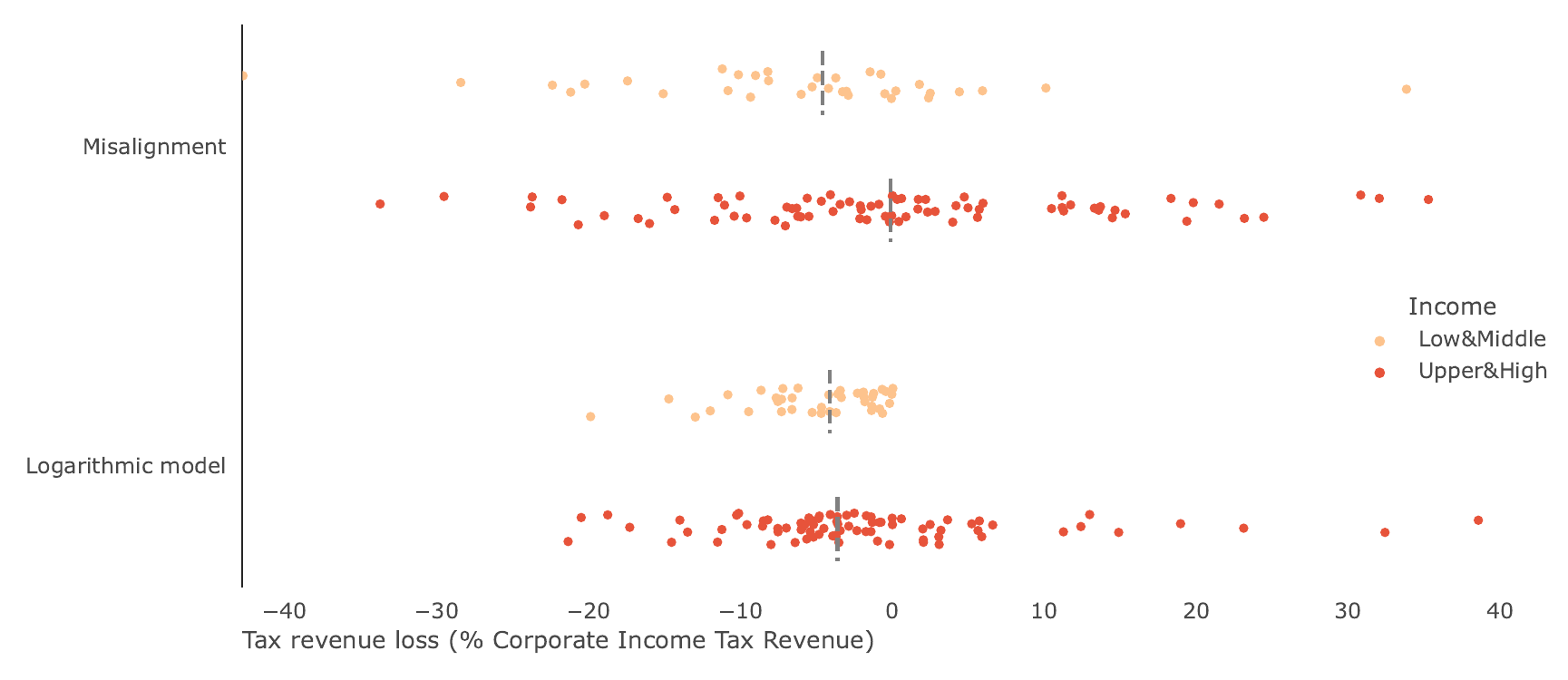}
  \label{fig:diffs_income_TRL_crev}
\footnotesize
Notes: Tax revenue loss as a percentage of corporate income tax revenue.  Only observations within a distance from the median of 5 interquartile ranges are shown. %Significance level: p<0.05 (*), p<0.01 (**), p<0.001 (***).
\end{figure}

\begin{figure}[ht!]
 \caption{Tax revenue loss as a percentage of GDP}
  \includegraphics[width=1\textwidth]{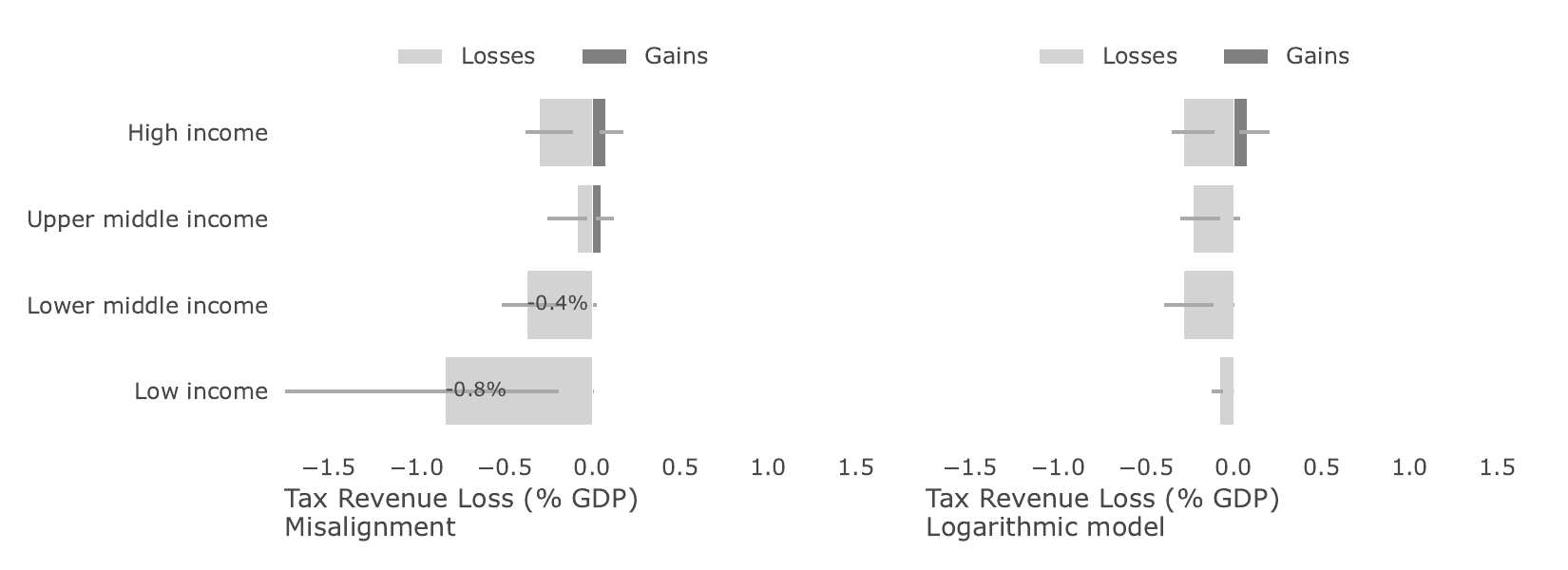}
  \includegraphics[width=1\textwidth]{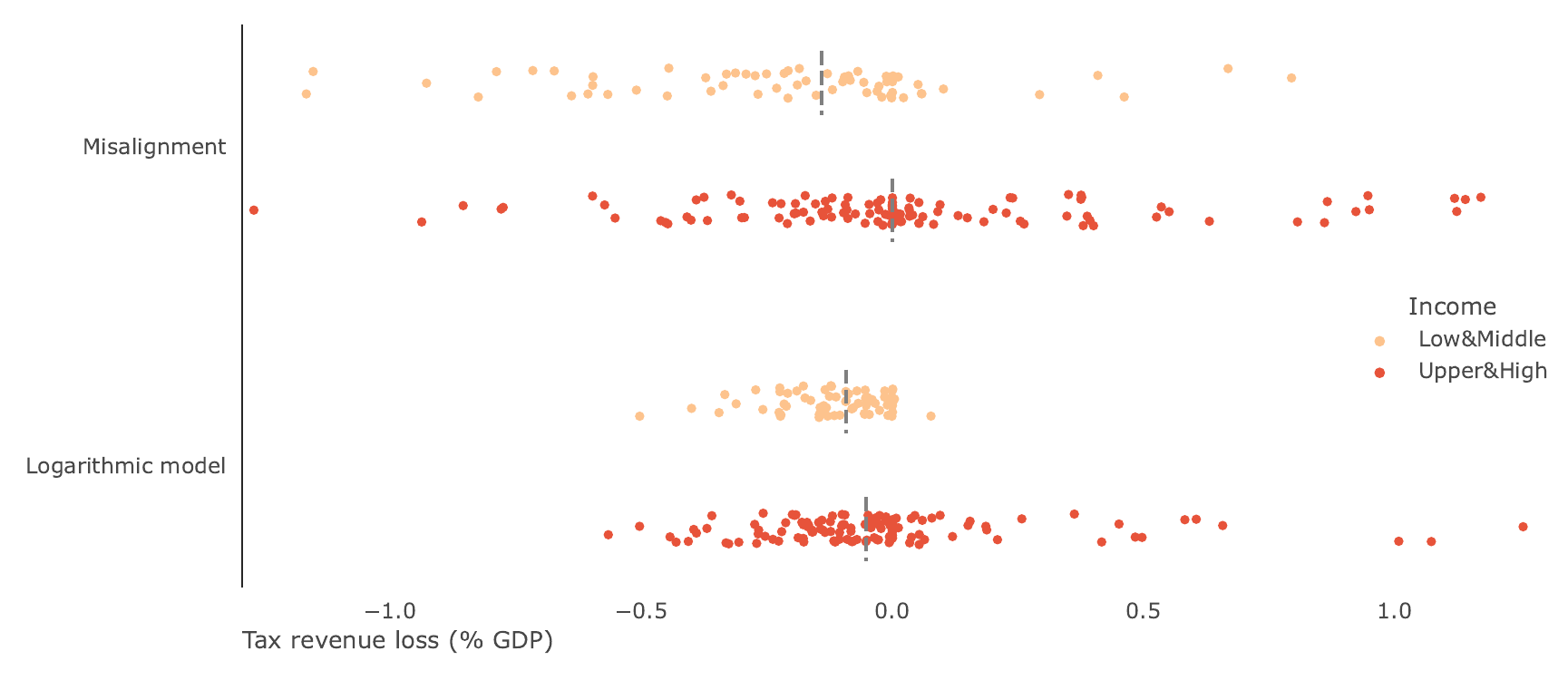}
  \label{fig:diffs_income_TRL_gdp}
\footnotesize
Notes: Tax revenue loss as a percentage of GDP. Only observations within a distance from the median of 5 interquartile ranges are shown. %Significance level: p<0.05 (*), p<0.01 (**), p<0.001 (***).
\end{figure}

\iffalse
\begin{figure}[ht!]
 \caption{Tax revenue loss as a percentage of health expenditure}
  \includegraphics[width=1\textwidth]{Figures/diffs_aggs_IncomeClass_TaxRevenueLoss_health.pdf}
  \includegraphics[width=1\textwidth]{Figures/diffs_income_TRL_health.pdf}
  \label{fig:diffs_income_TRL_health}
\footnotesize
Notes: Tax revenue loss as a percentage of health expenditure. Only observations within a distance from the median of 5 interquartile ranges are shown. % Significance level: p<0.05 (*), p<0.01 (**), p<0.001 (***). 
\end{figure}

\begin{figure}[ht!]
 \caption{Tax revenue loss as a percentage of government education expenditure}
  \includegraphics[width=1\textwidth]{Figures/diffs_aggs_IncomeClass_TaxRevenueLoss_educ.pdf}
  \includegraphics[width=1\textwidth]{Figures/diffs_income_TRL_educ.pdf}
  \label{fig:diffs_income_TRL_educ}
\footnotesize
Notes: Tax revenue loss as a percentage of government education expenditure. Only observations within a distance from the median of 5 interquartile ranges are shown. % Significance level: p<0.05 (*), p<0.01 (**), p<0.001 (***).
\end{figure}
\fi

\begin{figure}[ht!]
 \caption{Distribution of tax revenue losses for lower income countries}
  \includegraphics[width=1\textwidth]{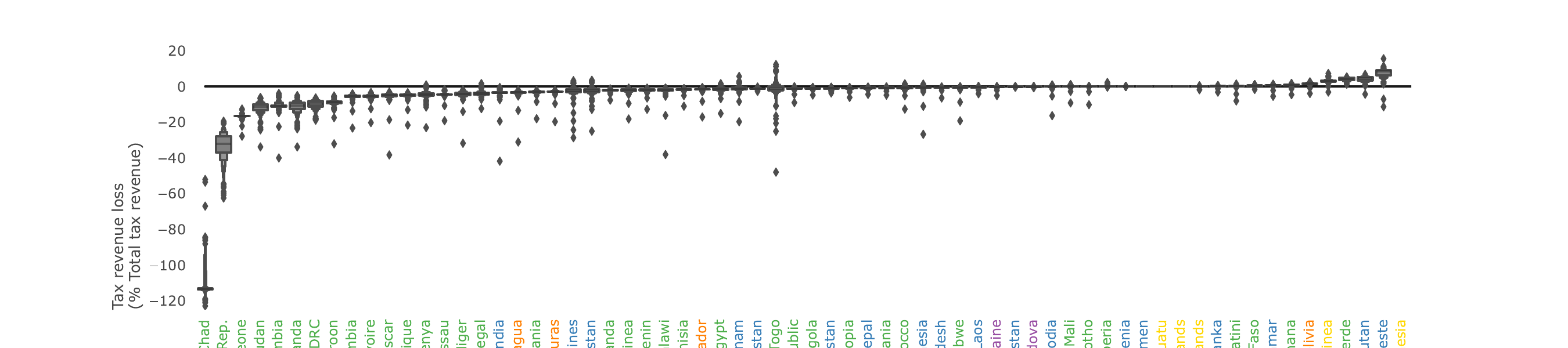}
  \label{fig:lower_boxen}
\footnotesize
Notes: Distribution of tax revenue losses for lower income countries, visualised with a boxen plot. Countries are colored by geographical region---green (Africa), blue (Asia), orange (South America), purple (Europe), yellow (Oceania), red (Caribbean Islands). A boxenplot is similar to a boxplot, but where the whiskers are replaced with smaller boxes. The central box visualises the interquartile range, leaving 25\% of the data at each end. The smaller boxes increasingly leave smaller fractions of the data at each end (12.5\%, 6.25\%, etc). Outliers are represented with dots.
\end{figure}

\clearpage

\subsection{Supplementary Tables} \label{sec:appendix_tables}

\begin{table}[ht]
\small
    \caption{Number of jurisdictions available per country using the OECD data}
    \label{tab:n_jur_country}
\begin{tabular}{lrrrrrrrrrr}
\toprule
 &  JPN &  IND &  DEU &  USA &  ZAF &  CHN &  CHE &  ESP &  DNK &  ITA \\
\midrule
All Sub-Groups               &  198 &  163 &  159 &  145 &  138 &  124 &  119 &  119 &  114 &  109 \\
Sub-Groups with pos. profits &  183 &  134 &  119 &   97 &  114 &   91 &   96 &   79 &  114 &   80 \\
\bottomrule
\end{tabular}

\begin{tabular}{lrrrrrrrrrr}
\toprule
 &  BMU &  MEX &  LUX &  FRA &  AUS &  IDN &  BRA &  MYS &  SGP &  BEL \\
\midrule
All Sub-Groups               &   98 &   96 &   90 &   89 &   80 &   41 &   41 &   35 &   28 &   20 \\
Sub-Groups with pos. profits &   67 &   60 &   90 &   39 &   55 &   29 &   29 &   26 &   37 &   20 \\
\bottomrule
\end{tabular}

\begin{tabular}{lrrrrrrrrrr}
\toprule
 &  PER &  ARG &  CAN &  LVA &  CHL &  ROU &  IMN &  NOR &  GBR &  SVN \\
\midrule
All Sub-Groups               &   19 &   18 &   13 &   13 &    9 &    6 &    5 &    5 &    5 &    5 \\
Sub-Groups with pos. profits &   11 &   14 &    7 &    0 &    6 &    6 &    5 &    5 &    5 &    4 \\
\bottomrule
\end{tabular}

\begin{tabular}{lrrrrrrrr}
\toprule
 &  SWE &  AUT &  GRC &  POL &  KOR &  FIN &  NLD &  IRL \\
\midrule
All Sub-Groups               &    5 &    5 &    5 &    3 &    1 &    1 &    1 &    1 \\
Sub-Groups with pos. profits &    5 &    4 &    5 &    0 &    1 &    1 &    1 &    1 \\
\bottomrule
\end{tabular}
% \begin{tabular}{lrrrrrrrrr}
% {} &  IND &  USA &  ZAF &  ITA &  MEX &  DNK &  CHN &  FRA &  LUX \\
% \midrule
% All sub-groups                   &  163 &  137 &  127 &  103 &  102 &   96 &   90 &   83 &   80 \\
% Sub-groups with positive profits &  134 &   84 &   87 &   77 &   73 &   93 &   53 &   32 &   79 \\
% \bottomrule
% \end{tabular}

% \begin{tabular}{lrrrrrrrrr}

% {} &  BMU &  AUS &  BEL &  IDN &  JPN &  CHL &  SGP &  CAN &  SVN \\
% \midrule
% All sub-groups                   &   72 &   56 &   40 &   33 &   29 &   14 &   12 &    8 &    5 \\
% Sub-groups with positive profits &   57 &   39 &   32 &   26 &   12 &   11 &   22 &    3 &    5 \\
% \bottomrule
% \end{tabular}

% \begin{tabular}{lrrrrrrrr}
% \toprule
% {} &  POL &  AUT &  NLD &  NOR &  IRL &  SWE &  FIN &  KOR \\
% \midrule
% All sub-groups                   &    4 &    1 &    1 &    1 &    1 &    1 &    1 &    1 \\
% Sub-groups with positive profits &    0 &    1 &    1 &    1 &    1 &    1 &    1 &    1 \\
% \bottomrule
% \end{tabular}
\footnotesize
Notes: Number of jurisdictions available per country (aggregates excluded) using the OECD data. Jurisdictions with 1 observation only report on domestic activities of MNCs.
JPN (Japan), IND (India), DEU (Germany), USA (United States), ZAF (South Africa), CHN (China), CHE (Switzerland), ESP (Spain), DNK (Denmark), ITA (Italy), BMU (Bermuda), MEX (Mexico), LUX (Luxembourg), FRA (France),  AUS (Australia), IDN (Indonesia), BRA (Brazil), MYS (Malaisia), SGP (Singapore), BEL (Belgium), PER (Peru), ARG (Argentia), CAN (Canada), LVA (Latvia), CHL (Chile),  ROU (Romania), IMN (Isle of Man), NOR (Norway), GBR (United Kingdom), SVN (Slovenia), SWE (Sweden), AUT (Austria), GRC (Greece), POL (Poland), KOR (South Korea), FIN (Finland), NLD (the Netherlands), IRL (Ireland).
\end{table}

{\scriptsize
    \begin{longtable}{llrrrrrrrll}
\caption{Summary statistics for the 36 countries of the sample used in the log-model, for the 2017 OECD data containing ``Sub-Groups with positive profits''} \label{tab:stats}\\
\toprule
&         &  Firms &  Profits &  Tax accrued &  Tax paid  &  Employees  &  Revenue &  Assets & ETR  & ETR  \\
Reporting & Partner & profits>0  &   US\$bn     &   US\$bn     &   US\$bn   & Thousands  &    US\$bn         &         US\$bn                 &       accrued          &    cash           \\
\midrule
Argentina & Dom. &               15 &          2.9 &              1.6 &           1.2 &           143.7 &         43.5 &                 34.8 &            54.9\% &         42.2\% \\
             & For. &               15 &          0.8 &              0.0 &           0.0 &             8.5 &          2.3 &                  3.4 &             4.0\% &          3.9\% \\
Australia & Dom. &              112 &         59.6 &             20.8 &          16.9 &          1079.0 &        438.9 &                447.9 &            34.9\% &         28.4\% \\
             & For. &              112 &         31.0 &              4.8 &           3.8 &           384.5 &        135.0 &                103.1 &            15.4\% &         12.2\% \\
Austria & For. &               76 &         13.4 &              2.5 &           1.6 &           577.6 &        172.6 &                 79.4 &            18.9\% &         11.6\% \\
\end{longtable}
}
\footnotesize
Notes: Summary statistics for the 36 countries of the sample used in the log-model, for the OECD data containing ``Sub-Groups with positive profits''. The aggregated number of firms (``Firms profits>0''), profits, tax accrued, tax paid, number of employees, unrelated-party revenue, tangible assets and ETRs (accrued and cash-based) are shown for domestic activities (financial reporting of MNCs in the reporting (i.e. headquarter) countries) and foreign activities (financial reporting in all other countries). Since we are using sub-groups with positive profits, the number of firms included in the domestic section can be lower than the number of firms reporting on foreign operations. Not that two countries (Latvia and Poland) only report data on all groups, and that some countries omit reporting on domestic (Austria, Chile) or total foreign (United Kingdom) activities.

\begin{table}
\caption{Expected number of large MNCs headquartered in the country and observed in the OECD CBCR data}
\footnotesize
\label{tab:ratio}
\centering
\begin{tabular}{lrrr}
\toprule
       Country &  \#Expected\textbackslash nOrbis &  \#Observed\textbackslash nCBCR &  Ratio \\
\midrule
         China &               538 &              264 &   2.04 \\
      Malaysia &                41 &               34 &   1.21 \\
       Bermuda &                49 &               48 &   1.02 \\
   South Korea &               221 &              221 &   1.00 \\
  South Africa &                50 &               51 &   0.98 \\
        Sweden &                94 &              102 &   0.92 \\
         Japan &               785 &              866 &   0.91 \\
\bottomrule
\end{tabular}
\footnotesize
\\Notes: Expected number of large MNCs headquartered in the country (according to Orbis), number of companies observed in the  OECD CBCR data (using all sub-groups), and the ratio between the two.
\end{table}

\newpage
\footnotesize{
\begin{longtable}{p{2cm}p{1.5cm}p{1.5cm}p{1.5cm}p{1cm}p{1cm}p{1cm}p{1cm}p{1cm}p{1cm}}
\caption{Countries acting as profit destination, estimated by the misalignment model}\label{tab:winners_mis}\\
\toprule
Country & Profits Shifted (M) & Profits booked (M) & TRG (M) & PS (\%GDP) & PS (\%Profits booked) & TRG (\% tax revenue) & TRG (\% corp. tax revenue) & TRG (\%health) & TRG (\%education) \\
\midrule
Cayman Islands           &         130,276 &              136,292 &              197 &    2580.36 &         95.59 &                  &                       &                 &               \\
Singapore                &          59,602 &              121,237 &            3,132 &      18.01 &         49.16 &             7.03 &                 24.43 &           50.89 &         31.95 \\
Netherlands              &          54,029 &              114,845 &            3,257 &       6.30 &         47.04 &             1.65 &                 13.57 &            5.59 &          7.03 \\
Canada                   &          51,011 &              162,820 &            6,600 &       3.00 &         31.33 &             1.41 &                 11.15 &            4.96 &               \\
Switzerland              &          47,639 &              107,476 &            3,840 &       6.64 &         44.33 &             2.58 &                 18.33 &           14.77 &         10.54 \\
Hong Kong                &          46,727 &               91,630 &            3,640 &      14.64 &         51.00 &             8.07 &                 19.79 &                 &         32.77 \\
\end{longtable}
Notes: Countries acting as profit destination, estimated by the misalignment model. Profits booked reflect all sub-groups. PS stands for profit shifted, TRG stands for tax revenue gain. We compare PS with the country's GDP. We compare the TRG with the total tax revenue, the corporate tax revenue, and the public health and education expenditures.
}

\newpage
{\footnotesize
\begin{longtable}{p{2cm}p{1.5cm}p{1.5cm}p{1.5cm}p{1cm}p{1cm}p{1cm}p{1cm}p{1cm}p{1cm}}
\caption{Countries acting as profit origins, estimated by the misalignment model}\label{tab:lossers_mis}\\
\toprule
Country & Profits Shifted (M) & Profits booked (M) & Tax Revenue Loss (M) & PS (\%GDP) & PS (\%Potential base) & TRL (\% tax revenue) & TRL (\% corp. tax revenue) & TRL (\%health) & TRL (\%education) \\
\midrule
United States            &         287,971 &              954,310 &           85,258 &        1.52 &        23.18 &             2.32 &                 23.70 &            5.43 &          9.12 \\
Germany                  &          82,717 &              175,971 &           15,117 &        2.24 &        31.98 &             1.76 &                 21.74 &            4.77 &          8.33 \\
France                   &          82,377 &               92,626 &           20,619 &        3.10 &        47.07 &             2.70 &                 33.70 &            9.44 &         14.21 \\
Japan                    &          71,961 &              346,261 &           19,731 &        1.42 &        17.21 &             2.15 &                 10.03 &            4.28 &         11.28 \\
India                    &          33,269 &               70,523 &           13,885 &        1.43 &        32.05 &             3.51 &                 17.43 &           64.86 &         15.46 \\
Venezuela                &          21,079 &              -20,977 &            1,750 &        6.90 &     20634.27 &             3.52 &                 62.04 &           22.89 &            \\
\end{longtable}
\noindent Notes: Countries acting as profit origins, estimated by the misalignment model. Profits booked reflect all sub-groups. PS stands for profit shifted, TRL stands for tax revenue loss. We compare PS with the country's GDP. We compare the TRL with the total tax revenue, the corporate tax revenue, and the public health and education expenditures. 
}

\newpage
\footnotesize{
\begin{longtable}{p{2cm}p{1.5cm}p{1.5cm}p{1.5cm}p{1cm}p{1cm}p{1cm}p{1cm}p{1cm}p{1cm}}
\caption{Countries acting as profit destination, estimated by the logarithmic model}\label{tab:winners_log}\\
\toprule
Country & Profits Shifted (M) & Profits booked (M) & TRG (M) & PS (\%GDP) & PS (\%Profits booked) & TRG (\% tax revenue) & TRG (\% corp. tax revenue) & TRG (\%health) & TRG (\%education) \\
\midrule
United Kingdom                 &         190,160 &                           367,449 &           13,933 &       6.78 &         51.75 &             1.90 &                 18.96 &            6.25 &          9.02 \\
Cayman Islands                 &         133,045 &                           133,045 &              225 &    2635.21 &        100.00 &                  &                       &                 &               \\
Luxembourg                     &          83,360 &                            96,860 &            1,459 &     128.86 &         86.06 &             8.34 &                 46.49 &           49.82 &         57.88 \\
Singapore                      &          79,058 &                           126,182 &            4,154 &      23.89 &         62.65 &             9.33 &                 32.41 &           67.51 &         42.38 \\
Canada                         &          59,342 &                           179,683 &            7,678 &       3.49 &         33.03 &             1.64 &                 12.98 &            5.77 &               \\
Netherlands                    &          59,256 &                           143,699 &            3,573 &       6.91 &         41.24 &             1.81 &                 14.88 &            6.13 &          7.71 \\
\end{longtable}
Notes: Countries acting as profit destination, estimated by the logarithmic model. Profits booked reflect all sub-groups. PS stands for profit shifted, TRG stands for tax revenue gain. We compare PS with the country's GDP. We compare the TRG with the total tax revenue, the corporate tax revenue, and the public health and education expenditures.
}

\newpage
{\footnotesize
\begin{longtable}{p{2cm}p{1.5cm}p{1.5cm}p{1.5cm}p{1cm}p{1cm}p{1cm}p{1cm}p{1cm}p{1cm}}
\caption{Countries acting as profit origins, estimated by the logarithmic model}\label{tab:lossers_log}\\
\toprule
Country & Profits Shifted (M) & Profits booked (M) & Tax Revenue Loss (M) & PS (\%GDP) & PS (\%Potential base) & TRL (\% tax revenue) & TRL (\% corp. tax revenue) & TRL (\%health) & TRL (\%education) \\
\midrule
United States            &         259,254 &                           830,405 &           76,756 &        1.37 &        23.79 &             2.09 &                 21.34 &            4.89 &          8.21 \\
China                    &         139,394 &                           568,916 &           36,044 &        1.18 &        19.68 &             1.57 &                  7.99 &           10.82 &          8.71 \\
Japan                    &         104,214 &                           345,212 &           28,575 &        2.06 &        23.19 &             3.11 &                 14.53 &            6.20 &         16.33 \\
Germany                  &          51,268 &                           186,474 &            9,369 &        1.39 &        21.56 &             1.09 &                 13.48 &            2.96 &          5.17 \\
France                   &          45,782 &                           102,368 &           11,459 &        1.72 &        30.90 &             1.50 &                 18.73 &            5.25 &          7.90 \\
India                    &          28,086 &                            73,589 &           11,722 &        1.21 &        27.62 &             2.96 &                 14.71 &           54.75 &         13.05 \\
\end{longtable}
\noindent Notes: Countries acting as profit origins, estimated by the logarithmic model. Profits booked reflect all sub-groups. PS stands for profit shifted, TRL stands for tax revenue loss. We compare PS with the country's GDP. We compare the TRL with the total tax revenue, the corporate tax revenue, and the public health and education expenditures. 
}

\newpage
{\footnotesize
\begin{longtable}{p{2.5cm}p{1.5cm}p{1.5cm}p{1cm}p{1cm}p{1cm}p{1cm}p{1cm}p{1cm}p{1cm}p{1cm}}
\caption{Profits and effective tax rates}\label{tab:n_ETR_country}\\
\toprule
 Country  &  Profits (+) &  Profits (all groups) &  ETRa (wmean) &  ETRc (wmean) &  ETRa (median) &  ETRc (median) &  Number of reporters &  ETRa for (wmean) &  ETRc for (wmean) & N. reporters  \\
\midrule
Japan                          &     342,267 &              336,551 &          30.2 &          27.4 &            0.3 &            0.2 &                 19.0 &              22.8 &              21.9 &           198 \\
India                          &      73,594 &               64,902 &          36.0 &          41.7 &            0.3 &            0.4 &                 18.0 &              32.5 &              34.1 &           163 \\
Germany                        &     186,474 &              165,736 &          18.6 &          18.2 &            0.2 &            0.2 &                 18.0 &              20.3 &              20.0 &           155 \\
United States                  &     830,459 &              722,804 &          35.4 &          29.6 &            0.1 &            0.1 &                 23.0 &              15.8 &              15.9 &           141 \\
South Africa                   &      24,629 &               19,773 &          16.2 &          19.7 &            0.2 &            0.2 &                 18.0 &              19.4 &              31.6 &           138 \\
China                          &     571,000 &              566,574 &          24.1 &          25.9 &            0.2 &            0.2 &                 21.0 &              21.6 &              21.7 &           120 \\
\bottomrule
\end{longtable}
Notes: Profits (in USD million) reported by groups with positive profits (Profits (+)) and all groups (Profits (all groups)). Effective tax rates (ETRs) accrued (signified with `a') and cash-based (signified with `c'). Three types of ETRs were calculated, the weighted ETR by profits (ETRx (wmean)), the median (ETRx (median) and the weighted ETR by foreign profits (ETRx for (wmean). The number of countries disclosing data on the country (N. reporters) and the number of countries reported by the country (Reporting on). 
}

\begin{table}[h]
    \caption{Estimates of profits shifted and tax revenue loss worldwide}
    \label{tab:trl}
%     \begin{tabular}{lllll}
% \toprule
% {} & Profits shifted & TRL (total ETR) & TLR (foreign ETR) & TRL (CIT) \\
% \midrule
% Misalignment &         \$994 bn &         \$205 bn &           \$214 bn &   \$307 bn \\
% Logarithmic  &         \$965 bn &         \$186 bn &            \$200 bn &    \$300 bn \\
% \bottomrule
%     \end{tabular}
\begin{tabular}{lllll}
\toprule
{} & Profits shifted & TRL (total ETR) & TRL (foreign ETR) & TRL (CIT) \\
\midrule
Misalignment &           854 B &           214 B &             173 B &     238 B \\
Logarithmic  &           862 B &           226 B &             177 B &     234 B \\
\bottomrule
\end{tabular}

\begin{flushleft}
\footnotesize
Notes: Estimates of profits shifted and tax revenue loss (TRL) for the misalignment and logarithmic models. Three different tax rates are used, the total ETR (both domestic and foreign MNCs), the foreign ETR (only foreign MNCs), and the statutory tax rate (CIT).
\end{flushleft}
\end{table}

\begin{table}[]
\footnotesize
    \caption{Comparing estimates of profits shifted and tax revenue loss worldwide}
    \label{tab:trl_comparison}
    %\centering
%\resizebox{\textwidth}{!}{
\begin{tabular}{@{}p{5.4cm}p{1.7cm}p{1.5cm}p{2cm}p{1.5cm}p{1.5cm}p{1cm}@{}}
\toprule
Study                       & Profit shifting & Revenue loss & Data (type) & Individual countries & Countries (number) & Year (data) \\ \midrule
\textcite{cobhamGlobalDistributionRevenue2018}         & -         & 90         & Revenue           & Yes  & 102   & 2013     \\
IMF’s \textcite{crivelliBaseErosionProfit2016}   & -         & 123         & Revenue           & No  & 173 & 2013      \\
\textcite{keenSpilloversInternationalCorporate2014}                      & -         & 180         & Revenue & Yes   & 46   & 2012 \\
OECD's \textcite{johanssonTaxPlanningMultinational2017} & -         & 100-240     & Orbis             & No   & 46   & 2010   \\
\textcite{fuestGlobalProfitShifting2022}   & 271         & 104         & CBCR           & No  & - & 2019      \\
\textcite{janskyEstimatingScaleProfit2019}      & 420         & 125         & FDI               & Yes  & 79   & 2016    \\
UNCTAD’s \textcite{bolwijnEstablishingBaselineEstimating2018}                   & 700         & 200         & FDI               & No   & 72   & 2012  \\
\textcite{brattaAssessingProfitShifting2021} & 786         & 217     & CBCR             & No   & -   & 2017   \\
\textbf{This paper}       & \textbf{862-867} & \textbf{177-257}        & \textbf{CBCR}               & \textbf{Yes}  & \textbf{214}   & \textbf{2017}  \\
\textcite{torslovMissingProfitsNations2023} & 946         & 243         & FDI & Yes  & 57   & 2018   \\ 
\textcite{wierGlobalProfitShifting2022} & 969         & 247         & FDI & Yes  & 57   & 2019   \\ 
\textcite{clausingEffectProfitShifting2016}               & 1076         & 279        & FDI               & Yes   & 25   & 2012 \\
\textcite{taxjusticenetworkStateTaxJustice2021} & 1163-1334         & 312        & CBCR               & Yes   & 200   & 2017 \\
 \bottomrule
\end{tabular}

\footnotesize
Notes: Profit shifting and tax revenue loss are annual estimates expressed in billion USD (using average exchange rates from the year of the data if needed, as in the case of \citealp{brattaAssessingProfitShifting2021}). The studies are listed in the table according to the highest estimated tax revenue loss. Some studies estimate only either profit shifting or tax revenue loss due to profit shifting. We focus on those providing tax revenue losses, some of which do not provide estimates of profit shifting scale (e.g. due to the methodology as in the case of \textcite{crivelliBaseErosionProfit2016} and \textcite{cobhamGlobalDistributionRevenue2018}. The country coverage differs a lot across studies and we focus on those aiming at a worldwide coverage or covering many countries. We thus omit, for example, some studies that investigated only profit shifting by US-headquartered MNCs such as \textcite{guvenenOffshoreProfitShifting2022} or \textcite{cobhamMeasuringMisalignmentLocation2019}. On country coverage, see, for example, \textcite{janskyEstimatingScaleProfit2019} for a detailed comparison of selected studies. \textcite{keenSpilloversInternationalCorporate2014}  refers to the analysis in their Appendix IV. Using Gross Operating Surplus to Explore Spillovers (rather than Appendix III., which has been published as \citealp{crivelliBaseErosionProfit2016}). Authors of \textcite{torslovMissingProfitsNations2023} publish estimates for 2016 on their website. While we refer to profit shifting here, each of the studies has its nuances and its own concepts and definitions, for example, aggressive tax planning \citep{johanssonTaxPlanningMultinational2017} or excess income booked in low-tax countries \citep{clausingEffectProfitShifting2016}. The data column lists the type of the main data source a study relies on and, for example, FDI can indicate both country-level foreign affiliate statistics \citep{torslovMissingProfitsNations2023} as well as  bilateral data on FDI stocks \citep{janskyEstimatingScaleProfit2019}. The individual countries column indicates whether the results have been published for individual countries. The year column refers to the year of data used (or the last year of data in case the data are used for a period of time).
\end{table}

% \begin{figure}[H]
%   \centering
%   \includegraphics[width=1\textwidth]{Figures/diffs_aggs_income_gdp.pdf}
%  \caption{Profit shifted as a percentage of GDP.}
%   \label{fig:diffs_aggs_income_rev}

% \end{figure}

\clearpage

\subsection{Data-driven calculation of the effect of tax on profit shifting} \label{sec:appendix_eureqa}

An alternative method of profit shifting analysis adopts a model-agnostic approach. Instead of fitting the data to a pre-determined model, we can use symbolic regression to search for models that fit the data well. The search space (i.e. potential models) in symbolic regression is, however, infinitely large. The state-of-the-art method uses evolutionary algorithms – algorithms inspired by biological evolution, widely used in optimisation problems, to find the best model, balancing the fitness of the model with its complexity in order to avoid overfitting. These algorithms start with a pool of solutions (in this case models), which are \textit{re-combined} and \textit{mutated}, increasing the pool of solutions. The best solutions found in this augmented pool are \textit{selected} and allowed to be combined and mutated in the next generation. The algorithms efficiently explore the search space and reach a near-optimum solution. The main challenge in symbolic regression is how to score the solutions in order to find the best model, since using an error term alone will lead to finding highly complex models (overfitting). Here, we use the software \textit{Eureqa} \citep{schmidtDistillingFreeFormNatural2009}. In order to avoid overfitting, the software keeps the best model for each level of complexity, where the complexity is determined by the number of variables included and the operations (e.g. a log-transformation costs 4 units, an addition cost 1 unit) and the best model by the R\textsuperscript{2}. The model ``profits = 20'' is very simple, but has a very low R\textsuperscript{2}. A model with 20 variables interacting with each other may be able to fit the data perfectly, but will most likely result in data overfitting.

The model tries to find the best model for the location of profits, based on the following variables: revenue from unrelated party, employment, tangible assets, ETR, log(ETR), wages, population, gross fixed capital formation, consumption, GDP, tax complexity, statistical capacity, and revenue resources. The first five variables come from CBCR, wages are approximated as employment*GDP/population, the last variable from the ICTD / UNU-WIDER Government Revenue Dataset 2018, and the rest from the World Bank. We do not include revenue from related parties since it is heavily affected by profit shifted. Employees and tangible assets are less relevant to tax avoidance structures. All variables except for the last three are log-transformed. While the evolutionary algorithm can transform variables, that transformation is costly and seldom found.

We let the algorithm run for 20 hours using eight CPU cores, until it reached a ``Percent Coverage'' of 100\%; ``Percent Coverage is actually designed to replace the older stability and maturity metrics by providing an improved estimate of how close the search is to the plateau point where continuing the search will most likely not turn up any better solutions. It is based on the early stopping rule of thumb estimate and tracks how long has it been since any significant improvement on the validation data set.'' (Eureqa documentation).

\begin{figure}[ht!]
 \caption{A frequency of the variables found in the models}
  \includegraphics[width=0.5\textwidth]{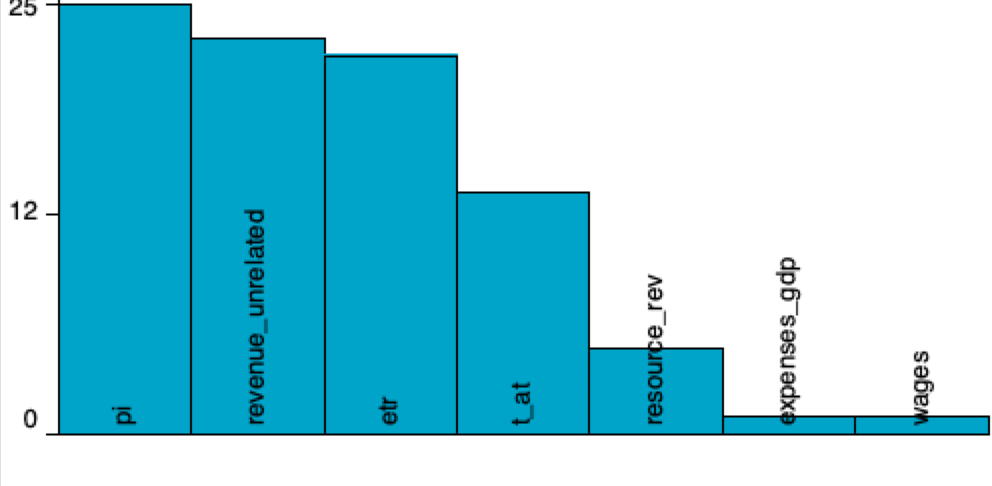}
  \label{fig:eureqa_variables}
\newline
\footnotesize
Notes: Variables found in the models. 25 different models were found, and unrelated party revenues, effective tax rates, and tangible assets were found in the majority of them.
\end{figure}

Figure ~\ref{fig:eureqa_variables} shows the variables included in at least one model. Our dependent variable (pi) is logically included in all models. Revenue from unrelated party and ETRs were almost always included, while tangible assets, revenue from natural resources, expenses/GDP and wages were only included in more complicated models. The absence of statutory corporate income tax in all models is perhaps surprising.

\begin{figure}[ht!]
 \caption{The relationship between the effective tax rates and profits}
  \includegraphics[width=0.6\textwidth]{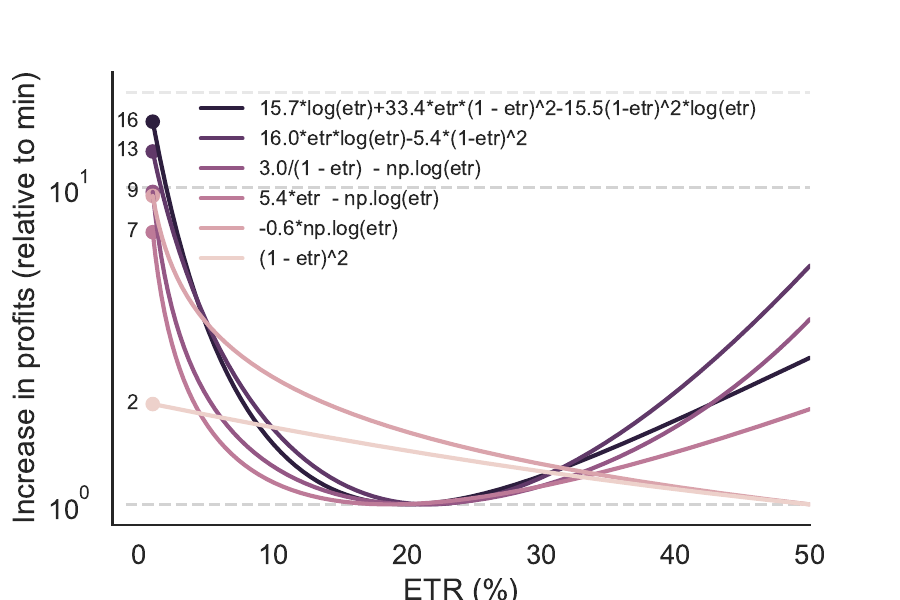}
  \label{fig:eureqa_tax}
\newline
\footnotesize
Notes: The found dependency between the ETR and profits is visualised. The complexity of the model is depicted with a number.
\end{figure}

Figure \ref{fig:eureqa_tax} shows the relationship between $\log{\pi}$ and the ETR. We see an extreme non-linear relationship for all models, independently of their complexity. In general, when the ETR is 1 per cent, the profits increase 6 to 14 times in comparison with the minimum. Interestingly, all models allowed for a U-curve, and the minimum effect found is achieved when ETR is 20 per cent. This could reflect the importance of resource-rich countries.

\clearpage

\subsection{Sensitivity analysis, additional information} \label{sec:appendix_sensit}

\subsubsection{(iii) Comparison with \textcite{torslovMissingProfitsNations2023}}

\begin{figure}[ht!]
 \caption{Comparison with the estimates of profit shifting by \textcite{torslovMissingProfitsNations2023}}
  \includegraphics[width=.9\textwidth]{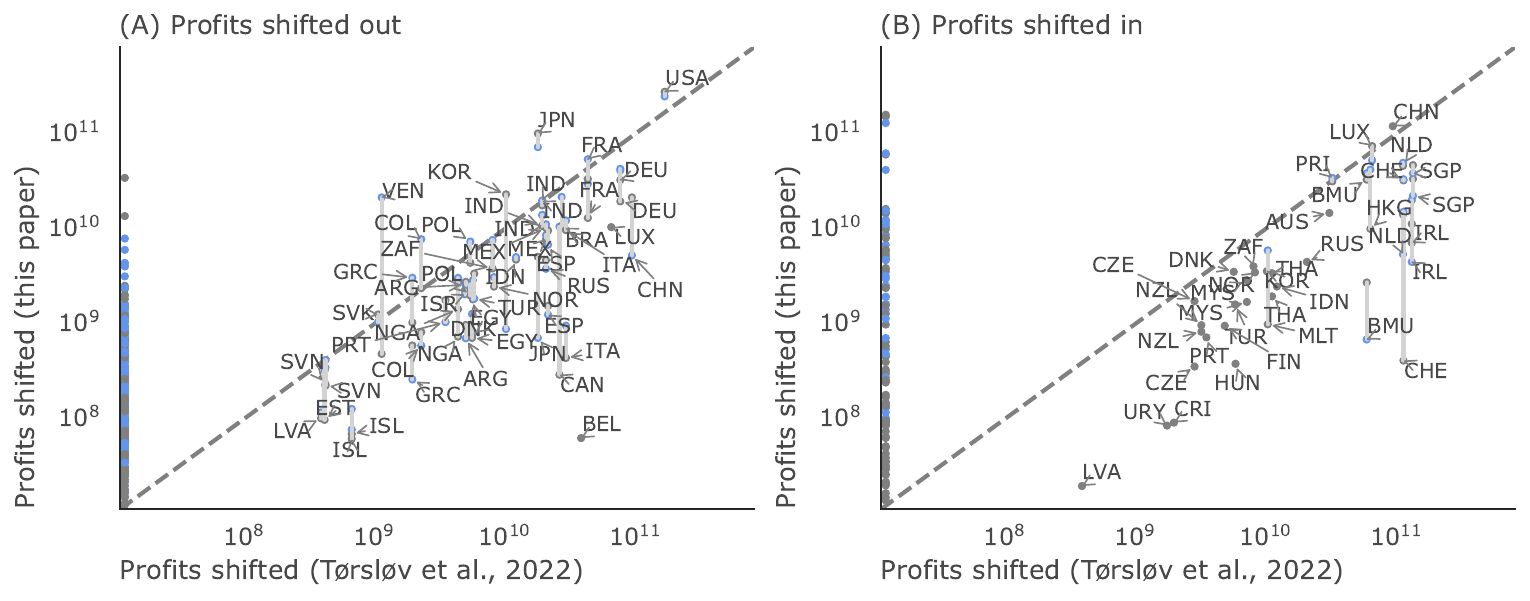}
  \label{fig:torslov}
\newline
\footnotesize
Notes: Comparison with the estimates of profit shifting by \textcite{torslovMissingProfitsNations2023}. Countries not included in the sample of \textcite{torslovMissingProfitsNations2023} appear in the left of the plots. Estimates of profits shifted using the misalignment model are visualised in blue. Estimates using the logarithmic model are visualised in grey. A light grey line is used to connect the blue and grey points corresponding to the same country.
\end{figure}

\subsubsection{(vi) Sensitivity to the offset in the logarithmic model}

\begin{figure}[ht!]
 \caption{Effect of the parameter $t$ in the logarithmic model}
  \includegraphics[width=1\textwidth]{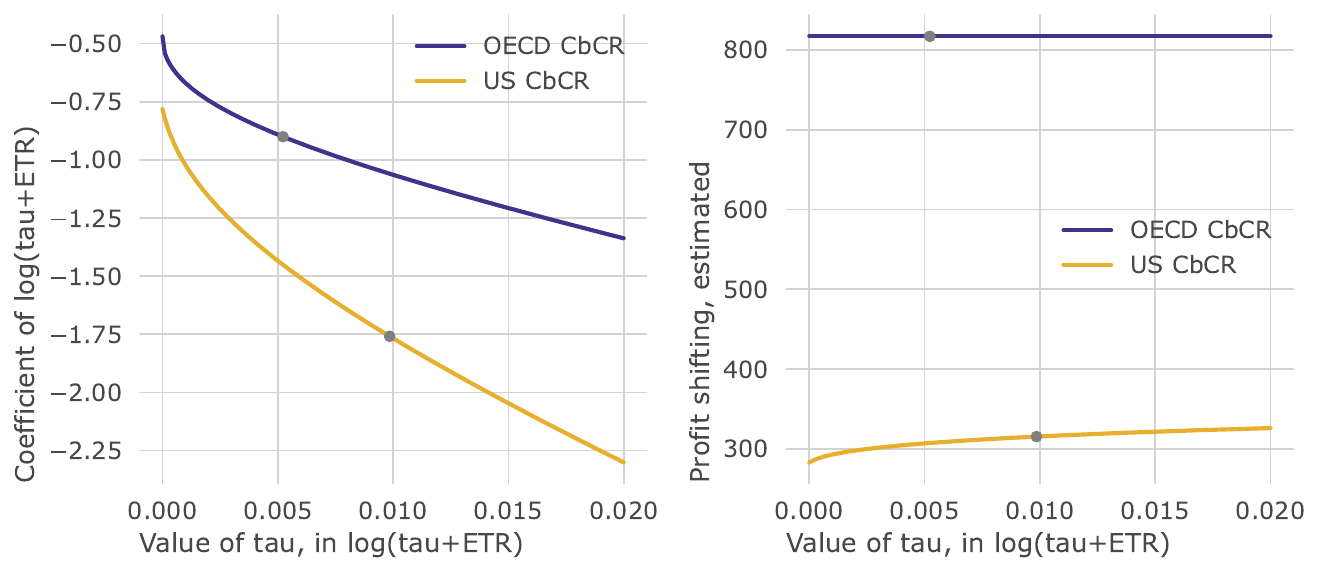}
  \label{fig:offset_robust}
\footnotesize
Notes: Effect of the parameter $t$ in the logarithmic model. (A) The coefficient of $log(t+ETR)$ is affected by the election of $t$. (B) The effect of $t$ on our estimate of  total profits shifted is small. Dots indicate the parameter and results of this paper.
\end{figure}

\newpage
\subsubsection{(vii) Other non-linear functions}

\begin{figure}[ht!]
 \caption{Graphical representation of Table \ref{tab:app:robustness_log} for the $log(\tau+ETR)$, $1/(\tau+ETR)$, $1/(\tau+ETR)^2$,$1/(\tau+ETR)^3$ and $coth(\tau+ETR)$ models}
  \includegraphics[width=1\textwidth]{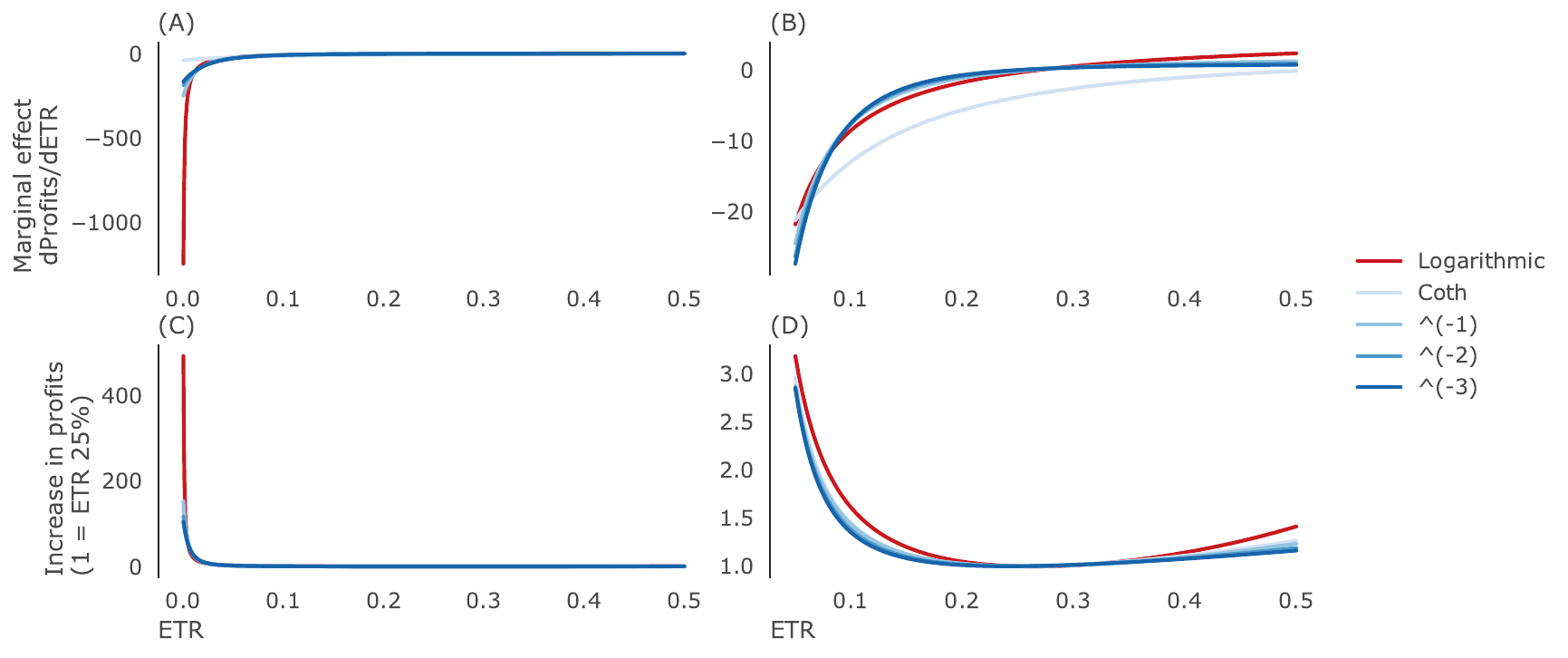}
  \label{fig:elasticity_robustness}
\footnotesize
Notes: Graphical representation of Table \ref{tab:app:robustness_log} for the $log(\tau+ETR)$, $1/(\tau+ETR)$, $1/(\tau+ETR)^2$,$1/(\tau+ETR)^3$ and $coth(\tau+ETR)$ models. Semi-elasticities calculated using US data. The $\tau$ offset is calculated independently for all models. (A, B) Marginal effect of ETR on profits. (C,D) Relative increase in profits due to profit shifting, compared with a country with an ETR of 25\%. Plots B and D are close-ups of plots A and C, constraining ETRs between 5 and 50\%.
\end{figure}

\begin{table}[ht!]
\caption{Regression table using the US data and applying other specifications that allow for extreme non-linearities}
\label{tab:app:robustness_log}
\begin{center}
\begin{tabular}{llllll}
\hline
                                        &    $log(ETR)$     &   $1/(ETR)$   &   $1/(ETR)^2$   &   $1/(ETR)^3$  &     $coth(ETR))$    \\
\midrule
$1/(0.024 + ETR)$ &             &  0.1445*** &            &            &            \\
                                       &             &   (0.0204) &            &            &            \\
$1/(0.056 + ETR)^2$ &             &            &  0.0164*** &            &            \\
                                       &             &            &   (0.0023) &            &            \\
$1/(0.090 + ETR)^3$ &             &            &            &  0.0036*** &            \\
                                       &             &            &            &   (0.0005) &            \\
$(oth(0.026 + ETR)$    &             &            &            &            &  0.1563*** \\
\bottomrule
\end{tabular}
\end{center}
\end{table}
\footnotesize
Notes: Regression table using the US data and applying other specifications that allow for extreme non-linearities. A graphical visualization is presented in Fig. \ref{fig:elasticity_robustness}.

\begin{figure}[ht!]
 \caption{Graphical representation of Table \ref{tab:app:robustness_log} for the model with extra dummy variables for categories of effective tax rates}
  \includegraphics[width=1\textwidth]{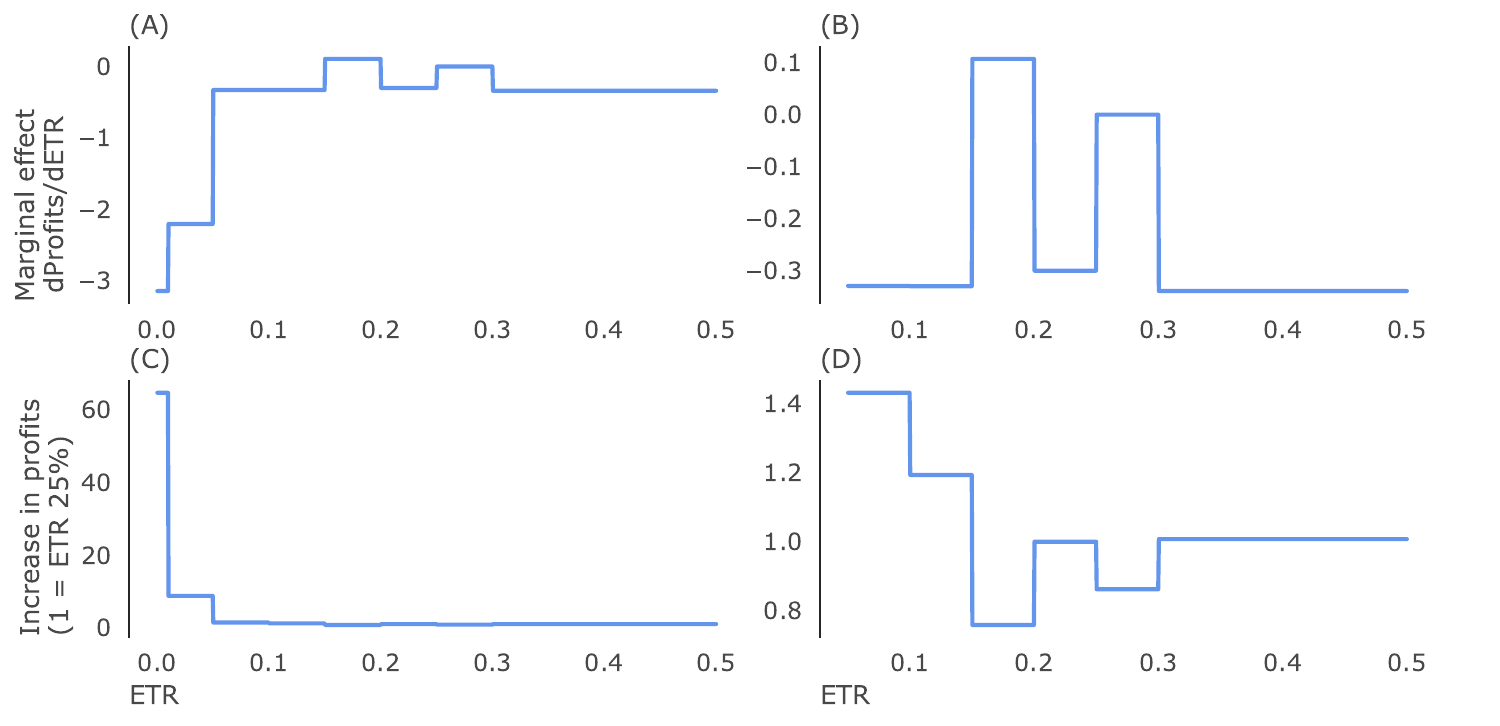}
  \label{fig:elasticity_robustness_dummies}
\footnotesize
Notes: Graphical representation of Table \ref{tab:app:robustness_log} for the model with extra dummy variables for the following categories of ETRs: <1\%,1-5\%,5-10\%,10-15\%,15-20\%,20-25\%,25-30\%,>30\%.  Semi-elasticities are calculated using US data. (A, B) Marginal effect of ETR on profits. (C,D) Relative increase in profits due to profit shifting, compared with a country with an ETR of 25\%. Plots B and D are close-ups of plots A and C, constraining ETRs between 5 and 50\%.
\end{figure}

\clearpage
\subsubsection{(viii) Effect of the redistribution formula }

\begin{figure}[ht!]
 \caption{Fit of the robust model}
  \includegraphics[width=0.6\textwidth]{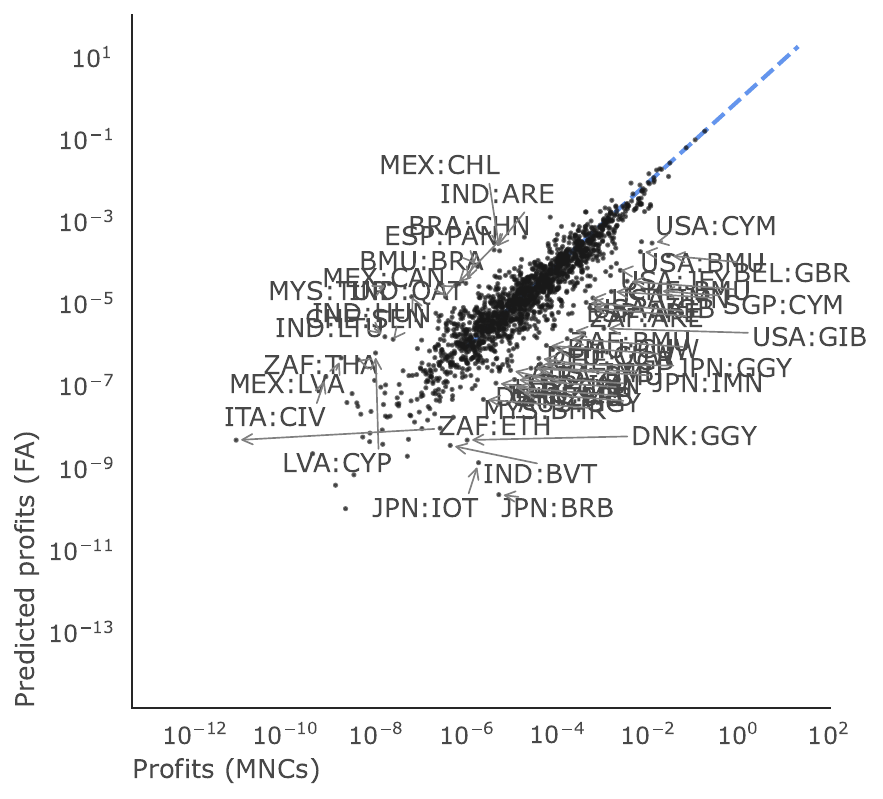}
  \label{fig:robustness_fit_formula}
\footnotesize
Notes: Each dot correspond to an observation (reporting country : partner country). Dotted line represent the fit of a robust linear model. Outliers are annotated. Note the large presence of tax havens among the outliers.
\end{figure}

\begin{table}[h!]
\caption{Results of a robust linear model, where the share of profits booked in a country is regressed against the shares of employees, capital, sales and wages}\label{tab:red_robust}
\begin{tabular}{lc}
\hline
             &     Share of profits     \\
\midrule
Share of employees   & 0.0733***  \\
             & (0.0005)   \\
Share of capital & 0.2683***  \\
             & (0.0004)   \\
Share of sales  & 0.4372***  \\
             & (0.0004)   \\
Share of wages & 0.0607***  \\
             & (0.0002)   \\
N            & 2199       \\
\hline
\end{tabular}
\end{table}
\footnotesize
\noindent Notes: Results of a robust linear model, where the share of profits booked in a country is regressed against the shares of employees, capital, sales and wages, using the dataset created for the misalignment method. See Fig. \ref{fig:robustness_fit_formula}.

\begin{figure}[ht!]
 \caption{Comparison of the main results and the results using a different reditribution formula}
  \includegraphics[width=1\textwidth]{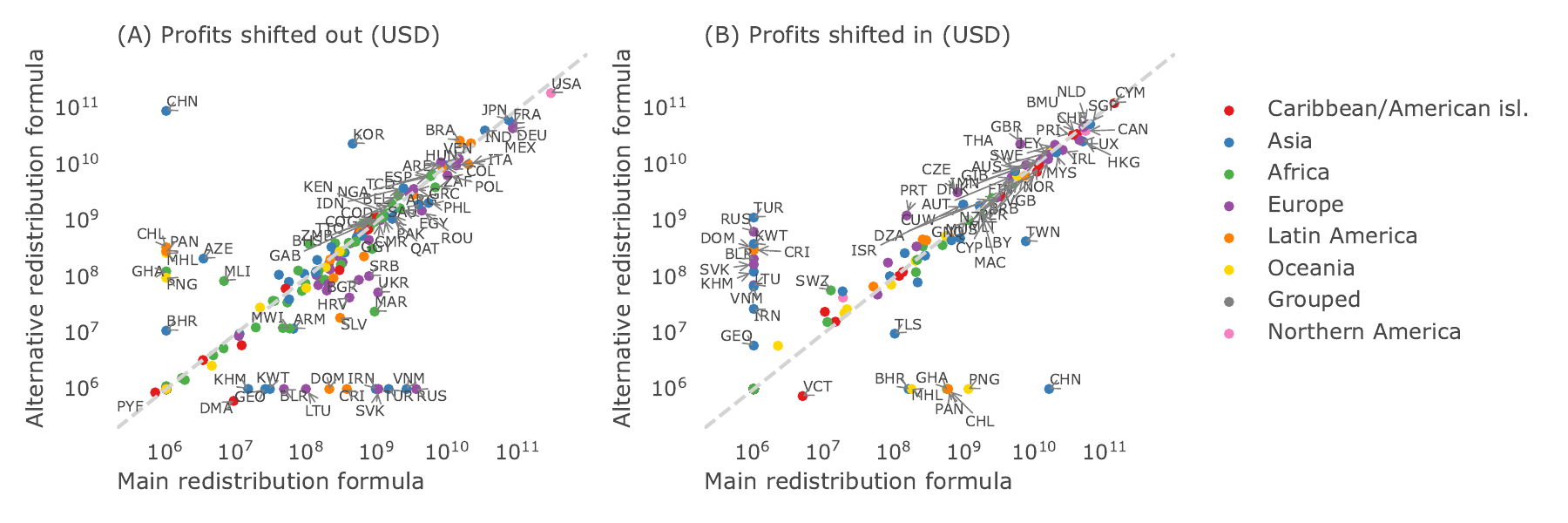}
  \label{fig:Origin_profits_OECD_ut_weights}
\footnotesize
Notes: Comparison of profits shifted out (A) and profits shifted in (B) for the misalignment models and the OECD data using the redistribution formula of the main results and the redistribution formula from Table \ref{tab:red_robust}. Each dot represents a country, coloured by region. 
\end{figure}

\subsubsection{Regressions using statutory corporate income tax rates (US data)} \label{sec:cit_etr}

\begin{table}[ht!]
\caption{Regressions using statutory corporate income tax rates}
\label{tab:cit_etr_us}

\begin{center}
\begin{tabular}{lllll}
\toprule
{} &  Logarithmic (CIT) & Quadratic (CIT) & Log + Quad (CIT) &  Linear (CIT) \\
\midrule
I(CIT ** 2)     &            &    10.1173 &           3.9315 &            \\
                &            &   (7.5504) &        (13.7265) &            \\
Intercept       &     2.9519 &   5.1495** &           3.6572 &   4.7595** \\
                &   (2.4645) &   (2.1322) &         (3.4936) &   (2.1223) \\
R-squared       &     0.8135 &     0.8130 &           0.8137 &     0.8090 \\
R-squared Adj.  &     0.8000 &     0.7995 &           0.7978 &     0.7976 \\
CIT             &     2.4111 &    -5.7256 &          -0.4890 &    -0.7712 \\
                &   (2.6177) &   (3.9366) &        (10.4620) &   (1.3576) \\
log(0.02 + CIT) &    -0.6100 &            &          -0.4231 &            \\
                &   (0.4299) &            &         (0.7826) &            \\
log(Population)    &    -0.1368 &    -0.1425 &          -0.1365 &   -0.1668* \\
                &   (0.1017) &   (0.1013) &         (0.1023) &   (0.1001) \\
log(GDPpc)      &     0.1237 &     0.1072 &           0.1173 &     0.1236 \\
                &   (0.1636) &   (0.1642) &         (0.1660) &   (0.1646) \\
log(Tangible assets)       &  0.6697*** &  0.6867*** &        0.6734*** &  0.7048*** \\
                &   (0.0887) &   (0.0864) &         (0.0902) &   (0.0857) \\
log(wages)      &    0.1826* &    0.1739* &          0.1814* &     0.1564 \\
                &   (0.1032) &   (0.1025) &         (0.1039) &   (0.1022) \\
N               &         90 &         90 &               90 &         90 \\
R2              &      0.814 &      0.813 &            0.814 &      0.809 \\
BIC             &     268.46 &     268.69 &           272.87 &     266.12 \\
\bottomrule
\end{tabular}
\end{center}
\end{table}

\newpage
\subsubsection{Regressions using statutory corporate income tax rates (OECD data)} \label{sec:cit_etr_oecd}

{\footnotesize
\begin{longtable}{llllll}
\caption{Regressions using statutory corporate income tax rates and OECD data}\label{tab:cit_etr_oecd}\\
\toprule
{} & Logarithmic &     Quadratic & Log*FE + Quad & Log + Quad*FE &       Linear \\
\midrule
CIT                    &     1.4400* &    -6.0996*** &     6.5629*** &       4.9875* &   -0.8741*** \\
                       &    (0.7389) &      (1.4248) &      (1.9323) &      (2.5680) &     (0.3089) \\
CIT\textsuperscript{2} &             &    12.9445*** &    -8.7991*** &       -3.8842 &              \\
                       &             &      (3.0646) &      (2.5242) &      (4.0846) &              \\
log(0.0005 + CIT)      &  -0.2870*** &               &    -0.3942*** &    -0.3273*** &              \\
                       &    (0.0422) &               &      (0.0579) &      (0.0856) &              \\
\bottomrule
\end{longtable}

Notes: The CIT was imputed using ETR for the following countries: Anguilla, Antigua and Barbuda, Cuba, Djibouti, French Guiana, Guadeloupe, Haiti, Kiribati, Kosovo, Kyrgyz Republic, Sao Tome and Principe, St. Lucia, St. Vincent \& Grenadines, Syria, Turkmenistan, Turks and Caicos Islands
}

\end{document}